\documentclass[12pt]{article}
\usepackage{amsmath}
\usepackage{graphicx}
\usepackage{enumerate}
\usepackage{natbib}
\usepackage{url} % not crucial - just used below for the URL 
\usepackage{times}
\usepackage{bm}
\usepackage{mathtools}
\usepackage{multirow}
\usepackage{amsfonts}
\usepackage{mathrsfs}
\usepackage{xcolor}
\usepackage{subfigure}
\usepackage{float}
\usepackage[normalem]{ulem}
\usepackage[plain,noend]{algorithm2e}
\def\T{{ \mathrm{\scriptscriptstyle T} }}

\DeclareMathOperator*{\argmax}{argmax}
\newcommand{\ind}{\perp\!\!\!\!\perp} 
\newcommand*{\QEDB}{\hfill\ensuremath{\square}}
\usepackage{hyperref}
\hypersetup{
	hidelinks
}

\usepackage{theorem}
\theoremstyle{definition}
\newtheorem{assumption}{Assumption}\newtheorem{theorem}{Theorem}\newtheorem{remark}{Remark}

\newcommand{\blind}{1}

\allowdisplaybreaks[2]

\begin{document}

\def\spacingset#1{\renewcommand{\baselinestretch}%
{#1}\small\normalsize} \spacingset{1}

%%%%%%%%%%%%%%%%%%%%%%%%%%%%%%%%%%%%%%%%%%%%%%%%%%%%%%%%%%%%%%%%%%%%%%%%%%%%%%

\if1\blind
{
  \title{\bf Targeted Optimal Treatment Regime Learning Using Summary Statistics}
  \author{Jianing Chu \thanks{jchu3@ncsu.edu} , Wenbin Lu \thanks{wlu4@ncsu.edu} , and Shu Yang \thanks{syang24@ncsu.edu}\\ Department of Statistics, North Carolina State University}
  \date{}
  \maketitle
} \fi

\if0\blind
{
  \bigskip
  \bigskip
  \bigskip
  \begin{center}
    {\LARGE\bf Title}
\end{center}
  \medskip
} \fi

\bigskip
\begin{abstract}
Personalized decision-making, aiming to derive optimal treatment regimes based on individual characteristics,  has recently attracted increasing attention in many fields, such as medicine, social services, and economics. Current literature mainly focuses on estimating treatment regimes from a single source population. In real-world applications, the distribution of a target population can be different from that of the source population. Therefore, treatment regimes learned by existing methods may not generalize well to the target population. Due to privacy concerns and other practical issues,  individual-level data from the target population is often not available, which makes treatment regime learning more challenging. We consider the problem of treatment regime estimation when the source and target populations may be heterogeneous,  individual-level data is available from the source population, and only the summary information of covariates, such as moments, is accessible from the target population. We develop a weighting framework that tailors a treatment regime for a given target population by leveraging the available summary statistics. Specifically, we propose a calibrated augmented inverse probability weighted estimator of the value function for the target population and estimate an optimal treatment regime by maximizing this estimator within a class of pre-specified regimes.  We show that the proposed calibrated estimator is consistent and asymptotically normal even with flexible semi/nonparametric models for nuisance function approximation, and the variance of the value estimator can be consistently estimated. We demonstrate the empirical performance of the proposed method using simulation studies and a real application to an eICU dataset as the source sample and a MIMIC-III dataset as the target sample.
\end{abstract}

\noindent%
\vfill

\newpage
\spacingset{1.8} % DON'T change the spacing!
\section{Introduction}
\label{sec:intro}

Personalized decision-making, a pseudo intelligence paradigm tailored to an individual's characteristics, has recently attracted a great deal of attention in many fields, such as precision medicine,  social services, economics,  and recommendation system. An individualized treatment rule (ITR) formalizes treatment decisions as a function mapping from patient information to a recommended treatment. An optimal ITR is the one that leads to the greatest expected outcome in the population of interest, known as the value function. 

A variety of approaches have been developed for estimating optimal ITRs. One class of approaches is model-based as they directly model the conditional mean outcome given covariates and treatment, known as the Q-function,  and then use the estimated Q-function to infer the optimal ITR. Such methods include Q-learning \citep{qian2011performance} 
and its semiparametric extension, A-leaning \citep{murphy2003optimal}, 
where only the contrast function is modeled while the baseline mean function is completely unspecified. Alternatively, direct value search methods have been developed and extensively studied recently \citep[e.g.][]{zhang2013robust,luckett2020estimating, athey2021policy}. 
These methods learn the optimal treatment regime by regime evaluation. They first establish a flexible estimator of the value function, such as the augmented inverse probability weighted (AIPW) estimator, and the optimal ITR is then estimated by maximizing the estimated value function within a class of pre-specified ITRs, such as linear decision rules and tree-based decision rules. The AIPW value estimator possesses the double robustness property, i.e., it is consistent for the value function if either the Q-function or the propensity score model is correctly specified. 

Though the double robustness of the AIPW value estimator is appealing, it's only maintained when the source and target populations are identical. In other words, when there exists heterogeneity between the source and target populations, the AIPW value estimator obtained based on the source sample may no longer be consistent for the value function of the target population. Thus, the optimal ITR learned from the source data may not be optimal for the target population. In many real-world applications, the value function of an ITR over the distribution of the target population is of significant interest, which can be different from that of the source population. For example, in medical studies, it is known that the results of a randomized controlled trial  cannot be directly transported because the covariate distribution in a target population may be different \citep{cole2010generalizing}. Due to study design and inclusion/exclusion criteria, the source sample can be unrepresentative of the target population we are interested in. When there is heterogeneity between the source and target populations, an estimated optimal ITR from the source sample may not generalize well to the target population \citep{lee2021improving}. Such problems gain increasing attention in the ITR learning fields recently. \cite{zhao2019robustifying} and \cite{mo2021learning} proposed different collections of possible target covariate distributions and estimated the optimal ITR by optimizing the worst-case quality assessment among the collection. \cite{uehara2020off} considered a nonparametric estimator for the density ratio of the covariate distributions of the source and target populations and constructed a weighted estimator for the target value function based on the estimated density ratio. However, all these methods require the availability of individual-level data from both the source and target populations, which may be unrealistic in many applications. For example, while large-population based databases, such as the Surveillance, Epidemiology and End Results database, can provide reliable summary statistics for covariates, such as means and medians, and overall survival statistics for the disease population, critical information about individual factors that influence the choice of treatment and clinical outcomes of interest may not be available \citep{huang2020unified,chen2021combining}. Moreover, due to privacy and confidentiality concerns, comprehensive individual-level data is often prohibited to share with researchers. In contrast, summary statistics of patient characteristics of the target population are often available and can be easily shared for research purposes.

In this paper, we consider the targeted optimal treatment regime learning where we have individual-level data from the source sample but only a few summary statistics of covariate distributions from the target population. As we alluded to previously, when there is heterogeneity in covariate distributions between the source and target populations, the estimated optimal ITR obtained by maximizing the value estimator constructed based on the source sample may not be optimal for the target population. One way to address this issue is to assign different subject weights to the source sample and calibrate the source covariate distributions to the target covariate distributions. Calibration weighting is widely used to integrate auxiliary information in survey sampling and causal inference, such as empirical likelihood based methods \citep{qin2007empirical}, entropy-based covariate balancing methods \citep{hainmueller2012entropy}, and quadratic loss based covariate balancing methods \citep{zubizarreta2015stable}. Such weighting methods allow adjusting covariate distributions of the source sample using various summary statistics of covariates in the target population, such as means, variances, correlations, and quantiles. We propose a calibrated AIPW estimator of the value function using summary statistics from the target population and then search for the optimal ITR for the target population by maximizing the calibrated AIPW value estimator over a pre-specified class of ITRs. Here, the subject weights for the source sample are estimated by solving a general convex optimization problem with constraints. The objective function in the optimization problem can be chosen from the Cressie-Read family \citep{cressie1984multinomial}, while the constraints force the weighted summary statistics of source covariates to be the same as that from the target population. We show that the calibrated AIPW estimator for the target value function is consistent, asymptotically normal, and has the double robustness property if the estimated weight function converges to the density ratio of covariate distributions between the two populations. 
The double robustness entails that the value estimator remains root-$n$ consistent if any one of the two parametric models for the propensity score and outcome mean is correctly specified or if both models are estimated nonparametrically satisfying a certain rate condition for convergence.   
Interestingly, if the source and target populations have the same covariate distribution, the calibrated optimal value estimator gains efficiency over the uncalibrated one by utilizing additional summary information. 
However, in general, the weights learned from the calibration methods may not consistently estimate the density ratio. Under such general cases, the proposed calibrated AIPW estimator can still converge to the value function of a pseudo population that may be closer to the target distribution compared with the source population. As such, it can give a more accurate estimator for the value function of the target population than the uncalibrated value estimator, and the optimal ITR obtained by maximizing the calibrated AIPW value estimator can be more favorable for the target population. 

\section{Statistical Framework}
\label{sec:framework}
\subsection{Value Function and Optimal ITR}

In a randomized trial or observational study, suppose there are two treatment options, labeled as control/treatment $0$ and experimental treatment/treatment $1$. Let $A$ taking values $0$ or $1$ in accordance with the two options, denote the treatment received. Let $X\in \mathbb{R}^{p}$ be a vector of baseline covariates and $Y$ be the observed outcome of interest. We assume larger values of $Y$ are preferred by convention. The observed data are then $\{O_i=(Y_i,A_i,X_i), i=1,\dots,n\}$, which are independent and identically distributed. 
Define the potential outcomes $Y^*(0)$ and $Y^*(1)$ as the outcomes that would be observed if a subject received treatment $0$ or $1$, respectively. As is customary in causal inference \citep{rubin1978bayesian}, we make the following assumptions. 
\begin{assumption}\label{assump:causal}
	(A1) $Y=Y^*(1)A+Y^*(0)(1-A)$, (A2) $\{Y^*(1),Y^*(0)\} \ind A \mid X$, and (A3) $0< {\rm pr}(A = 1\mid X = x)<1$ for all $x$ such that ${\rm pr}(X=x)>0$.	
\end{assumption}
An ITR is a function $d(\cdot)$ that maps values of $X$ to $\{0,1\}$, so that a subject with covariates value $X=x$ would receive treatment $1$ if $d(x)=1$ and treatment $0$ if $d(x)=0$. For any  arbitrary ITR $d(\cdot)$, we can define the potential outcome as 
$Y^*(d) = Y^*(1)d(X) + Y^*(0)\{1-d(X)\},$ 
which would be observed if a randomly chosen individual had been assigned a treatment according to $d(\cdot)$, where we suppress the dependence of $Y^*(d)$ on $X$.  We then define the value function under $d(\cdot)$ as the expectation of the potential outcome as $V(d) = {E}\{Y^*(d)\}={E}\left[Y^*(1)d(X) + Y^*(0)\{1-d(X)\}\right].$

Suppose $\mathcal{D}$ is a class of ITRs of interest. Then we define the optimal ITR as $d^{\rm{opt}}(X) = \argmax_{d\in \mathcal{D}} V(d)$. In clinical practice, it may be desirable to consider a class of ITRs indexed by a vector of parameters $\beta$ for feasibility and interpretability. We denote such a class of rules as $\mathcal{D}_{\beta}$ and its element as $d(X;\beta)$. For example, we can consider a class of linear ITRs \{$d(X;\beta)={I}(\beta^{\T}\tilde{X}>0):\beta\in\mathbb{R}^{p+1},\|\beta\|_2=1\}$, where $\tilde{X}=(1,X^{\T})^{\T}$. Given a linear ITR $d(X;\beta)$, we use a shorthand to write its value function $V(d)$ as $V(\beta)$. Let $\beta^*=\argmax_{\beta} V(\beta)$. Then, the optimal linear ITR is $d_{\beta}^{\rm{opt}}=d(X;\beta^*)$. The true optimal ITR $d^{\rm{opt}}$ may not be in $\mathcal{D}_{\beta}$. Thus, $d^{\rm{opt}}_\beta$ may not be the same as $d^{\rm{opt}}$. However, when attention focuses on the feasible class $\mathcal{D}_\beta$, estimation of $d^{\rm{opt}}_\beta$ is still of considerable interest. In this paper, we focus on linear ITRs. 

\subsection{Source and Target Populations}
The difference between covariate distributions in the source and target populations is called a covariate shift \citep{sugiyama2012machine}. 
In this paper, we assume that there is a pooled population $\mathbb{P}$ consisting of both the source population $\mathbb{P}^{\rm{s}}$ and the target population $\mathbb{P}^{\rm{t}}$. Let $S$ be a binary indicator for selection action: $S=1$ if the individual comes from the source population and $S=0$ if the individual comes from the target population. A covariate shift results from the situation where ${\rm pr}(S=1\mid X)\not = {\rm pr}(S=0\mid X)$. 

For the source population $\mathbb{P}^{\rm{s}}$, we observe individual-level data $\{O_i=(Y_i,A_i,X_i), i=1,\dots,n\}$. For the target population $\mathbb{P}^{\text{t}}$, only summary statistics of covariate distributions, such as mean, variance or quantiles are available. For $\mathbb{P}^{\rm{s}}$, denote the density or probability mass function of covariates as $f^{\rm{s}}(X)$ and its associated expectation as ${E}$; for $\mathbb{P}^{\text{t}}$, we use the notation  $f^{\text{t}}(X)$  and ${E}^{\text{t}}$ correspondingly. The summary statistics from the target population $\mathbb{P}^{\rm{t}}$ are denoted as $\mu_{g0}= {E}^{\rm{t}}\{g(X)\}$, where $g(X)= \left\{g_1(X),g_2(X),\dots,g_q(X)\right\}^{\T}$ is a $q\times1$ specified function. For example, a common choice is $g(X)=(X_1,X_2,\dots,X_p)$, and $\mu_{g0}$ gives the mean of all covariates in the target population. We assume that summary statistics from the target population are derived from large databases so that their uncertainty are negligible. With only the summary statistics, targeted ITR learning is impossible without further assumptions in order to borrow information from the source sample.  
\begin{assumption}\label{assump:transport}
	(A4) ${E}\{Y(a)\mid X\}={E}^{\rm t}\{Y(a)\mid X\}$, and 
	(A5) $ {\rm pr}(S=1\mid X) >0$.
\end{assumption}

Assumption (A4) implies that true Q-functions are identical in both the source and target populations \citep{dahabreh2019generalizing}. A stronger version of this assumption is the ignorability assumption that $\{Y(1),Y(0)\}\ind S \mid X$ \citep{buchanan2018generalizing}. Assumption (A5) implies that the support of the target $X$ distribution must be covered by the support of the source $X$ distribution. 

\section{Proposed Method}
\label{sec:method}
\subsection{Calibrated AIPW Estimator}
For the source population, define the propensity score as $\pi(X)={\rm pr}(A=1\mid X,S=1)$ and 
conditional mean outcome model as $\mu(X,A)={E}(Y\mid X,A)$.
In practice, $\pi(\cdot)$ and $\mu(\cdot)$ can be estimated from the observed source data based on some posited parametric models $\pi(X;\eta)$ and $\mu(X,A;\theta)$, respectively. Alternatively, they can also be estimated nonparametrically, e.g. using kernel regression or random forest. Given an ITR $d(X;\beta)$, \cite{zhang2012robust} proposed an AIPW estimator of the value function $V(\beta)$ as
$$\widehat{V}^{\rm{o}}(\beta)=\frac{1}{n}\sum\limits_{i=1}^n\left[\frac{{I}\{A_i=d(X_i;\beta)\}}{\varrho(A_i\mid X_i;\widehat{\eta})}\{Y_i-\mu_d(X_i;\beta,\widehat{\theta})\}+\mu_d(X_i;\beta,\widehat{\theta})\right],$$
where the superscript $\rm o$ is a shorthand for original,   $\varrho(A_i\mid X_i;\widehat{\eta})=\pi(X_i;\widehat{\eta})A_i+\{1-\pi(X_i;\widehat{\eta})\}(1-A_i)$, $\mu_d(X_i;\beta,\widehat{\theta})=\mu(X_i,1;\widehat{\theta})I\{d(X_i;\beta)=1\}+\mu(X_i,0;\widehat{\theta})I\{d(X_i;\beta)=0\}$, and $\widehat{\eta}$ and $\widehat{\theta}$ are the estimates of $\eta$ and $\theta$, respectively, based on some posited parametric models.  

Denote the value function of the target population under the ITR $d(X;\beta)$ as $V^{\rm{t}}(\beta)$. If the source and target populations have the same covariate distributions, i.e. $f^{\rm{s}}(X)=f^{\text{t}}(X)$, then $V^{\rm{t}}(\beta)$ can be consistently estimated by $\widehat{V}^{\rm{o}}(\beta)$  based on the source sample. However, since covariate distributions between the source and target populations often differ in practice,  $\widehat{V}^{\rm{o}}(\beta)$ may be biased for $V^{\rm{t}}(\beta)$. To reduce the bias, a natural approach is to consider calibration weights, i.e., to assign different weights to individual data points in the source sample so that the weighted data is more representative of the target distribution. Specifically, we consider the following calibrated AIPW estimator 
$$\widehat{V}^{\rm{c}}(\beta)=\sum\limits_{i=1}^nw_i\left[\frac{{I}\{A_i=d(X_i;\beta)\}}{\varrho(A_i\mid X_i;\widehat{\eta})}\{Y_i-\mu_d(X_i;\beta,\widehat{\theta})\}+\mu_d(X_i;\beta,\widehat{\theta})\right],$$ 
where the superscript $\rm c$ is a shorthand for calibrated, $w_i$'s are calibration weights satisfying $\sum_{i=1}^n w_i = 1$ and other constraints. Using the summary statistics $\mu_{g0}$ from the target population, we can utilize methods, such as empirical likelihood \citep{qin2007empirical} and entropy balancing \citep{hainmueller2012entropy,zhao2017entropy}, to learn the weights. In the next section, we propose a general framework to estimate the weights.

\subsection{Weights Estimation}
Let $h\left(w\right)$ denote a generic convex distance function between a scalar $w$ and ${n}^{-1}$. We consider the following optimization problem 
$$\min_{w_1,\ldots,w_n}\sum_{i=1}^n h\left(w_i\right),$$
under the constraints
$\sum_{i=1}^n w_i \{g(X_i)-\mu_{g0}\}=0$, $ \sum_{i=1}^n w_i=1$, and $w_i\geq0.$

In the considered optimization problem, the function $h(w)$ plays a role in quantifying the discrepancy of calibration weights and the uniform distribution ${n}^{-1}$. We choose the function $h(w)$  from the Cressie-Read family of discrepancies \citep{cressie1984multinomial}. 
The Cressie-Read family is defined through a class of additive convex functions that encompasses a broad family of distance functions. 
Specifically, 
\[
CR(\gamma)=\sum_{i=1}^n h\left(w_i\right)=\sum_{i=1}^n  \{\gamma(\gamma+1)\}^{-1}\{(nw_i)^{\gamma+1}-1\}.
\] 
Three special cases with $\gamma \in \{-1,0,1\}$  are popular. In particular, 
$CR(-1) =\sum_{i=1}^n-\log(nw_i)$ and $CR(0) =\sum_{i=1}^n(nw_i)\log(nw_i)$. 
Minimizing $CR(-1)$ is equivalent to maximizing $\sum_{i=1}^n\log(w_i)$, leading to the maximum empirical log-likelihood objective function. Minimizing $CR(0)$ is equivalent to maximizing $-\sum_{i=1}^n w_i\log(w_i)$, leading to the maximum empirical exponential likelihood or entropy. Finally, minimizing $CR(1)$ is equivalent to minimizing the sum of squares $\sum_{i=1}^n(w_i-{n}^{-1})^2$. To be consistent with the existing literature, we call the weight estimation method as the empirical likelihood method for $\gamma=-1$ and the entropy balancing method for $\gamma=0$. We summarize the correspondence between $\gamma$ and the form of $h(w)$ in Table \ref{table1}.

The first constraint is referred to as the balancing constraint, which calibrates the covariate distribution of the source sample to the target population in terms of $g(X)$. As a common premise to solve the above optimization problem, $\mu_{g0}$ should fall within the convex hull of $\{g(X_i),i= 1,...,n\}$.  Then, the optimization problem can be solved using the method of Lagrangian multipliers with the loss function 
\begin{align}\label{equa1}
	L=\{\gamma(\gamma+1)\}^{-1}\sum_{i=1}^n\{(nw_i)^{\gamma+1}-1\}-n\lambda^{\T}\sum_{i=1}^nw_i\left\{g(X_i)-\mu_{g0}\right\}+n\varphi\left(1-\sum_{i=1}^nw_i\right).
\end{align}
As noted in \cite{newey2004higher}, by minimizing (\ref{equa1}),  the estimator for $w_i$ is
\begin{align}\label{equa2}
	w(X_i;\widehat{\lambda})=\rho\left[\widehat{\lambda}^{\T}\{g(X_i)-\mu_{g0}\}\right] \bigg/ \sum_{j=1}^n \rho\left[\widehat{\lambda}^{\T}\{g(X_j)-\mu_{g0}\}\right],
\end{align}
where the function $\rho(x)$ for different $\gamma$ values are summarized in Table \ref{table1}, and $\widehat{\lambda}$ solves the equation  $\sum_{i=1}^n\rho\left[\lambda^{\T}\{g(X_i)-\mu_{g0}\}\right]\{g(X_i)-\mu_{g0}\} = 0$. 
\begin{table}
	\def~{\hphantom{0}}
	\caption{The formulation of $\rho(x)$ for the empirical likelihood, entropy balancing, and Cressie-Read  family}
		\centering
		\spacingset{1.45}
		\begin{tabular}{ccccc}
			Method & Empirical Likelihood & Entropy Balancing & Cressie-Read\\
			$\gamma$ & -1 & 0 & $\gamma$\\
			$h(w)$ & $-\ln(nw)$  & $nw\ln(nw)$  & $\frac{(nw)^{\gamma+1}-1}{\gamma(\gamma+1)}$\\
			$\rho(x)$ & $(1-x)^{-1}$ & $\exp(x)$ & $(1+\gamma x)^{1/\gamma}$
	\end{tabular}
	\label{table1}
\end{table}

Let $W(X;\lambda)=n w(X;\lambda)$. The proposed calibrated AIPW estimator is then 
$$\widehat{V}^{\rm{c}}(\beta)=\frac{1}{n}\sum\limits_{i=1}^n W(X_i;\widehat{\lambda})\left[\frac{{I}\{A_i=d(X_i;\beta)\}}{\varrho(A_i\mid X_i;\widehat{\eta})}\{Y_i-\mu_d(X_i;\beta,\widehat{\theta})\}+\mu_d(X_i;\beta,\widehat{\theta})\right].$$
Thus, the regime learning procedure can be summarized as a 3-step algorithm:

\noindent 	\textit{Step 1:} Estimate calibration weights, e.g., using the empirical likelihood method or entropy balancing method.

\noindent 	\textit{Step 2:} Estimate the propensity score  $\pi(\cdot)$ and the conditional outcome mean $\mu(\cdot)$ using either parametric models or nonparametric models.

\noindent 	\textit{Step 3:} Construct the calibrated AIPW estimator with the components estimated in Steps 1 and 2, and obtain the optimal  ITR by maximizing the calibrated AIPW estimator within a class of pre-specified ITRs, such as linear decision rules.

Before delving into theoretical analysis, it is important to define the underlying population for which $\widehat{V}^{\rm{c}}(\beta)$ is targeting unambiguously. Toward this end, let $\lambda^*$ be the limit of $\widehat{\lambda}$ and 
$$W^*(X;\lambda)=\rho\left[\lambda^{\T}\{g(X)-\mu_{g0}\}\right] \big/ E\left( \rho\left[\lambda^{\T}\{g(X)-\mu_{g0}\}\right]\right).$$
In general, the calibration weights are not guaranteed to be non-negative. As pointed out in \cite{schennach2007point}, when $\gamma\leq 0$, the estimated weights are non-negative by construction. It can be shown that $f^+(X)\propto f^{\rm{s}}(X)W^*(X;\lambda^*)$ is a valid density or probability mass function when $\gamma=-1$ or $0$. Therefore, it defines a pseudo population $\mathbb{P}^+$. While for $\gamma>0$, the calibration weights can take on negative values, and thus the corresponding $f^+(X)$ is not always a valid density or probability mass function. Therefore, we focus on $\gamma=-1,0$ for illustration.  It is expected that $\widehat{V}^{\rm{c}}(\beta)$ will converge to the value function under the ITR $d(X;\beta)$ for the pseudo population $\mathbb{P}^+$, when either the propensity score or the conditional mean outcome model is correctly specified.  

Moreover, when $W(X;\lambda^*) \propto f^{\text{t}}(X)/f^{\text{s}}(X)$, we have $f^+(X) = f^{\text{t}}(X)$. Then, $\widehat{V}^{\rm{c}}(\beta)$ is also a consistent estimator of the value function for the target population. 
Denote the density or probability mass function of covariates in the pooled population $\mathbb{P}$ as $q(X)$. Then $f^{\rm{s}}(X)$ and $f^{\rm{t}}(X)$ can be described as %the conditional distributions 
$q(X\mid S=1)$ and $q(X\mid S=0)$, respectively. By Bayesian Theorem, we have,
\begin{align*}
	\frac{f^{\rm{t}}(X)}{f^{\rm{s}}(X)}=\frac{q(X\mid S=0)}{q(X\mid S=1)}\propto \frac{{\rm pr}(S=0\mid X)}{{\rm pr}(S=1\mid X)}.
\end{align*}
If ${\rm pr}(S=0\mid X)$ follows a logistic regression with covariates $g(X)$, we have ${\rm pr}(S=0\mid X)/{\rm pr}(S=1\mid X) \propto \exp\{\alpha^{\T}g(X)\}$. Moreover, based on (\ref{equa2}), we have $W(X;\widehat{\lambda}) \propto \exp\{\widehat{\lambda}^{\T}g(X)\}$ when $\gamma=0$.  Therefore, the weights obtained by the entropy balancing method satisfy $W(X;\lambda^*) \propto f^{\text{t}}(X)/f^{\text{s}}(X)$ under the logistic regression model for ${\rm pr}(S=0\mid X)$. 
Similarly, if ${\rm pr}(S=0\mid X)$ can be represented by the following model
$$
{\rm pr}(S=0\mid X) = \frac{\kappa_0}{1 - \alpha^{\T}\{g(X)-\mu_{g0}\}}\bigg/\left[1+\frac{\kappa_0}{1 - \alpha^{\T}\{g(X)-\mu_{g0}\}}\right],
$$
where $\kappa_0$ is a positive constant and $\alpha$ satisfies $1 - \alpha^{\T}\{g(X)-\mu_{g0}\} > 0$, we have ${\rm pr}(S=0\mid X)/{\rm pr}(S=1\mid X) \propto [1 - \alpha^{\T}\{g(X)-\mu_{g0}\}]^{-1}$. Therefore, under the above model, the weights obtained by the empirical likelihood method satisfy $W(X;\lambda^*) \propto f^{\text{t}}(X)/f^{\text{s}}(X)$.  

In general, the calibration weights 
can not lead to a pseudo population with exactly the same covariate distribution as for the target population. However, it is expected that with more constraints based on summary statistics from the target population, the covariate distribution of the pseudo population will get closer to that of the target population. Therefore, the optimal ITR obtained based on $\widehat{V}^{\rm{c}}(\beta)$ would be better than that obtained based on $\widehat{V}^{\rm{o}}(\beta)$. Let $V^+(\beta)$ denote the value function under the ITR $d(X;\beta)$ for the pseudo population $\mathbb{P}^+$ and define $\beta^* = \argmax_{\beta}V^+(\beta)$. 
Then, the true optimal linear ITR for $\mathbb{P}^+$ is $d(X;\beta^*)$ and the estimated optimal linear ITR is  $d(X;\widehat{\beta}^{\rm{c}})$, where $\widehat{\beta}^{\rm{c}}=\argmax_{\beta}\widehat{V}^{\rm{c}}(\beta)$. Similarly, define $\widehat{\beta}^{\rm{o}}=\argmax_{\beta}\widehat{V}^{\rm{o}}(\beta)$. The estimated optimal linear ITR without calibration is $d(X;\widehat{\beta}^{\rm{o}})$.  

\section{Theoretical Properties}
\label{sec:theorem}
In this section, we establish the asymptotic 
properties of the calibrated AIPW estimator $\widehat{V}^{\rm{c}}(\widehat{\beta}^{\rm{c}})$. The proofs of all theorems  are given in the supplementary material.    

We first consider the case when the propensity score model $\pi(x)$ and conditional mean outcome model $\mu(x,a)$ are estimated based on some posited parametric models $\pi(X;\eta)$ and $\mu(X,A;\theta)$, respectively. Denote the estimating equations for $\lambda$, $\theta$, $\eta$ and $V^+(\beta^*)$ as 
\begin{align*}
	\frac{1}{n}\sum_{i=1}^n\left(\begin{array}{cc}
		\rho\left[\lambda^{\T}\{g(X_i)-\mu_{g0}\}\right]\left\{g(X_i)-\mu_{g0}\right\}\\
		C(X_i,A_i,Y_i;\theta)\\
		S(X_i,A_i;\eta)\\
		W(X_i;\lambda)\psi(X_i,A_i,Y_i;\beta,\theta,\eta)-V^+(\beta^*)\\
	\end{array}\right)=0,
\end{align*}
where $$\psi(X_i,A_i,Y_i;\beta,\theta,\eta)=\frac{{I}\{A_i=d(X_i;\beta)\}}{\varrho(A_i\mid X_i;\eta)}\{Y_i-\mu_d(X_i;\beta,\theta)\}+\mu_d(X_i;\beta,\theta).$$ Let $\widehat{\lambda}$, $\widehat{\theta}$ and $\widehat{\eta}$ denote the estimators of $\lambda$, $\eta$ and $\theta$ obtained from the above equations and let $\lambda^*$, $\theta^*$ and $\eta^*$ denote the limits of $\widehat{\lambda}$, $\widehat{\theta}$ and $\widehat{\eta}$, respectively. To establish the asymptotic properties of $\widehat{V}^{\rm{c}}(\widehat{\beta}^{\rm{c}})$, we impose the following regularity conditions.  
\begin{assumption}Assume the following regularity conditions hold:
	(A6) The supports of $X$ and $Y$ are bounded.
	(A7) The function $\mu(x, a)$ is smooth and bounded for all $(x,a)$.
	(A8) The weight function $W(x;\lambda)$ is smooth and bounded away from $\infty$, and it has bounded first derivatives with respect to $\lambda$.
	(A9) The value function $V^+(\beta)$ is twice continuously differentiable in a neighborhood  of $\beta^*$.
	(A10) There exist some constants $\delta_0 > 0$ such that
	${\rm pr}(|\tilde{X}^{\T}\beta^*| \le \delta) = O(\delta),$
	where the big-O term is uniform in $0 < \delta \le \delta_0$.
	(A11)  (i)  $\sqrt{n}(\widehat{\lambda}- \lambda^*) = O_p(1)$,  (ii) $\sqrt{n}(\widehat{\theta}- \theta^*) = O_p(1)$, and (iii) $\sqrt{n}(\widehat{\eta}- \eta^*) = O_p(1)$.
\end{assumption}
Conditions (A6) - (A9) are standard regularity conditions used to establish the uniform convergence results. Condition (A10) excludes the situation with ${\rm pr}( \tilde{X}^{\T} \beta^* = 0) > 0$ and 
ensures the true targeted optimal ITR is uniquely defined, known as the margin condition, which is often assumed to derive a sharp convergence rate for the value function under the estimated optimal ITR \citep[e.g.][]{luedtke2016statistical}. 
Condition (A11) assumes the $\sqrt{n}$-convergence rates of parameter estimates in the calibration weight function, propensity score model, and conditional mean outcome model, which usually hold under mild conditions for posited parametric models, for example, a logistic or probit regression model for the propensity score, a linear model for the conditional mean outcome, and weights obtained using the empirical likelihood method or entropy balancing method.       

Define 
\begin{align*}
	& \xi_{i1}=W(X_{i};\lambda^{*})\psi(Y_{i},A_{i},X_{i};\beta^{*},\theta^{*},\eta^{*})-V^{+}(\beta^{*}), &  & \xi_{i3}=H_{\theta}^{\T}G_{\theta}^{-1}C(X_{i},A_{i},Y_{i};\theta^{*}),\\
	& \xi_{i2}=H_{\lambda}^{\T}G_{\lambda}^{-1}\rho\left[(\lambda^{*})^{\T}\{g(X_{i})-\mu_{g0}\}\right]\{g(X_{i})-\mu_{g0}\}, &  & \xi_{i4}=H_{\eta}^{\T}G_{\eta}^{-1}S(X_{i},A_{i};\eta^{*}),
\end{align*}
where 
\begin{align*}
	&H_{\lambda}=\lim\limits _{n\to\infty}\frac{1}{n}\sum_{i=1}^{n}\left\{ \frac{\partial W(X_{i};\lambda^{*})}{\partial\lambda}\right\} \psi(Y_{i},A_{i},X_{i};\beta^{*},\theta^{*},\eta^{*}),\\
	&H_{s}=\lim\limits _{n\to\infty}\frac{1}{n}\sum_{i=1}^{n}W(X_{i};\lambda^{*})\frac{\partial\psi(Y_{i},A_{i},X_{i};\beta^{*},\theta^{*},\eta^{*})}{\partial s}\ \ (s=\theta,\eta),\\
	&G_{\lambda}=-\mathbb{E}\left(\rho'\left[(\lambda^{*})^{\T}\{g(X)-\mu_{g0}\}\right]\{g(X)-\mu_{g0}\}\{g(X)-\mu_{g0}\}^{\T}\right),\\
	&G_{\theta}=-\mathbb{E}\left\{ {\partial C(X,A,Y;\theta^{*})}/{\partial\theta^{\T}}\right\} ,G_{\eta}=-\mathbb{E}\left\{ {\partial S(X,A;\eta^{*})}/{\partial\eta^{\T}}\right\}.
\end{align*}
Note that $\xi_{i2}$, $\xi_{i3}$ and $\xi_{i4}$ are the terms in the inference function of $\widehat{V}^{\rm{c}}(\widehat{\beta}^{\rm{c}})$ due to estimators $\widehat{\lambda}$, $\widehat{\theta}$ and $\widehat{\eta}$, respectively.  

\begin{theorem}
	\label{thm1}
	Assume either $\pi(X;\eta)$ or $\mu(X, A; \theta)$ is correctly specified. Under (A1)-(A11), we have, as $n \rightarrow \infty$, 
	$\sqrt{n}\{\widehat{V}^{\rm{c}}(\widehat{\beta}^{\rm{c}})-V^+(\beta^*)\}\longrightarrow N(0,\sigma_1^2),$
	in distribution, \\where
	$\sigma^2_1={E}\left\{(\xi_{i1}+\xi_{i2}+\xi_{i3}++\xi_{i4})^2\right\}$. In addition, $\sigma_1^2$ can be estimated by replacing expectation with empirical sum and true values $V^+(\beta^*)$, $\lambda^*$, $\theta^*$, and $\eta^*$ with $\widehat{V}^{\rm{c}}(\widehat{\beta}^{\rm{c}})$, $\widehat{\lambda}$, $\widehat{\theta}$, and $\widehat{\eta}$, respectively.
\end{theorem}

Next, we consider the case when both propensity score model $\pi(x)$ and conditional mean outcome model $\mu(x,a)$ are estimated by flexible semi/nonparametric models with certain convergence rates. For example, $\pi(x)$ and/or $\mu(x,a)$ are estimated using kernel regression or random forest. Let $\widehat{\pi}(x)$ and $\widehat{\mu}(x,a)$ denote the corresponding estimators. The calibrated AIPW estimator $\widehat{V}^{\rm{c}}(\beta)$ can be similarly defined by replacing $\pi(x;\widehat{\eta})$ and $\mu(x,a;\widehat{\theta})$ with $\widehat{\pi}(x)$ and $\widehat{\mu}(x,a)$, respectively. To derive the asymptotic distribution of $\widehat{V}^{\rm{c}}(\beta)$, we need the following modified condition. 

\noindent (A11') (i) $\sqrt{n}(\widehat{\lambda}- \lambda^*) = O_p(1)$; (ii) 
$
\left[{P}\{\widehat{\pi}(X)-\pi(X)\}^2\right]^{\frac{1}{2}}\sum_{a=0}^1\left[{P}\{\widehat{\mu}(X,a)-\mu(X,a)\}^2\right]^{\frac{1}{2}}=o_p(n^{-1/2}) 
$, where $P\{f(X)\} = \int f(x)^2dF_X(x)$.

Condition (A11') (ii) is commonly imposed in the causal inference
literature to derive the asymptotic distribution of the AIPW estimators when the nuisance functions are estimated with certain convergence rates \citep{kennedy2016semiparametric,farrell2021deep}. For example, if $\pi(x)$ is estimated based on a correctly specified parametric model, $\widehat{\pi}(x)$ is $\sqrt{n}$-consistent. Then it only requires $\widehat{\mu}(x,a)$ to be consistent for (A11') to hold. This can be easily achieved by most nonparametric regression methods. However, when both $\mu(x,a)$ and $\pi(x)$ are estimated nonparametrically, it usually requires both terms to be estimated with the rate $o_p(n^{-1/4})$. This can be achieved by some nonparametric methods, such as kernel regression or random forest under certain conditions. With Condition (A11') (ii), we can establish the $\sqrt{n}$-consistency of $\widehat{V}^{\rm{c}}(\widehat{\beta}^{\rm{c}})$. In addition, the asymptotic variance of $\widehat{V}^{\rm{c}}(\widehat{\beta}^{\rm{c}})$ will not depend on the variances of estimates $\widehat{\pi}(x)$ and $\widehat{\mu}(x,a)$. The results are summarized in the following theorem.  

\begin{theorem}
	\label{thm2}
	Under (A1)-(A10) and (A11'), we have, as $n \rightarrow \infty$,  
	$\sqrt{n}\{\widehat{V}^{\rm{c}}(\widehat{\beta}^{\rm{c}})-V^+(\beta^*)\}\longrightarrow N(0,\sigma_2^2),$
	in distribution, where $\sigma_2^2={E}\left\{(\xi_{i1}+\xi_{i2})^2\right\}$. Here, $\xi_{i1}$ and $\xi_{i2}$ are defined the same as in Theorem \ref{thm1} but replacing $\pi(x;\eta)$ and $\mu(x,a;\theta)$ with $\pi(x)$ and $\mu(x,a)$, respectively. In addition, $\sigma_2^2$ can be estimated by replacing expectation with empirical sum and true values  $V^+(\beta^*)$, $\lambda^*$, $\pi(x)$ and $\mu(x,a)$ with $\widehat{V}^{\rm{c}}(\widehat{\beta}^{\rm{c}})$, $\widehat{\lambda}$, $\widehat{\pi}(x)$ and $\widehat{\mu}(x,a)$, respectively.
\end{theorem}

\begin{remark}
	The theorems established  above focus on the inference for the optimal value function. In the proof of Theorems \ref{thm1} and \ref{thm2}, we show that $\widehat{\beta}^{\rm{c}}$ has the cubic root convergence rate. In addition, the asymptotic distribution of $\widehat{\beta}^{\rm{c}}$ can be established and its associated inference can be done by bootstrap-based methods \citep[e.g.][]{cattaneo2020bootstrap}.
\end{remark}

Finally, we compare the efficiency of $\widehat{V}^{\rm{o}}(\beta)$ and $\widehat{V}^{\rm{c}}(\beta)$ when the source and target populations have the same covariate distributions, i.e. $\mathbb{P}^{\rm{s}}=\mathbb{P}^{\rm{t}}$. Under such case, $V^+(\beta) = V^{\rm{t}}(\beta)$, the value function under the ITR $d(X;\beta)$ for the target population.   

\begin{theorem}
	\label{thm3}
	Assume (A1)-(A10) and (A11') hold. When $\mathbb{P}^{\rm{s}}=\mathbb{P}^{\rm{t}}$, we have that both $\sqrt{n}\{\widehat{V}^{\rm{o}}(\beta)-V^{\rm{t}}(\beta)\}$ and $\sqrt{n}\{\widehat{V}^{\rm{c}}(\beta)-V^{\rm{t}}(\beta)\}$ are asymptotically normal with mean zero, while the latter one has the same or smaller asymptotic variance.
\end{theorem}

Theorem \ref{thm3} implies that even when the source and target populations have the same covariate distributions, the calibrated AIPW value estimator can be more efficient than the original AIPW value estimator without calibration. The efficiency gain of the calibrated estimator comes from the constraints imposed based on available summary statistics of the covariate distribution for the target population.

\section{Simulation Studies}
\label{sec:simulation}

We have carried out extensive simulation studies to evaluate the performance of the proposed methods. Here we focus on two methods for computing the weights: empirical likelihood ($\gamma = -1$) and entropy balancing ($\gamma = 0$). The results for $\gamma=1$ are provided in the supplementary material. For illustration, we only considered means of all covariates as the summary statistics from the target population. 
The corresponding pseudo populations are denoted as $\mathbb{P}^+_{EB}$ for $\gamma=0$ and $\mathbb{P}^+_{EL}$ for $\gamma=-1$, respectively. 
Table \ref{table2} defines additional notation for the simulation. 
Since estimated value functions are non-smooth and non-convex in $\beta$, following \cite{zhang2012robust}, we used the genetic algorithm \citep{whitley1994genetic} to find $\widehat{\beta}^{\rm{o}}$, $\widehat{\beta}^{\rm{c}}_{EB}$, and $\widehat{\beta}^{\rm{c}}_{EL}$. The optimization was implemented using the function $\mathtt{genoud}$ in the R package $\mathtt{rgenoud}$ \citep{mebane2011genetic}. 

For the source sample, outcomes are generated from the model $Y=\mu(X,A)+\epsilon$,
where $X = (X_1, X_2, X_3)^{\T}$, 
\begin{align*}
	\mu(X,A)=\exp\left\{2-0.1X_{1}-0.2X_{2}+0.2X_{3}+A\frac{2\hbox{sign}(X_{3}-X_{2}^2+1)}{2+|X_{3}-X_{2}^2+1|}\right\},
\end{align*}
and $\epsilon$ is generated from a normal distribution with mean 0 and variance 0.25. In addition, we considered two different propensity score models for 
the treatment indicator $A$: $\pi(X)=0.5$, which represents a randomization study;  $\text{logit}\{\pi(X)\}=0.5X_{1}-0.5X_{2}+0.5X_{3}$, which represents an observational study. 

We considered four different scenarios of the covariate distributions for $\mathbb{P}^{\rm{s}}$ and $\mathbb{P}^{\text{t}}$, which are summarized in Table \ref{table3} and Table \ref{table4}. In Scenario 1, the covariate distributions of the source and target populations are the same. We have $\mathbb{P}^{\rm{s}} = \mathbb{P}^+_{EB} = \mathbb{P}^+_{EL} = \mathbb{P}^{\rm{t}}$. In Scenario 2, the ratio $f^{\rm{t}}(X)/f^{\rm{s}}(X)$ can be written as $\exp\{\ln(0.4)+\ln(4)X_1\}$ or $1/\{1-1.875(X_1-0.8)\}$.
It can be shown that $W(X;\lambda^*) \propto f^{\text{t}}(X)/f^{\text{s}}(X)$ for both calibration methods. Therefore, we have $\mathbb{P}^+_{EB} = \mathbb{P}^+_{EL} = \mathbb{P}^{\rm{t}}$ even if we only use means of covariates as the summary statistics from the target population. This implies $V^+_{EB}(\beta) = V^+_{EL}(\beta) = V^{\rm{t}}(\beta)$, and both $\widehat{V}^{\rm{c}}_{EB}(\widehat{\beta}^{\rm{c}}_{EB})$ and $\widehat{V}^{\rm{c}}_{EL}(\widehat{\beta}^{\rm{c}}_{EL})$ are consistent estimators of $V^{\rm{t}}(\beta^{\rm{t}})$ when either the propensity score or conditional mean outcome model is correctly specified. 
However, in Scenarios 3 and 4, $W(X;\lambda^*)$ is no longer proportional to $f^{\rm{t}}(X)/f^{\rm{s}}(X)$. Thus, $\widehat{V}^{\rm{c}}_{EB}(\widehat{\beta}^{\rm{c}}_{EB})$ and $\widehat{V}^{\rm{c}}_{EL}(\widehat{\beta}^{\rm{c}}_{EL})$ are doubly robust estimators only for the value functions of their corresponding pseudo populations, but not for that of the target population. 

\begin{table}
	\def~{\hphantom{0}}
		\spacingset{1.45}
		\centering
	\caption{Additional notation used in the simulation studies.}
	\vspace{6pt}
		\begin{tabular}{ccccc}
			Population & Value & Optimal ITR & \multicolumn{2}{c}{Estimators}\tabularnewline[5pt]
			$\mathbb{P}^{{\rm {t}}}$ & $V^{{\rm {t}}}(\beta)$ & $d(X;\beta^{{\rm {t}}})$; $\beta^{{\rm {t}}}=\argmax_{\beta}V^{{\rm {t}}}(\beta)$ &  & \tabularnewline
			$\mathbb{P}_{EB}^{+}$ & $V_{EB}^{+}(\beta)$ & $d(X;\beta_{EB}^{*})$; $\beta_{EB}^{*}=\argmax_{\beta}V_{EB}^{+}(\beta)$ & $\widehat{V}_{EB}^{{\rm {c}}}(\beta)$; $\widehat{\beta}_{EB}^{{\rm {c}}}=\argmax_{\beta}\widehat{V}_{EB}^{{\rm {c}}}(\beta)$ \tabularnewline
			$\mathbb{P}_{EL}^{+}$ & $V_{EL}^{+}(\beta)$ & $d(X;\beta_{EL}^{*})$; $\beta_{EL}^{*}=\argmax_{\beta}V_{EL}^{+}(\beta)$ & $\widehat{V}_{EL}^{{\rm {c}}}(\beta)$; $\widehat{\beta}_{EL}^{{\rm {c}}}=\argmax_{\beta}\widehat{V}_{EL}^{{\rm {c}}}(\beta)$\tabularnewline
	\end{tabular}
	\label{table2}
\end{table}

\begin{table}[H]
	\spacingset{1.45}
	\def~{\hphantom{0}}
	\caption{Covariate distributions for $\mathbb{P}^{\rm{s}}$ and $\mathbb{P}^{\rm{t}}$ used in the simulation studies.}
	\vspace{6pt}
		\begin{tabular}{ccccccc}
			\hline
			Scenario & $f^{\rm{s}}(X)$ & $f^{\text{t}}(X)$\\
			\hline
			\multirow{2}*{1}&$X_1\sim Bernoulli(0.5)$ & $X_1 \sim Bernoulli(0.5)$\\
			~ & $(X_2,X_3)^{\T}\sim N((-1,0)^{\T},\Sigma_1)$& $(X_2,X_3)^{\T}\sim N((-1,0)^{\T},\Sigma_1)$\\
			\hline
			\multirow{3}*{2} & $X_1\sim Bernoulli(0.5)$ & $X_1 \sim Bernoulli(0.8)$\\
			~ & $(X_2,X_3)^{\T}\mid X_1=1\sim N((1,-1)^{\T},\Sigma_1)$ & $(X_2,X_3)^{\T}\mid X_1=1\sim N((1,-1)^{\T},\Sigma_1)$\\
			~ & $(X_2,X_3)^{\T}\mid X_1=0\sim N((-1,1)^{\T},\Sigma_2)$ & $(X_2,X_3)^{\T}\mid X_1=0\sim N((-1,1)^{\T},\Sigma_2)$\\
			\hline
			\multirow{3}*{3} & \multirow{3}*{\shortstack{$X_1\sim Bernoulli(0.7)$\\$(X_2,X_3)^{\T}\sim N((0.1,-0.2)^{\T},\Sigma_1)$}} & $X_1 \sim Bernoulli(0.8)$\\
			~ & ~ & $(X_2,X_3)^{\T}\mid X_1=1\sim N((1,-1)^{\T},\Sigma_1)$\\
			~ & ~ & $(X_2,X_3)^{\T}\mid X_1=0\sim N((-1,1)^{\T},\Sigma_2)$\\
			\hline
			\multirow{3}*{4} & \multirow{3}*{\shortstack{$X_1\sim Bernoulli(0.6)$\\$(X_2,X_3)^{\T}\sim N((0,0)^{\T},\Sigma_1)$}} & $X_1 \sim Bernoulli(0.8)$\\
			~ & ~ & $(X_2,X_3)^{\T}\mid X_1=1\sim N((1,-1)^{\T},\Sigma_1)$\\
			~ & ~ & $(X_2,X_3)^{\T}\mid X_1=0\sim N((-1,1)^{\T},\Sigma_2)$\\
			\hline
	\end{tabular}
	\label{table3}
	
	\vspace{6pt}
		\centering $\Sigma_1 = \left(\begin{array}{cc}
			1&-0.25\\
			-0.25&1
		\end{array}
		\right), \quad \Sigma_2=\left(\begin{array}{cc}
			1&-0.3\\
			-0.3&1
		\end{array}
		\right).$
\end{table}

\begin{table}[H]
	\spacingset{1.45}
	\def~{\hphantom{0}}
	\caption{Summary statistics of $X_1, X_2, X_3$  in different scenarios.}
	\vspace{6pt}
	\centering
		\begin{tabular}{cc|c|c|c|c}
			\hline 
			\multirow{2}{*}{Population} & \multirow{2}{*}{Statistics} & \multicolumn{4}{c}{Scenario}\tabularnewline
			&  & \multicolumn{1}{c}{1} & \multicolumn{1}{c}{2} & \multicolumn{1}{c}{3} & 4\tabularnewline
			\hline 
			\multirow{2}{*}{$\mathbb{P}^{{\rm {s}}}$} & Mean & $0.5,-1,0$ & $0.5,0,0$ & $0.7,0.1,-0.2$ & $0.6,0,0$\tabularnewline
			& Variance & $0.25,1,1$ & $0.25,2,2$  & $0.21,1,1$ & $0.24,1,1$\tabularnewline
			\hline 
			\multirow{2}{*}{$\mathbb{P}^{{\rm {t}}}$} & Mean & $0.5,-1,0$ & \multicolumn{3}{c}{$0.8,0.6,-0.6$}\tabularnewline
			& Variance & $0.25,1,1$ & \multicolumn{3}{c}{$0.16,1.64,1.64$}\tabularnewline
			\hline 
	\end{tabular}
	\label{table4}
\end{table}

We considered a source sample with size $n=250, 1000$. For each setting, we conducted 500 replications. In our implementation, the propensity score and conditional mean outcome models are estimated using two methods: 
\begin{itemize}
	\item[(I)] Both are estimated based on posited parametric models. In particular, the propensity score is estimated using a correctly specified logistic regression model, while the conditional mean outcome is estimated using a linear model with all the covariates and covariate-treatment interactions, which is a misspecified model. 
	\item[(II)] The propensity score is estimated nonparametrically  using a generalized additive model, and the conditional mean outcome model $\mu(x,a)$ is estimated nonparametrically using the random forest for $a=0$ and 1, separately. 
\end{itemize}

\begin{figure}
	\centering
	\subfigure[Implementation method I]{
		\begin{minipage}[b]{0.5\textwidth}
			\includegraphics[width=1\textwidth]{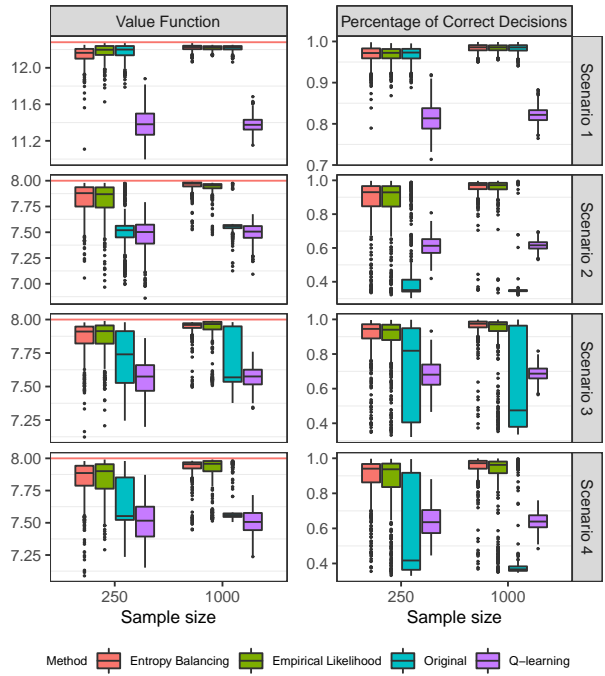}
		\end{minipage}
		\label{fig1}
	}
	\subfigure[Implementation method II]{
		\begin{minipage}[b]{0.5\textwidth}
			\includegraphics[width=1\textwidth]{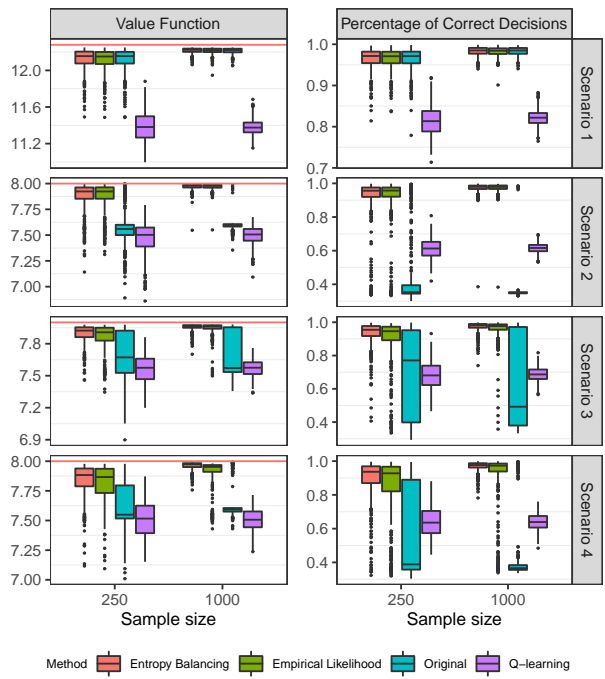}
		\end{minipage}
		\label{fig2}
	}
	\caption{The value and percentage of correct decisions results of estimated optimal ITRs for the observational study with implementation methods I and II. The red lines are the values of the true optimal ITRs for the target population.}
	\label{fig:simulation}
\end{figure}

We also implemented Q-learning as a benchmark for comparison. Specifically, we fitted linear models for Q-functions and inferred optimal linear ITRs from the estimated Q-functions. An ITR estimated by Q-learning is denoted as $d(x;\widehat{\beta}_Q)$. To evaluate and compare the performance of estimated optimal ITRs obtained from the original AIPW estimator, proposed calibrated AIPW estimators, and Q-learning, we compute the corresponding value functions and percentages of correct decisions for the target population. Specifically, we generate covariates $X^{\rm{t}}$ for a large sample with size $N=10^5$ from the target population. The value function of an estimated ITR $d(x;\widehat{\beta})$, where $\widehat{\beta} = \widehat{\beta}^{\rm{o}}$, $\widehat{\beta}^{\rm{c}}_{EB}$, $\widehat{\beta}^{\rm{c}}_{EL}$, or $\widehat{\beta}_Q$ is computed by  
$V^{\rm{t}}(\widehat{\beta})={N}^{-1}\sum_{i=1}^N \mu\{X_i^{\rm{t}}, d(X_i^{\rm{t}};\widehat{\beta})\},$
and its associated percentage of correct decisions is $1-N^{-1}\sum_{i=1}^N |d(X_i^{\rm{t}};\widehat{\beta})-d(X_i^{\rm{t}};\beta^{\rm{t}})|$. 
Here, the true optimal ITR $d(X;\beta^{\rm{t}})$ for the target population is obtained by maximizing $V^{\rm{t}}(\beta)$ over $\beta$ using the grid-search method. 
We report the values and percentages of correct decisions results of $d(x;\widehat{\beta}^{\rm{o}})$, $d(x;\widehat{\beta}^{\rm{c}}_{EB})$, $d(x;\widehat{\beta}^{\rm{c}}_{EL})$, and $d(x;\widehat{\beta}_Q)$ for the observational study in Figure \ref{fig1} (method I) and Figure \ref{fig2} (method II). Similar results for the randomization study are provided in the supplementary material.

We have the following observations. In Scenario 1, the optimal ITR estimated by Q-learning  has poor performance in terms of value and percentage of correct decisions, due to the misspecification of Q-function. All other three estimated optimal ITRs have good and comparable performance in terms of values and percentages of correct decisions, which is expected since $\mathbb{P}^{\rm{s}} = \mathbb{P}^+_{EB} = \mathbb{P}^+_{EL} = \mathbb{P}^{\rm{t}}$. In addition, as the sample size increases, the means of value functions become closer to the true optimal value for the target population, percentages of correct decisions get closer to 1, and the standard deviations of value functions and percentages of correct decisions become smaller. However, in Scenarios 2-4 where $\mathbb{P}^{\rm{s}} \neq \mathbb{P}^{\rm{t}}$, the estimated optimal ITR obtained using the original method has poor performance: the means of value functions are much smaller than the true optimal value for the target population and percentages of correct decisions are also much smaller than 1. This implies that the estimated optimal ITR obtained using the original method may not generalize well to the target population when $\mathbb{P}^{\rm{s}} \neq \mathbb{P}^{\rm{t}}$. The optimal ITR estimated by Q-learning still yields poor performance. However, the estimated optimal ITRs obtained using the proposed calibration methods still have competitive performance similar to those observed in Scenario 1. This supports that the proposed calibration using summary statistics can improve the treatment decision for the target population.  

Next, we study the estimation and inference results of $\widehat{V}^{\rm{c}}_{EB}(\widehat{\beta}^{\rm{c}}_{EB})$ and $\widehat{V}^{\rm{c}}_{EL}(\widehat{\beta}^{\rm{c}}_{EL})$.  For implementation method I, the asymptotic variances of $\widehat{V}^{\rm{c}}_{EB}(\widehat{\beta}^{\rm{c}}_{EB})$ and $\widehat{V}^{\rm{c}}_{EL}(\widehat{\beta}^{\rm{c}}_{EL})$ were estimated using the results established in Theorem \ref{thm1}, while for implementation method II, the corresponding asymptotic variances were estimated using the results established in Theorem \ref{thm2} because Condition (A11') holds.    
In our simulations, we observed that the empirical likelihood method may produce a few extreme calibration weights in Scenarios 3 and 4. These extreme weights usually do not inflate the biases of  $\widehat{V}^{\rm{c}}_{EL}(\widehat{\beta}^{\rm{c}}_{EL})$, but they do lead to overestimated standard errors due to the instability in variance estimation. 
To control the effects of these extreme weights, we
stabilize the weights by reducing the large weights $\widehat{w}_{i}$
$(>a_{n}^{-1})$ to $\tilde{w}_{i}$ according to $(\tilde{w}_{i})^{-1}=(\widehat{w}_{i})^{-1}+a_{n}$
for $a_{n}=o_{p}(n)$. Such a stabilization leads all weights to be no
larger than $a_{n}^{-1}.$ The rationale for considering $a_{n}$
to be $o_{p}(n)$ is that because $w_{i}\propto n^{-1}$, the stabilization
does not affect the weights asymptotically. In the simulation study,
we take $a_{n}=12\log n$.
Based on our numerical studies, such a stabilization doesn't affect the biases of $\widehat{V}^{\rm{c}}_{EL}(\widehat{\beta}^{\rm{c}}_{EL})$ much but can give a reasonable standard error estimate. On the other hand, the calibration weights computed using the entropy balancing method do not have any extreme values in all four scenarios. Such an observation is consistent with the findings in the literature since the entropy balancing loss tends to penalize the deviation of the estimated weights $\widehat{w}_i$'s from $n^{-1}$ more than the empirical likelihood method. We report the mean and standard deviation of $\widehat{V}^{\rm{c}}_{EB}(\widehat{\beta}^{\rm{c}}_{EB})$ and $\widehat{V}^{\rm{c}}_{EL}(\widehat{\beta}^{\rm{c}}_{EL})$, the mean of estimated standard errors and the empirical coverage probability (CP) of 95\% Wald-type confidence intervals. The true optimal values $V^+_{EB}(\beta^*_{EB})$ and $V^+_{EL}(\beta^*_{EL})$ are computed using the grid-search method based on a large dataset generated from the corresponding pseudo populations similar to the computation of $V^{\rm{t}}(\beta^{\rm{t}})$. In addition, we consider two types of CP: (1) $\text{CP}^+$  for the optimal values $V^+_{EB}(\beta^*_{EB})$ or $V^+_{EL}(\beta^*_{EL})$ of the corresponding pseudo population; (2) $\text{CP}^{\text{t}}$ for the optimal values $V^{\rm{t}}(\beta^{\rm{t}})$ of the target population. Simulation results for the observational study  are summarized in Table \ref{table5} with implementation method I and Table \ref{table6} with implementation method II. Similar results for the randomization study are provided in the supplementary material.

\begin{table}
	\def~{\hphantom{0}}
	\spacingset{1.45}
	\caption{Simulation results for the observational study with implementation method I. Mean, the average of estimates; SD, the empirical standard deviation of estimates; SE, the mean of estimated standard errors; $\rm{CP}^+ (\%)$, the empirical coverage probability of a 95\% confidence interval for  $V^{+}_{EB}(\beta^*_{EB})$ or $V^{+}_{EL}(\beta^*_{EL})$; $\rm{CP}^{\rm{t}}  (\%)$, the empirical coverage probability of a 95\% confidence interval for  $V^{\rm{t}}(\beta^{\rm{t}})$.}
	\vspace{6pt}
	\centering
		\begin{tabular}{cccccccccc}
			\hline 
			\multirow{3}{*}{Method} & Scenario & \multicolumn{2}{c}{1} & \multicolumn{2}{c}{2} & \multicolumn{2}{c}{3} & \multicolumn{2}{c}{4}\tabularnewline
			& $n$ & 250 & 1000 & 250 & 1000 & 250 & 1000 & 250 & 1000\tabularnewline
			& $V^{{\rm {t}}}(\beta^{{\rm {t}}})$ & \multicolumn{2}{c}{12.28} & \multicolumn{2}{c}{8.00} & \multicolumn{2}{c}{8.00} & \multicolumn{2}{c}{8.00}\tabularnewline
			\hline 
			\multirow{6}{*}{\shortstack{Entropy\\Balancing}} & $V_{EB}^{+}(\beta_{EB}^{*})$ & \multicolumn{2}{c}{12.28} & \multicolumn{2}{c}{8.00} & \multicolumn{2}{c}{7.99} & \multicolumn{2}{c}{7.99}\tabularnewline
			& Mean & 12.34 & 12.32 & 8.20 & 8.06 & 8.22 & 8.08 & 8.24 & 8.08\tabularnewline
			& SD & 0.42 & 0.21 & 0.36 & 0.17 & 0.31 & 0.16 & 0.45 & 0.17\tabularnewline
			& SE & 0.47 & 0.24 & 0.41 & 0.19 & 0.35 & 0.17 & 0.43 & 0.19\tabularnewline
			& $\text{CP}^{+}$ & 96.8 & 95.6 & 95.4 & 95.6 & 95.0 & 95.2 & 94.6 & 94.8\tabularnewline
			& $\text{CP}^{\text{t}}$ & 96.8 & 95.6 & 95.4 & 95.6 & 95.2 & 96.0 & 94.8 & 96.0\tabularnewline
			\hline 
			\multirow{6}{*}{\shortstack{Empirical\\Likelihood}} & $V_{EL}^{+}(\beta_{EL}^{*})$ & \multicolumn{2}{c}{12.28} & \multicolumn{2}{c}{8.00} & \multicolumn{2}{c}{8.14} & \multicolumn{2}{c}{8.16}\tabularnewline
			& Mean & 12.36 & 12.31 & 8.17 & 8.06 & 8.09 & 8.19 & 7.90 & 8.22\tabularnewline
			& SD & 0.41 & 0.22 & 0.37 & 0.16 & 0.35 & 0.20 & 0.43 & 0.26\tabularnewline
			& SE & 0.38 & 0.20 & 0.41 & 0.19 & 0.32 & 0.20 & 0.37 & 0.27\tabularnewline
			& $\text{CP}^{+}$ & 93.6 & 96.0 & 96.4 & 97.0 & 92.0 & 96.8 & 81.2 & 94.0\tabularnewline
			& $\text{CP}^{\text{t}}$ & 93.6 & 96.0 & 96.4 & 97.0 & 93.2 & 88.4 & 87.0 & 93.4\tabularnewline
			\hline 
		\end{tabular}
	\label{table5}
\end{table}

\begin{table}
	\def~{\hphantom{0}}
		\spacingset{1.45}
	\caption{Simulation results  for the observational study with implementation method II. Mean, the average of estimates; SD, the empirical standard deviation of estimates; SE, the mean of estimated standard errors; $\rm{CP}^+ (\%)$, the empirical coverage probability of a 95\% confidence interval for  $V^{+}_{EB}(\beta^*_{EB})$ or $V^{+}_{EL}(\beta^*_{EL})$; $\rm{CP}^{\rm{t}}  (\%)$, the empirical coverage probability of a 95\% confidence interval for  $V^{\rm{t}}(\beta^{\rm{t}})$.}
	\vspace{6pt}
	\centering
		\begin{tabular}{cccccccccc}
			\hline 
			\multirow{3}{*}{Method} & Scenario & \multicolumn{2}{c}{1} & \multicolumn{2}{c}{2} & \multicolumn{2}{c}{3} & \multicolumn{2}{c}{4}\tabularnewline
			& $n$ & 250 & 1000 & 250 & 1000 & 250 & 1000 & 250 & 1000\tabularnewline
			& $V^{{\rm {t}}}(\beta^{{\rm {t}}})$ & \multicolumn{2}{c}{12.28} & \multicolumn{2}{c}{8.00} & \multicolumn{2}{c}{8.00} & \multicolumn{2}{c}{8.00}\tabularnewline
			\hline 
			\multirow{6}{*}{\shortstack{Entropy\\Balancing}} & $V_{EB}^{+}(\beta_{EB}^{*})$ & \multicolumn{2}{c}{12.28} & \multicolumn{2}{c}{8.00} & \multicolumn{2}{c}{7.99} & \multicolumn{2}{c}{7.99}\tabularnewline
			& Mean & 12.42 & 12.32 & 8.12 & 8.03 & 8.15 & 8.06 & 8.19 & 8.06\tabularnewline
			& SD & 0.50 & 0.22 & 0.33 & 0.15 & 0.29 & 0.13 & 0.34 & 0.15\tabularnewline
			& SE & 0.54 & 0.25 & 0.38 & 0.18 & 0.30 & 0.14 & 0.37 & 0.17\tabularnewline
			& $\text{CP}^{+}$ & 97.4 & 97.0 & 96.8 & 96.2 & 93.8 & 95.4 & 94.6 & 95.4\tabularnewline
			& $\text{CP}^{\text{t}}$ & 97.4 & 97.0 & 96.8 & 96.2 & 94.6 & 95.6 & 95.0 & 96.0\tabularnewline
			\hline 
			\multirow{6}{*}{\shortstack{Empirical\\Likelihood}} & $V_{EL}^{+}(\beta_{EL}^{*})$ & \multicolumn{2}{c}{12.28} & \multicolumn{2}{c}{8.00} & \multicolumn{2}{c}{8.14} & \multicolumn{2}{c}{8.16}\tabularnewline
			& Mean & 12.42 & 12.32 & 8.11 & 8.03 & 8.06 & 8.18 & 7.86 & 8.16\tabularnewline
			& SD & 0.50 & 0.23 & 0.33 & 0.15 & 0.31 & 0.17 & 0.43 & 0.23\tabularnewline
			& SE & 0.46 & 0.21 & 0.37 & 0.17 & 0.29 & 0.17 & 0.35 & 0.24\tabularnewline
			& $\text{CP}^{+}$ & 95.4 & 95.4 & 96.6 & 96.0 & 91.6 & 95.4 & 74.4 & 93.2\tabularnewline
			& $\text{CP}^{\text{t}}$ & 95.4 & 95.4 & 96.6 & 96.0 & 94.0 & 84.0 & 83.4 & 96.2\tabularnewline
			\hline 
		\end{tabular}
	\label{table6}
\end{table}

We have the following observations. In Scenarios 1 and 2, since $\mathbb{P}^+_{EB} = \mathbb{P}^+_{EL} = \mathbb{P}^{\rm{t}}$, we have $V^{+}_{EB}(\beta^*_{EB}) = V^{+}_{EL}(\beta^*_{EL}) = V^{\rm{t}}(\beta^{\rm{t}})$. Both calibrated value estimators are nearly unbiased. The mean of estimated standard errors is close to the standard deviation of the estimators, and the empirical coverage probability of 95\% confidence intervals is close to the nominal level for all settings. In Scenarios 3 and 4, $V^{+}_{EB}(\beta^*_{EB})$ or $V^{+}_{EL}(\beta^*_{EL})$ is no longer equal to $V^{\rm{t}}(\beta^{\rm{t}})$. 
However, we can see that both $V^{+}_{EB}(\beta^*_{EB})$ and $V^{+}_{EL}(\beta^*_{EL})$ are close to $V^{\rm{t}}(\beta^{\rm{t}})$. In particular, the difference between $V^+_{EB}(\beta^*_{EB})$ and $V^{\rm{t}}(\beta^{\rm{t}})$ is nearly negligible. This implies that both calibration methods give good approximation of the target population, while the entropy balancing method is better than the empirical likelihood method for the considered Scenarios 3 and 4. A possible explanation is that the probability ${\rm pr}(S=0\mid X)$ can be well approximated by a logistic regression under Scenarios 3 and 4 so that the entropy balancing calibration method can approximate the target population very well. Moreover, the entropy balancing estimators are nearly unbiased, the mean of estimated standard errors is close to the standard deviation of the estimators, and the empirical coverage probabilities of 95\% confidence intervals for both $V^+_{EB}(\beta^*_{EB})$ and $V^{\rm{t}}(\beta^{\rm{t}})$ are close to the nominal level for all settings. For the empirical likelihood method, as $n$ increases, the mean of estimators get closer to its true value $V^{+}_{EL}(\beta^*_{EL})$, the mean of estimated standard errors get closer to the standard deviation of estimators, and the empirical coverage probability of 95\% confidence intervals for $V^+_{EL}(\beta^*_{EL})$ get closer to the nominal level as expected. However, because of the difference between $V^+_{EL}(\beta^*_{EL})$ and $V^{\rm{t}}(\beta^{\rm{t}})$, the empirical coverage probability of 95\% confidence intervals for $V^{\rm{t}}(\beta^{\rm{t}})$ is lower than the nominal level for some settings even when $n$ increases to 1000. Finally, standard deviations of the estimators reported in Table \ref{table6} for implementation method II are generally smaller than the corresponding values reported in Table \ref{table5} for implementation method I. Such efficiency gains are mainly due to the nonparametric fit of the conditional mean outcome model in implementation method II
compared with the misspecified parametric conditional mean outcome model used in implementation method I.   

We also compared the original AIPW estimator without calibration with the calibrated AIPW estimators under Scenario 1, where the source and target populations are identical. As expected, all three estimators are consistent for the optimal value of the target population. The standard deviations of the original AIPW estimator for observational study with implementation method II are $0.55,0.27$ for $n=250, 1000$.  These values are  larger than the corresponding values of calibrated estimators in Table \ref{table6}, which supports the results established in Theorem \ref{thm3}.

\section{Real Data Analysis}
\label{sec:realdata}
We illustrate the proposed method using an application to data from the eICU collaborative research database (eICU-CRD) \citep{goldberger2000physiobank,pollard2018eicu,pollard2019icu} and the MIMIC-III
clinical database \citep{goldberger2000physiobank,johnson2016mimic,johnson2019mimic}. Specifically, we use the eICU dataset as the source population, while treating the MIMIC-III dataset as the target population. Both MIMIC-III and eICU data consist of patients who suffered from sepsis. The eICU-CRD is a multi-center ICU database comprising de-identified health-related data associated with over 200,000 admissions to ICUs across the US between 2014-2015. The MIMIC-III database is a single-center ICU database comprising de-identified health-related data associated with over 40,000 patients who stayed in critical care units of the Beth Israel Deaconess Medical Center between 2001 and 2012. It is likely that the populations in the two databases have some heterogeneity. 

Both eICU and MIMIC-III data collected information from ICU patients with sepsis disease, and thus contain common baseline covariates and treatment. In our study, we consider $p=7$ baseline covariates in both samples: age (years), admission weights (kg), admission temperature (Celsius), glucose level (mg/dL), blood urea nitrogen (BUN) amount (mg/dL), creatinine amount (mg/dL), white blood cell (WBC) count (K/uL). Here, treatment is coded as 1 if receiving the vasopressor, and 0 if receiving other medical supervisions such as IV fluid resuscitation. We consider the cumulative balance (mL) as the outcome of interest. A positive cumulative balance indicates that a patient’s fluid input is higher than their output. The condition describing excess fluid is known as hypervolaemia or fluid overload. In critically ill patients, fluid overload is related to increased mortality and also leads to several complications like pulmonary edema, cardiac failure, tissue breakdown, and impaired bowel function \citep{claure2016fluid}. A negative cumulative balance indicates that a patient’s fluid output is higher than their input. The condition describing inadequate fluid is known as hypovolaemia. Patients with severe hypovolemia can develop ischemic injury of vital organs, leading to multi-system organ failure \citep{taghavi2021hypovolemic}.  We use $Y=-|{\text{cumulative 
		balance}}|$ as the outcome, so a larger value of the outcome is better. After removing abnormal values, the MIMIC-III dataset consists of 10746 subjects, among which 2242 patients were treated with the vasopressor, while the rest were treated with other medical supervisions. The MIMIC-III data is treated as the target population. We sample $n=1000$ subjects from the eICU dataset as the source sample, among which 271 patients were treated with the vasopressor, while the rest were treated with other medical supervisions. Table \ref{table7} summarizes the mean and standard deviation of the outcome and covariates in the source and target samples. We can see some differences in the means of some covariates, such as glucose level, blood urea nitrogen amount, and WBC count. 

\begin{table}
	\def~{\hphantom{0}}
	\caption{Mean and standard deviations (in parenthesis) of baseline characteristics in the source and target datasets.}
		\spacingset{1.45}
		\centering
		\begin{tabular}{lll}
			& Source & Target\\[5pt]
			$-|\text{Cumulative Balance}|$ $(Y)$ & -1746.6 (1561.3) & -1785.0 (1246.6) \\
			Age $(X_1)$ & 65.7 (15.1) & 66.5 (16.5) \\
			Admission Weights $(X_2)$ & 80.0 (22.9) & 79.7 (20.7) \\
			Admission Temperature $(X_3)$ & 36.5 (1.1) & 36.8 (0.8)\\
			Glucose $(X_4)$ & 158.6 (102.7) & 145.6 (72.4)\\
			BUN $(X_5)$ & 31.7 (20.3) & 27.9 (18.4) \\
			Creatinine $(X_6)$  & 1.8 (1.5) & 1.5 (1.4)\\
			WBC $(X_7)$ & 14.4 (8.4) & 12.0 (6.5) \\
	\end{tabular}
	\label{table7}
\end{table}

We used the means of all seven covariates of the target population as the summary statistics to estimate the calibration weights by the entropy balancing and empirical likelihood methods.  We computed three optimal linear ITRs, $d(x;\widehat{\beta}^{\rm{o}})$, $d(x;\widehat{\beta}^{\rm{c}}_{EB})$, and $d(x;\widehat{\beta}^{\rm{c}}_{EL})$ by maximizing the original and calibrated AIPW value function estimators based on the source sample. In our implementation, the propensity score model was estimated using a logistic regression with all covariates and the conditional mean outcome model was estimated using the random forest for treatments 0 and 1 separately. To assess the performance of these three estimated optimal ITRs for treatment decisions in the target population, we apply them to random samples drawn from the target population. Specifically, we randomly sample $N=1000$ subjects from the MIMIC-III data as the target sample and repeat this sampling procedure 100 times. 
We have individual-level data from the target population, which can be used as the benchmark for evaluation. For a given ITR $d(x;\beta)$, we computed the AIPW estimator of its value function based on the target sample by  
$$\widehat{V}^{\rm{t}}(\beta)=\frac{1}{N}\sum\limits_{i=1}^N\left[\frac{I\{A_i^{\rm{t}}=d(X_i^{\rm{t}};\beta)\}}{\varrho^{\rm{t}}(A_i^{\rm{t}}\mid X_i^{\rm{t}};\widehat{\eta})}\{Y_i^{\rm{t}}-\widehat{\mu}_d^{\rm{t}}(X_i^{\rm{t}};\beta)\}+\widehat{\mu}_d^{\rm{t}}(X_i^{\rm{t}};\beta)\right],$$
where  $\varrho^{\rm{t}}(A_i^{\rm{t}}\mid X_i^{\rm{t}};\widehat{\eta})=\pi^{\rm{t}}(X_i^{\rm{t}};\widehat{\eta})A_i^{\rm{t}}+\{1-\pi^{\rm{t}}(X_i^{\rm{t}};\widehat{\eta})\}(1-A_i^{\rm{t}})$, $\widehat{\mu}_d^{\rm{t}}(X_i^{\rm{t}};\beta)=\widehat{\mu}^{\rm{t}}(X_i^{\rm{t}},1)I\{d(X_i^{\rm{t}};\beta)=1\}+\widehat{\mu}^{\rm{t}}(X_i^{\rm{t}},0)I\{d(X_i^{\rm{t}};\beta)=0\}$,   the propensity score $\pi^{\rm{t}}(X_i^{\rm{t}};\widehat{\eta})$ was estimated using a logistic regression model, and the conditional mean outcome models $\widehat{\mu}^{\rm{t}}(X_i^{\rm{t}},a)$, $a=0,1$, were estimated using random forest. 

Let $\widehat{\beta}^{\rm{oracle}} = \argmax_{\beta}\widehat{V}^{\rm{t}}(\beta)$. Then, $d(x;\widehat{\beta}^{\rm{oracle}})$ is the optimal linear ITR for the target sample and $\widehat{V}^{\rm{t}}(\widehat{\beta}^{\rm{oracle}})$ is the associated optimal value, which can serve as the benchmark. We also computed the estimated values of three estimated ITRs $d(x;\widehat{\beta}^{\rm{o}})$, $d(x;\widehat{\beta}^{\rm{c}}_{EB})$, and $d(x;\widehat{\beta}^{\rm{c}}_{EL})$ by $\widehat{V}^{\rm{t}}(\widehat{\beta}^{\rm{o}})$, $\widehat{V}^{\rm{t}}(\widehat{\beta}^{\rm{c}}_{EB})$, and $\widehat{V}^{\rm{t}}(\widehat{\beta}^{\rm{c}}_{EL})$, respectively, and their associated percentages of correct decisions, defined as $1-N^{-1}\sum_{i=1}^N |d(X_i^{\rm{t}};\widehat{\beta})-d(X_i^{\rm{t}};\widehat{\beta}^{\rm{oracle}})|$ for an estimated ITR $d(x;\widehat{\beta})$.  Table \ref{table8} summarize the means and standard deviations of the value estimators and percentages of correct decisions over 100 replications. We can see that the ITRs obtained using the proposed calibration methods have much better performance than the original AIPW estimator without calibration. Their estimated values are much closer to the optimal value computed using the target samples and the associated percentages of correct decisions are much closer to 1. Moreover, the ITR obtained using the entropy balancing method has slightly better performance than the one obtained using the  empirical likelihood method in terms of both value and percentage of correct decisions.   

\begin{table}
	\def~{\hphantom{0}}
	\caption{Mean and standard deviations (in parenthesis) of the value estimators and percentages of correct decisions (PCD).}
		\spacingset{1.45}
	\centering
		\begin{tabular}{ccccccccc}   
			& Oracle & Entropy Balancing & Empirical Likelihood & Original\\[5pt]
			Value & -1674.1 (47.2) & -1752.0 (49.5) & -1773.8 (49.3) & -1945.3 (75.6)\\
			PCD & / & 0.80 (0.1) & 0.76 (0.1) & 0.31 (0.1)\\
	\end{tabular}
	\label{table8}
\end{table}

\bibliographystyle{agsm}
\bibliography{main}

\newpage
 \spacingset{1}
\begin{center}
	\textbf{\Large{}{}Supplementary Material for ``Targeted Optimal Treatment Regime Learning Using Summary Statistic"}{\Large{} }{\Large\par}
	\par\end{center}
\begin{center}
	\large{Jianing Chu, Wenbin Lu, Shu Yang\\Department of Statistics, North Carolina State University}
	\par\end{center}
\appendix
\pagenumbering{arabic} %reset page counter to 1
\renewcommand*{\thepage}{S\arabic{page}}
\setcounter{figure}{0} 
\renewcommand\thefigure{S\arabic{figure}}
\setcounter{table}{0} 
\renewcommand\thetable{S\arabic{table}}
\renewcommand\thesection{S\arabic{section}}

 \spacingset{1.8}
	\section{Proof of Theorem \ref{thm1}}
\label{sec:thm1}

The proof of Theorem \ref{thm1} consists of three steps as follows.

\noindent\textit{Step 1:} We show that $\widehat{V}^{\rm{c}}(\beta)=V^+(\beta)+o_p(1)$. 
Let 
$$\psi(Y,A,X;\beta,\theta,\eta)=\frac{I\{A=d(X;\beta)\}}{\pi(X;\eta)A+\{1-\pi(X;\eta)\}(1-A)}\{Y-\mu_d(X;\beta,\theta)\}+\mu_d(X;\beta,\theta).$$

By the definition of $V^+(\beta)$,
\begin{align*}
	V^+(\beta)&=E^+\left\{\psi(Y,A,X;\beta,\theta^*,\eta^*)\right\}\tag{S1}\label{S1}\\
	&=E^+\left[E\{\psi(Y,A,X;\beta,\theta^*,\eta^*)\mid X\}\right]\\
	&=\int E\{\psi(Y,A,X;\beta,\theta^*,\eta^*)\mid X\} \{W^*(X,\lambda^*)f^{\rm{s}}(X)\}dX\\
	&=E\left[W^*(X,\lambda^*)E\{\psi(Y,A,X;\beta,\theta^*,\eta^*)\mid X\}\right]\\
	&=E\left[E\left\{W^*(X;\lambda^*)\psi(Y,A,X;\beta,\theta^*,\eta^*)\mid X\right\}\right]\\
	&=E\left\{W^*(X;\lambda^*)\psi(Y,A,X;\beta,\theta^*,\eta^*)\right\},
\end{align*}
where $E^+$ denotes the expectation of the pseudo population $\mathbb{P}^+$, and (\ref{S1}) follows from the double robustness property of the original AIPW estimator \citep{zhang2012robust}. By the strong law of large numbers and uniform consistency, we have $\widehat{V}^{\rm{c}}(\beta)=V^+(\beta)+o_p(1)$.

\noindent\textit{Step 2:} We show that $n^{1/3}\|\widehat{\beta}^{\rm{c}}-\beta^*\|_2=O_p(1)$, where $\|\cdot\|_2$ is the $L_2$ norm.

(A.) First, we show that $\widehat{\beta}^{\rm{c}}$ converges in probability to $\beta^*$ as $n \to \infty$, by checking three conditions for the Argmax Theorem:

(a1.) By  (A9), the true value function $V^+(\beta)$ is twice continuously differentiable in a neighborhood of $\beta^*$.

(a2.) In Step 1, we have shown that  for any $\beta$,
$$\widehat{V}^{\rm{c}}(\beta)=V^+(\beta)+o_p(1).$$

(a3.) Since $\widehat{\beta}^{\rm{c}}=\argmax\limits_{\beta:\|\beta\|_2=1} \widehat{V}^{\rm{c}}(\beta)$, we have the estimated ITR as $d(X,\widehat{\beta}^{\rm{c}})=I(\tilde{X}^{\T}\widehat{\beta}^{\rm{c}}>0)$ and the corresponding value function $\widehat{V}^{\rm{c}}(\widehat{\beta}^{\rm{c}})$ such that
$$\widehat{V}^{\rm{c}}(\widehat{\beta}^{\rm{c}})\geq\sup\limits_{\beta:\|\beta\|_2=1}\widehat{V}^{\rm{c}}(\beta).$$
Thus we have $\widehat{\beta}^{\rm{c}} \rightarrow \beta^*$, in probability, as $n\to\infty$.

(B.) Next, we show that the convergence rate of $\widehat{\beta}^{\rm{c}}$ is $n^{1/3}$, i.e., $n^{1/3}\|\widehat{\beta}^{\rm{c}}-\beta^*\|_2=O_p(1)$. We check three conditions of the Theorem 14.4: Rate of convergence in \cite{kosorok2008introduction}:

(b1.) For every $\beta$ in a neighborhood of $\beta^*$, i.e., $\| \beta-\beta^*\|_2 < \varepsilon$ for some $\varepsilon>0$, by  (A9), we take the second order Taylor expansion on $V^+(\beta)$ at $\beta=\beta^*$,
\begin{align*}
	V^+(\beta)-V^+(\beta^*)&=(V^+)'(\beta^*)\| \beta-\beta^*\|_2+\frac{1}{2}(V^+)''(\beta^*)\| \beta-\beta^*\|_2^2+o\left(\|\beta-\beta^*\|_2^2\right)\\
	&=\frac{1}{2}(V^+)''(\beta^*)\| \beta-\beta^*\|_2^2+o\left(\|\beta-\beta^*\|_2^2\right) ~\ (\text{by}~\ (V^+)'(\beta^*)=0).
\end{align*}
Since $(V^+)''(\beta^*)<0$, there exists  $c_0=-\frac{1}{2}(V^+)''(\beta^*)>0$ such that $V^+(\beta)-V^+(\beta^*)<c_0\|\beta-\beta^*\|^2_2$ holds. 

(b2.)  Define
$$V_n^*(\beta)=\frac{1}{n}\sum_{i=1}^nW(\lambda^*;X_i)\left[\frac{I\{A=d(X;\beta)\}}{\pi(X;\eta^*)A+\{1-\pi(X;\eta^*)\}(1-A)}\{Y-\mu_d(X;\beta,\theta^*)\}+\mu_d(X;\beta,\theta^*)\right].$$

For all $n$ large enough and sufficiently small $\varepsilon$, the centered process $\widehat{V}^{\rm{c}}-V^+$ satisfies 
\begin{align*}
	&E^*\left[n^{1/2} \sup\limits_{\|\beta-\beta^*\|_2<\varepsilon}\bigg|\widehat{V}^{\rm{c}}(\beta)-V^+(\beta)-\{\widehat{V}^{\rm{c}}(\beta^*)-V^+(\beta^*)\}\bigg|\right]\\
	=&E^*\left[n^{1/2} \sup\limits_{\|\beta-\beta^*\|_2<\varepsilon}\bigg|\widehat{V}^{\rm{c}}(\beta)-V_n^*(\beta)+V_n^*(\beta)-V^+(\beta)-\{\widehat{V}^{\rm{c}}(\beta^*)-V_n^*(\beta^*)+V_n^*(\beta^*)-V^+(\beta^*)\}\bigg|\right]\\
	\leq&  \underbrace{E^*\left[n^{1/2} \sup\limits_{\|\beta-\beta^*\|_2<\varepsilon}\bigg|\widehat{V}^{\rm{c}}(\beta)-V_n^*(\beta)       -\left\{\widehat{V}^{\rm{c}}(\beta^*)-V_n^*(\beta^*)\right\}\bigg|\right]}_{\tau_1}\\
	&+\underbrace{E^*\left[n^{1/2} \sup\limits_{\|\beta-\beta^*\|_2<\varepsilon}\bigg|V_n^*(\beta)-V^+(\beta)-\left\{V_n^*(\beta^*)-V^+(\beta^*)\right\}\bigg|\right]}_{\tau_2},\tag{S2}\label{S2}
\end{align*}
where $E^*$ is the outer expectation. 

We first derive two results (b2.1) and (b2.2) to bound $\tau_1$ and $\tau_2$, respectively, and then we are able to show that the second condition of Theorem 14.4 in \cite{kosorok2008introduction} is satisfied. 

%With the fact that 
%\begin{align*}
%I\{A=d(X;\beta)\}-I\{A=d(X;\beta^*)\}=(2A-1)\left\{I\left(\tilde{x}^{\T}\beta>0\right)-I\left(\tilde{x}^{\T}\beta^*>0\right)\right\},\\
%\mu_d(X;\beta,\theta^*)-\mu_d(X;\beta^*,\theta^*)=\{\mu(X,1;\theta^*)-\mu(X,0;\theta^*)\}\left\{I\left(\tilde{x}^{\T}\beta>0\right)-I\left(\tilde{x}^{\T}\beta^*>0\right)\right\},
%\end{align*}
%\begin{align*}
%&\mu_d(X;\beta,\theta^*)I\{A=d(X;\beta)\}-\mu_d(X;\beta^*,\theta^*)I\{A=d(X;\beta^*)\}\\
%=&\left\{\mu(X,1;\theta^*)A-\mu(X,0;\theta^*)(1-A)\right\}\left\{I\left(\tilde{x}^{\T}\beta>0\right)-I\left(\tilde{x}^{\T}\beta^*>0\right)\right\},
%\end{align*}
%we have 
(b2.1) 
\begin{align*}
	&V_n^*(\beta)-V_n^*(\beta^*)\\
	=&\frac{1}{n}\sum_{i=1}^nW(\lambda^*;X_i)\Bigg[\frac{I\{A=d(X_i;\beta)\}}{\pi(X_i;\eta^*)A_i+\{1-\pi(X_i;\eta^*)\}(1-A_i)}\{Y-\mu_d(X_i;\beta,\theta^*)\}+\mu_d(X_i;\beta,\theta^*)\\
	&-\frac{I\{A=d(X_i;\beta^*)\}}{\pi(X_i;\eta^*)A_i+\{1-\pi(X_i;\eta^*)\}(1-A_i)}\{Y-\mu_d(X_i;\beta^*,\theta^*)\}-\mu_d(X_i;\beta^*,\theta^*)\Bigg]\\
	=&\frac{1}{n}\sum_{i=1}^nW(\lambda^*;X_i)\Bigg[\frac{(2A_i-1)Y_i-\mu(X_i,1;\theta^*)A_i+\mu(X_i,0;\theta^*)(1-A_i)}{\pi(X_i;\eta^*)A_i+\{1-\pi(X_i;\eta^*)\}(1-A_i)}+\mu(X_i,1;\theta^*)-\mu(X_i,0;\theta^*)\Bigg]\\
	& \{I(\tilde{X}_i^{\T}\beta>0)-I(\tilde{X}_i^{\T}\beta^*>0)\}.
\end{align*}

We define a class of functions 
\begin{align*}
	\mathcal{F}^1_{\beta}(y,a,x)=&\Bigg\{W(\lambda^*;x)\left[\frac{(2a-1)y-\mu(x,1;\theta^*)a+\mu(x,0;\theta^*)(1-a)}{\pi(x;\eta^*)a+\{1-\pi(x;\eta^*)\}(1-a)}+\mu(x,1;\theta^*)-\mu(x,0;\theta^*)\right]\\
	&\left\{I\left(\tilde{x}^{\T}\beta>0\right)-I\left(\tilde{x}^{\T}\beta^*>0\right)\right\}: \|\beta-\beta^*\|_2<\varepsilon\Bigg\},
\end{align*}
where $\tilde{x}=(1,x^{\T})^{\T}$.

Let $M_1=\sup \left|W(\lambda^*;x)\left[\frac{(2a-1)y-\mu(x,1;\theta^*)a+\mu(x,0;\theta^*)(1-a)}{\pi(x;\eta^*)a+\{1-\pi(x;\eta^*)\}(1-a)}+\mu(x,1;\theta^*)-\mu(x,0;\theta^*)\right]\right|$. By (A6), (A8) and (A11), we have $M_1<\infty$. By (A6), there exists a constant $0<k_0<\infty$ s.t. $|\tilde{x}^{\T}(\beta-\beta^*)|<k_0\varepsilon$ when $\|\beta-\beta^*\|_2<\varepsilon$. For the indicator function $I\left(-k_0\varepsilon\leq\tilde{x}^{\T}\beta^*\leq k_0\varepsilon\right)$, 

(i) when $-k_0\varepsilon\leq\tilde{x}^{\T}\beta^*\leq k_0\varepsilon$, $$I\left(-k_0\varepsilon\leq\tilde{x}^{\T}\beta^*\leq k_0\varepsilon\right)=1\geq \left|I\left(\tilde{x}^{\T}\beta>0\right)-I\left(\tilde{x}^{\T}\beta^*>0\right)\right|;$$ 

(ii) when $\tilde{x}^{\T}\beta^*> k_0\varepsilon$, $\tilde{x}^{\T}\beta=\tilde{x}^{\T}(\beta-\beta^*)+\tilde{x}^{\T}\beta^*> -k_0\varepsilon+k_0\varepsilon>0$, $$I\left(-k_0\varepsilon\leq\tilde{x}^{\T}\beta^*\leq k_0\varepsilon\right)=0= \left|I\left(\tilde{x}^{\T}\beta>0\right)-I\left(\tilde{x}^{\T}\beta^*>0\right)\right|;$$

(iii) when $\tilde{x}^{\T}\beta^* < -k_0\varepsilon$, $\tilde{x}^{\T}\beta=\tilde{x}^{\T}(\beta-\beta^*)+\tilde{x}^{\T}\beta^*< k_0\varepsilon+(-k_0\varepsilon)<0$, $$I\left(-k_0\varepsilon\leq\tilde{x}^{\T}\beta^*\leq k_0\varepsilon\right)=0= \left|I\left(\tilde{x}^{\T}\beta>0\right)-I\left(\tilde{x}^{\T}\beta^*>0\right)\right|.$$ 

Therefore, we always have $I\left(-k_0\varepsilon\leq\tilde{x}^{\T}\beta^*\leq k_0\varepsilon\right)\geq \left|I\left(\tilde{x}^{\T}\beta>0\right)-I\left(\tilde{x}^{\T}\beta^*>0\right)\right|$ when $\|\beta-\beta^*\|_2<\varepsilon$. 

We then define the envelope of $\mathcal{F}^1_{\beta}(y,a,x)$ as $F_1=M_1I\left(-k_0\varepsilon\leq\tilde{x}^{\T}\beta^*\leq k_0\varepsilon\right)$. By (A10), there exits a positive constant $k_1$ such that
$$\|F_1\|_{P,2}=M_1\sqrt{{\rm pr}\left(-k_0\varepsilon\leq\tilde{x}^{\T}\beta^*\leq k_0\varepsilon\right)}\leq M_1\sqrt{k_1\cdot 2k_0\varepsilon}=M_1\sqrt{2k_0k_1}\varepsilon^{1/2}<\infty.$$

Since $\mathcal{F}^1_{\beta}$ is a class of indicator functions, by the conclusion of Lemma 9.6 and Lemma 9.9 in \cite{kosorok2008introduction}, $\mathcal{F}_{\beta}^1$ is a VC class of functions. Thus, the entropy of $\mathcal{F}^1_{\beta}$, denoted as $J^*_{[]}(1,\mathcal{F}^1)$, is finite, i.e., $J^*_{[]}(1,\mathcal{F}^1)<\infty$.
Next, we consider the following empirical process indexed by $\beta$,
$$\mathbb{G}_n\mathcal{F}^1_{\beta}=n^{-1/2}\sum_{i=1}^n\left[\mathcal{F}^1_{\beta}\left(Y_i,A_i,X_i\right)-E\left\{\mathcal{F}^1_{\beta}\left(Y_i,A_i,X_i\right)\right\}\right].$$
Note that $\mathbb{G}_n\mathcal{F}^1_{\beta}=n^{1/2}\left[V_n^*(\beta)-V_n^*(\beta^*)-\left\{V^+(\beta)-V^+(\beta^*)\right\}\right]$. By applying Theorem 11.2 in \cite{kosorok2008introduction}, we have 
\begin{align*}
	\tau_2&=E^*\left[n^{1/2} \sup\limits_{\|\beta-\beta^*\|_2<\varepsilon}\bigg|V_n^*(\beta)-V^+(\beta)-\left\{V_n^*(\beta^*)-V^+(\beta^*)\right\}\bigg|\right]\\
	&=E^*\left( \sup\limits_{\|\beta-\beta^*\|_2<\varepsilon}\left|\mathbb{G}_n\mathcal{F}^1_{\beta}\right|\right)\leq c_1J^*_{[]}(1,\mathcal{F}^1)\|F_1\|_{P,2}\leq c_1J^*_{[]}(1,\mathcal{F}^1)M_1\sqrt{2k_0k_1}\varepsilon^{1/2},
\end{align*}
where $c_1$ is a finite constant.

Let $C_1^*\equiv c_1J^*_{[]}(1,\mathcal{F}^1)M_1\sqrt{2k_0k_1}$, since $c_1, J^*_{[]}(1,\mathcal{F}^1)$, $M_1$, $k_0$ and $k_1$ are bounded, we have $C_1^*<\infty$, i.e.,
\begin{align*}
	\tau_2\leq C_1^*\varepsilon^{1/2}.\tag{S3}\label{S3}
\end{align*}

(b2.2) We rewrite the form of $\widehat{V}^{\rm{c}}(\beta)-V_n^*(\beta)       -\left\{\widehat{V}^{\rm{c}}(\beta^*)-V_n^*(\beta^*)\right\}$ as 
\begin{align*}
	&\widehat{V}^{\rm{c}}(\beta)-V_n^*(\beta)-\left\{\widehat{V}^{\rm{c}}(\beta^*)-V_n^*(\beta^*)\right\}=\widehat{V}^{\rm{c}}(\beta)-\widehat{V}^{\rm{c}}(\beta^*)-\left\{V_n^*(\beta)-V_n^*(\beta^*)\right\}\\
	=&\frac{1}{n}\sum_{i=1}^n\Bigg(W(\widehat{\lambda};X_i)\bigg[\frac{(2A_i-1)Y_i-\mu(X_i,1;\widehat{\theta})A_i+\mu(X_i,0;\widehat{\theta})(1-A_i)}{\pi(X_i;\widehat{\eta})A_i+\{1-\pi(X_i;\widehat{\eta})\}(1-A_i)}+\mu(X_i,1;\widehat{\theta})-\mu(X_i,0;\widehat{\theta})\bigg]\\
	&\qquad\quad- W(\lambda^*;X_i)\bigg[\frac{(2A_i-1)Y_i-\mu(X_i,1;\theta^*)A_i+\mu(X_i,0;\theta^*)(1-A_i)}{\pi(X_i;\eta^*)A_i+\{1-\pi(X_i;\eta^*)\}(1-A_i)}+\mu(X_i,1;\theta^*)-\mu(X_i,0;\theta^*)\bigg]\Bigg)\\
	& \qquad ~\ \{I(\tilde{X}_i^{\T}\beta>0)-I(\tilde{X}_i^{\T}\beta^*>0)\}.
\end{align*}

We take the Taylor expansion on the above equation at $(\lambda^*,\theta^*,\eta^*)$,
\begin{align*}
	&\widehat{V}^{\rm{c}}(\beta)-V_n^*(\beta)-\left\{\widehat{V}^{\rm{c}}(\beta^*)-V_n^*(\beta^*)\right\}\\
	=&\frac{1}{n}\sum_{i=1}^n \Bigg(\bigg[\frac{(2A_i-1)Y_i-\mu(X_i,1;\theta^*)A_i+\mu(X_i,0;\theta^*)(1-A_i)}{\pi(X_i;\eta^*)A_i+\{1-\pi(X_i;\eta^*)\}(1-A_i)}+\mu(X_i,1;\theta^*)-\mu(X_i,0;\theta^*)\bigg]\\
	&\qquad \quad \left\{\frac{\partial W(X_i;\lambda^*)}{\partial\lambda}\right\}^{\T}\left(\widehat{\lambda}-\lambda^*\right)+\\
	&\qquad \quad  W(\lambda^*;X_i)\bigg[\frac{-\{\partial\mu(X_i,1;\theta^*)/\partial\theta\}^{\T}A_i+\{\partial\mu(X_i,0;\theta^*)/\partial\theta\}^{\T}(1-A_i)}{\pi(X_i;\eta^*)A_i+\{1-\pi(X_i;\eta^*)\}(1-A_i)}+\\
	&\qquad \quad \left\{\frac{\partial\mu(X_i,1;\theta^*)}{\partial \theta}\right\}^{\T}
	-\left\{\frac{\partial\mu(X_i,0;\theta^*)}{\partial \theta}\right\}^{\T}\bigg]\left(\widehat{\theta}-\theta^*\right)-\\
	&\qquad  \quad W(\lambda^*;X_i)(2A_i-1)\frac{(2A_i-1)Y_i-\mu(X_i,1;\theta^*)A_i+\mu(X_i,0;\theta^*)(1-A_i)}{\left[\pi(X_i;\eta^*)A_i+\{1-\pi(X_i;\eta^*)\}(1-A_i)\right]^2} \\
	&\qquad \quad \left\{\frac{\partial \pi(X_i;\eta^*)}{\partial\eta}\right\}^{\T}
	\left(\widehat{\eta}-\eta^*\right)\Bigg)+o_p(n^{-1/2}).
\end{align*}

Next, we define three classes of functions,
\begin{align*}\mathcal{F}^2_{\beta}(y,a,x)=\Bigg\{&\bigg[\frac{(2a-1)y-\mu(x,1;\theta^*)a+\mu(x,0;\theta^*)(1-a)}{\pi(x;\eta^*)a+\{1-\pi(x;\eta^*)\}(1-a)}+\mu(x,1;\theta^*)-\mu(x,0;\theta^*)\bigg]\\
	&\left\{\frac{\partial W(x;\lambda^*)}{\partial\lambda}\right\}^{\T}\left\{I\left(\tilde{x}^{\T}\beta>0\right)-I\left(\tilde{x}^{\T}\beta^*>0\right)\right\}: \|\beta-\beta^*\|_2<\varepsilon\Bigg\},
\end{align*}
\begin{align*}
	\mathcal{F}^3_{\beta}(a,x)=\Bigg\{&W(\lambda^*;x)\Bigg[\frac{-\{\partial\mu(x,1;\theta^*)/\partial\theta\}^{\T}a+\{\partial\mu(x,0;\theta^*)/\partial\theta\}^{\T}(1-a)}{\pi(x;\eta^*)a+\{1-\pi(x;\eta^*)\}(1-a)}+\left\{\frac{\partial\mu(x,1;\theta^*)}{\partial \theta}\right\}^{\T}\\
	&-\left\{\frac{\partial\mu(x,0;\theta^*)}{\partial \theta}\right\}^{\T}\Bigg]\left\{I\left(\tilde{x}^{\T}\beta>0\right)-I\left(\tilde{x}^{\T}\beta^*>0\right)\right\}: \|\beta-\beta^*\|_2<\varepsilon\Bigg\},
\end{align*}
\begin{align*}
	\mathcal{F}^4_{\beta}(y,a,x)=\Bigg\{&-W(\lambda^*;x)(2a-1)\frac{(2a-1)y-\mu(x,1;\theta^*)a+\mu(x,0;\theta^*)(1-a)}{[\pi(x;\eta^*)a+\{1-\pi(x;\eta^*)\}(1-a)]^2}\\
	& \left\{\frac{\partial \pi(x;\eta^*)}{\partial\eta}\right\}^{\T}
	\left\{I\left(\tilde{x}^{\T}\beta>0\right)-I\left(\tilde{x}^{\T}\beta^*>0\right)\right\}: \|\beta-\beta^*\|_2<\varepsilon\Bigg\}.
\end{align*}

Let 
$$M_2=\sup\left|\bigg[\frac{(2a-1)y-\mu(x,1;\theta^*)a+\mu(x,0;\theta^*)(1-a)}{\pi(x;\eta^*)a+\{1-\pi(x;\eta^*)\}(1-a)}+\mu(x,1;\theta^*)-\mu(x,0;\theta^*)\bigg]\left\{\frac{\partial W(x;\lambda^*)}{\partial\lambda}\right\}^{\T}\right|,$$
\begin{align*}
	M_3=\sup\Bigg|&W(\lambda^*;x)\Bigg[\frac{-\{\partial\mu(x,1;\theta^*)/\partial\theta\}^{\T}a+\{\partial\mu(x,0;\theta^*)/\partial\theta\}^{\T}(1-a)}{\pi(x;\eta^*)a+\{1-\pi(x;\eta^*)\}(1-a)}+\left\{\frac{\partial\mu(x,1;\theta^*)}{\partial \theta}\right\}^{\T}\\
	&-\left\{\frac{\partial\mu(x,0;\theta^*)}{\partial \theta}\right\}^{\T}\Bigg]\Bigg|,
\end{align*}
\begin{align*}
	M_4=\sup\left|-W(\lambda^*;x)(2a-1)\frac{(2a-1)y-\mu(x,1;\theta^*)a+\mu(x,0;\theta^*)(1-a)}{[\pi(x;\eta^*)a+\{1-\pi(x;\eta^*)\}(1-a)]^2} \left\{\frac{\partial \pi(x;\eta^*)}{\partial\eta}\right\}^{\T}\right|.
\end{align*}

By (A6), (A8) and (A11), $M_2,M_3,M_4<\infty$.  Define the envelop of $\mathcal{F}_{\beta}^j$ as $F_j=M_jI\left(-k_0\varepsilon\leq\tilde{x}^{\T}\beta^*\leq k_0\varepsilon\right)$, for $j=2,3,4$. Similarly to (b2.1), we have 
$$\|F_j\|_{P,2}=M_j\sqrt{2k_0k_1}\varepsilon^{1/2}<\infty,$$
and $\mathcal{F}_{\beta}^j$ is a VC class of functions. Thus, the entropy of  $\mathcal{F}^j_{\beta}$, denoted as $J^*_{[]}(1,\mathcal{F}^j)$, is finite, i.e., $J^*_{[]}(1,\mathcal{F}^j)<\infty$, for $j=2,3,4$.
We construct three empirical processes indexed by $\beta$,
$$\mathbb{G}_n\mathcal{F}^j_{\beta}=n^{-1/2}\sum_{i=1}^n\left[\mathcal{F}^j_{\beta}\left(Y_i,A_i,X_i\right)-E\left\{\mathcal{F}^j_{\beta}\left(Y_i,A_i,X_i\right)\right\}\right].$$
By applying Theorem 11.2 in \cite{kosorok2008introduction}, we have 
$$E^*\left( \sup\limits_{\|\beta-\beta^*\|_2<\varepsilon}\left\|\mathbb{G}_n\mathcal{F}^j_{\beta}\right\|_1\right)\leq c_jJ^*_{[]}(1,\mathcal{F}^j)\|F_j\|_{P,2},    \quad j=2,3,4,$$
where $c_2$, $c_3$ and $c_4$ are finite constants. 

By Theorem 2.14.5 in \cite{van1996weak}, 
\begin{align*}
	\left\{E^*\left( \sup\limits_{\|\beta-\beta^*\|_2<\varepsilon}\left\|\mathbb{G}_n\mathcal{F}^j_{\beta}\right\|_2^2\right)\right\}^{1/2}&\leq l_j\left\{E^*\left( \sup\limits_{\|\beta-\beta^*\|_2<\varepsilon}\left\|\mathbb{G}_n\mathcal{F}^j_{\beta}\right\|_1\right)+\|F_j\|_{P,2}\right\}\\
	&\leq l_j\left\{c_jJ^*_{[]}(1,\mathcal{F}^j)+1\right\}\|F_j\|_{P,2}\\
	&= l_j\left\{c_jJ^*_{[]}(1,\mathcal{F}^j)+1\right\}M_j\sqrt{2k_0k_1}\varepsilon^{1/2}.
\end{align*}
where  $l_2$, $l_3$ and $l_4$ are finite constants. 
Let 
$$C_j^*\equiv l_j\left\{c_jJ^*_{[]}(1,\mathcal{F}^j)+1\right\}M_j\sqrt{2k_0k_1}<\infty,$$
and we have 
$$\left\{E^*\left( \sup\limits_{\|\beta-\beta^*\|_2<\varepsilon}\left\|\mathbb{G}_n\mathcal{F}^j_{\beta}\right\|_2^2\right)\right\}^{1/2}\leq C_j\varepsilon^{1/2}, \quad \quad j=2,3,4.$$
Note that
\begin{align*}
	\tau_1&=E^*\left[n^{1/2} \sup\limits_{\|\beta-\beta^*\|_2<\varepsilon}\bigg|\widehat{V}^{\rm{c}}(\beta)-V_n^*(\beta)       -\left\{\widehat{V}^{\rm{c}}(\beta^*)-V_n^*(\beta^*)\right\}\bigg|\right]\\
	&=E^*\left\{ \sup\limits_{\|\beta-\beta^*\|_2<\varepsilon}\left|(\widehat{\lambda}-\lambda^*)^\T\mathbb{G}_n\mathcal{F}^2_{\beta}+(\widehat{\theta}-\theta^*)^\T\mathbb{G}_n\mathcal{F}^3_{\beta}+(\widehat{\eta}-\eta^*)^\T\mathbb{G}_n\mathcal{F}^4_{\beta}+o_p(1)\right|\right\}\\
	&\leq n^{-1/2}\Bigg[E^*\left\{ \sup\limits_{\|\beta-\beta^*\|_2<\varepsilon}\left|n^{1/2}(\widehat{\lambda}-\lambda^*)^\T\mathbb{G}_n\mathcal{F}^2_{\beta}\right|\right\}+E^*\left\{\sup\limits_{\|\beta-\beta^*\|_2<\varepsilon}\left|n^{1/2}(\widehat{\theta}-\theta^*)^\T\mathbb{G}_n\mathcal{F}^3_{\beta}\right|\right\}\\
	&\quad +E^*\left\{ \sup\limits_{\|\beta-\beta^*\|_2<\varepsilon}\left|n^{1/2}(\widehat{\eta}-\eta^*)^\T\mathbb{G}_n\mathcal{F}^4_{\beta}\right|\right\}\Bigg].
\end{align*}

By the Cauchy-Schwarz Inequality, we have
\begin{align*}
	\tau_1&\leq n^{-1/2}\left\{E\left(n\|\widehat{\lambda}-\lambda^*\|_2^2\right)\right\}^{1/2}\left\{E^*\left( \sup\limits_{\|\beta-\beta^*\|_2<\varepsilon}\left\|\mathbb{G}_n\mathcal{F}^2_{\beta}\right\|_2^2\right)\right\}^{1/2}\\
	&+n^{-1/2}\left\{E\left(n\|\widehat{\theta}-\theta^*\|_2^2\right)\right\}^{1/2}\left\{E^*\left( \sup\limits_{\|\beta-\beta^*\|_2<\varepsilon}\left\|\mathbb{G}_n\mathcal{F}^3_{\beta}\right\|_2^2\right)\right\}^{1/2}\\
	&+n^{-1/2}\left\{E\left(n\|\widehat{\eta}-\eta^*\|_2^2\right)\right\}^{1/2}\left\{E^*\left( \sup\limits_{\|\beta-\beta^*\|_2<\varepsilon}\left\|\mathbb{G}_n\mathcal{F}^4_{\beta}\right\|_2^2\right)\right\}^{1/2}.
\end{align*}

By  (A11), $\sqrt{n}(\widehat{\lambda}-\lambda^*)=O_p(1)$,  $\sqrt{n}(\widehat{\theta}-\theta^*)=O_p(1)$ and $\sqrt{n}(\widehat{\eta}-\eta^*)=O_p(1)$, we have $M_{\lambda}\equiv \left\{E\left(n\|\widehat{\lambda}-\lambda^*\|_2^2\right)\right\}^{1/2}<\infty$ ,  $M_{\theta}\equiv \left\{E\left(n\|\widehat{\theta}-\theta^*\|_2^2\right)\right\}^{1/2}<\infty$, and  $M_{\eta}\equiv \left\{E\left(n\|\widehat{\eta}-\eta^*\|_2^2\right)\right\}^{1/2}<\infty$, then
\begin{align*}
	\tau_1\leq n^{-1/2}\left(M_\lambda C_2^*+ M_\theta C_3^*+ M_\eta C_4^*\right)\varepsilon^{1/2}.\tag{S4}\label{S4}
\end{align*}

By (\ref{S2}), (\ref{S3}) and (\ref{S4}), we have the centered process $\widehat{V}^{\rm{c}}-V^+$ satisfies 
\begin{align*}
	&E^*\left[n^{1/2} \sup\limits_{\|\beta-\beta^*\|_2<\varepsilon}\bigg|\widehat{V}^{\rm{c}}(\beta)-V^+(\beta)-\{\widehat{V}^{\rm{c}}(\beta^*)-V^+(\beta^*)\}\bigg|\right]\\
	\leq &\tau_1+\tau_2\leq C_1^*\varepsilon^{1/2}+n^{-1/2}\left(M_\lambda C_2^*+ M_\theta C_3^*+ M_\eta C_4^*\right)\varepsilon^{1/2}.
\end{align*}

Let $n$ goes infinite, we have 
\begin{align*}
	E^*\left[n^{1/2} \sup\limits_{\|\beta-\beta^*\|_2<\varepsilon}\bigg|\widehat{V}^{\rm{c}}(\beta)-V^+(\beta)-\left\{\widehat{V}^{\rm{c}}(\beta^*)-V^+(\beta^*)\right\}\bigg|\right]\leq C_1^*\varepsilon^{1/2}.\tag{S5}\label{S5}
\end{align*}

Let $\phi_n(\varepsilon)=\varepsilon^{1/2}$, and $\alpha=\frac{3}{2}<2$, check $\frac{\phi_n(\varepsilon)}{\varepsilon^\alpha}=\varepsilon^{-1}$ is decreasing not depending on $n$. Therefore, the second condition holds.

(b3.) By $\widehat{\beta}^{\rm{c}} \rightarrow \beta^*$, in probability, as $n\to\infty$, and $\widehat{V}^{\rm{c}}(\widehat{\beta}^{\rm{c}})\geq\sup\limits_{\beta:\|\beta\|_2=1}\widehat{V}(\beta)$ shown previously, choose $r_n=n^{1/3}$, then $r_n$ satisfies
\begin{align*}
	&\quad r_n^2\phi_n(r_n^{-1})=n^{2/3}\phi_n(n^{-1/3})\\
	&=n^{2/3}(n^{-1/3})^{1/2}=n^{1/2}.
\end{align*}
Thus, the third condition  holds. 

By the Theorem 14.4 in \cite{kosorok2008introduction}, we have $n^{1/3}\|\widehat{\beta}^{\rm{c}}-\beta^*\|_2=O_p(1)$.

\textit{Step 3:} We derive the asymptotic distribution of $\widehat{V}^{\rm{c}}(\widehat{\beta}^{\rm{c}})$ in this step. Note that
\begin{align*}
	&\sqrt{n}\left\{\widehat{V}^{\rm{c}}(\widehat{\beta}^{\rm{c}})-V^+(\beta^*)\right\}= \sqrt{n}\left\{\widehat{V}^{\rm{c}}(\widehat{\beta}^{\rm{c}})-\widehat{V}^{\rm{c}}(\beta^*)+\widehat{V}^{\rm{c}}(\beta^*)-V^+(\beta^*)\right\}\\
	=&\sqrt{n}\left\{\widehat{V}^{\rm{c}}(\widehat{\beta}^{\rm{c}})-\widehat{V}^{\rm{c}}(\beta^*)\right\}+\sqrt{n}\left\{\widehat{V}^{\rm{c}}(\beta^*)-V^+(\beta^*)\right\}.
\end{align*}

(C.) First, we show 
$$\sqrt{n}\left\{\widehat{V}^{\rm{c}}(\widehat{\beta}^{\rm{c}})-\widehat{V}^{\rm{c}}(\beta^*)\right\}=o_p(1),$$
which is sufficient to show $\sqrt{n}\left[\left\{\widehat{V}^{\rm{c}}(\widehat{\beta}^{\rm{c}})-\widehat{V}^{\rm{c}}(\beta^*)\right\}-\left\{V^+(\widehat{\beta}^{\rm{c}})-V^+(\beta^*)\right\}\right]=o_p(1)$ and\\ $\sqrt{n}\left\{V^+(\widehat{\beta}^{\rm{c}})-V^+(\beta^*)\right\}=o_p(1)$.

(c1.) First, by $n^{1/3}\|\widehat{\beta}-\beta^*\|_2=O_p(1)$ and  (A10), we take the Taylor expansion on 
$V^+(\widehat{\beta}^{\rm{c}})$ at $\beta^*$, 
\begin{align*}
	&\sqrt{n}\left\{V^+(\widehat{\beta}^{\rm{c}})-V^+(\beta^*)\right\}\\
	=&\sqrt{n}\left\{(V^+)'(\beta^*)\| \widehat{\beta}^{\rm{c}}-\beta^*\|_2+\frac{1}{2}(V^+)''(\beta^*)\| \widehat{\beta}^{\rm{c}}-\beta^*\|_2^2+o\left(\|\widehat{\beta}^{\rm{c}}-\beta^*\|_2^2\right)\right\}\\
	=&\sqrt{n}\left\{\frac{1}{2}(V^+)''(\beta^*)\| \widehat{\beta}^{\rm{c}}-\beta^*\|_2^2+o\left(\|\widehat{\beta}^{\rm{c}}-\beta^*\|_2^2\right)\right\} ~\ (\text{by}~\ (V^+)'(\beta^*)=0)\\
	=&\sqrt{n}\left\{\frac{1}{2}(V^+)''(\beta^*)O_p(n^{-2/3})+o_p(n^{-2/3})\right\}\\
	=&\frac{1}{2}(V^+)''(\beta^*)O_p(n^{-1/6})=o_p(1). \tag{S6}\label{S6}
\end{align*}

(c2.) Next, recall the result (\ref{S5}) that
\begin{align*}
	E^*\left[n^{1/2} \sup\limits_{\|\beta-\beta^*\|_2<\varepsilon}\bigg|\widehat{V}^{\rm{c}}(\beta)-V^+(\beta)-\left\{\widehat{V}^{\rm{c}}(\beta^*)-V^+(\beta^*)\right\}\bigg|\right]\leq C_1^*\varepsilon^{1/2},
\end{align*}
where $C_1^*$ is a finite constant. Since $\|\widehat{\beta}^{\rm{c}}-\beta^*\|_2=O_p(n^{-1/3})$, i.e., $\|\widehat{\beta}^{\rm{c}}-\beta^*\|_2=c_5n^{-1/3}$, where $c_5$ is a finite constant. We have
\begin{align*}
	&\sqrt{n}\left[\left\{\widehat{V}^{\rm{c}}(\widehat{\beta}^{\rm{c}})-\widehat{V}^{\rm{c}}(\beta^*)\right\}-\left\{V^+(\widehat{\beta}^{\rm{c}})-V^+(\beta^*)\right\}\right]\\
	\leq & E^*\left[n^{1/2} \sup\limits_{\|\beta-\beta^*\|_2<c_5n^{-1/3}}\bigg|\widehat{V}^{\rm{c}}(\beta)-V^+(\beta)-\left\{\widehat{V}^{\rm{c}}(\beta^*)-V^+(\beta^*)\right\}\bigg|\right]\\
	\leq& C_1^*\sqrt{c_5n^{-1/3}}=C_1^*\sqrt{c_5}n^{-1/6}=o_p(1).\tag{S7}\label{S7}
\end{align*}

(c3.) By (\ref{S6}) and (\ref{S7}), we have
\begin{align*}
	&\sqrt{n}\left\{\widehat{V}^{\rm{c}}(\widehat{\beta}^{\rm{c}})-\widehat{V}^{\rm{c}}(\beta^*)\right\}\\
	=&\sqrt{n}\left[\left\{\widehat{V}^{\rm{c}}(\widehat{\beta}^{\rm{c}})-\widehat{V}^{\rm{c}}(\beta^*)\right\}-\left\{V^+(\widehat{\beta}^{\rm{c}})-V^+(\beta^*)\right\}\right]+\sqrt{n}\left\{V^+(\widehat{\beta}^{\rm{c}})-V^+(\beta^*)\right\}\\
	=&o_p(1)+o_p(1)=o_p(1).\tag{S8}\label{S8}
\end{align*}

(D.) Next, we only need to show the asymptotic distribution of $\sqrt{n}\left\{\widehat{V}^{\rm{c}}(\beta^*)-V^+(\beta^*)\right\}$.
By taking the Taylor expansion on $\widehat{V}^{\rm{c}}(\beta^*)$ at $(\lambda^*,\theta^*,\eta^*)$, we have 
\begin{align*}
	\widehat{V}^{\rm{c}}(\beta^*)=V^*_n(\beta^*)+H_{\lambda}^{\T}(\widehat{\lambda}-\lambda^*)+H_{\theta}^{\T}(\widehat{\theta}-\theta^*)+H^{\T}_{\eta}(\widehat{\eta}-\eta^*)+o_p(n^{-1/2}),
\end{align*}
where 
\begin{align*}
	&H_\lambda=\lim\limits_{n\to\infty}\frac{1}{n}\sum_{i=1}^n\left\{\frac{\partial W(X_i;\lambda^*)}{\partial \lambda}\right\}\psi(Y_i,A_i,X_i;\beta^*,\theta^*,\eta^*),\\
	&H_{\theta}=\lim\limits_{n\to\infty}\frac{1}{n}\sum_{i=1}^nW(X_i;\lambda^*)\frac{\partial \psi(Y_i,A_i,X_i;\beta^*,\theta^*,\eta^*)}{\partial \theta},\\
	&H_{\eta}=\lim\limits_{n\to\infty}\frac{1}{n}\sum_{i=1}^nW(X_i;\lambda^*)\frac{\partial \psi(Y_i,A_i,X_i;\beta^*,\theta^*,\eta^*)}{\partial \eta}.
\end{align*}
Note that
\begin{align*}
	\frac{1}{n}\sum_{i=1}^n\left(\begin{array}{cc}
		\rho\left[\widehat{\lambda}^{\T}\{g(X_i)-\mu_{g0}\}\right]\left\{g(X_i)-\mu_{g0}\right\}\\
		C(X_i,A_i,Y_i;\widehat{\theta})\\
		S(X_i,A_i;\widehat{\eta})\\
	\end{array}\right)=0.
\end{align*}
By taking the Taylor expansion on these equations at $\lambda^*$, $\theta^*$, $\eta^*$, respectively, we have
\begin{align*}
	&\sqrt{n}(\widehat{\lambda}-\lambda^*)=G_\lambda^{-1}\frac{1}{\sqrt{n}}\sum_{i=1}^n \rho'\left[(\lambda^*)^{\T}\{g(X_i)-\mu_{g0}\}\right]\{g(X_i)-\mu_{g0}\}+o_p(1),\\
	&\sqrt{n}(\widehat{\theta}-\theta^*)=G_\theta^{-1}\frac{1}{\sqrt{n}}\sum_{i=1}^n C(X_i,A_i,Y_i;\theta^*)+o_p(1),\\
	&\sqrt{n}(\widehat{\eta}-\eta^*)=G_\eta^{-1}\frac{1}{\sqrt{n}}\sum_{i=1}^n S(X_i,A_i;\eta^*)+o_p(1),
\end{align*}
where 
\begin{align*}
	&G_\lambda=-E\left(\rho'\left[(\lambda^*)^{\T}\{g(X)-\mu_{g0}\}\right]\{g(X)-\mu_{g0}\}\{g(X)-\mu_{g0}\}^{\T}\right),\\
	&G_\theta=-E\left\{\frac{\partial C(X,A,Y;\theta^*)}{\partial \theta^{\T}}\right\}, G_\eta=-E\left\{\frac{\partial S(X,A;\eta^*)}{\partial \eta^{\T}}\right\}.
\end{align*}
Thus,
$$\sqrt{n}\left\{\widehat{V}^{\rm{c}}(\beta^*)-V^+(\beta^*)\right\}=\frac{1}{\sqrt{n}}\sum_{i=1}^n\left(\xi_{i1}+\xi_{i2}+\xi_{i3}+\xi_{i4}\right)+o_p(1),$$
where 
\begin{align*}
	\xi_{i1}&=W(X_i;\lambda^*)\psi(Y_i,A_i,X_i;\beta^*,\theta^*,\eta^*)-V^+(\beta^*),\\
	\xi_{i2}&=H_\lambda^{\T}G_\lambda^{-1}\rho\left[(\lambda^*)^{\T}\{g(X_i)-\mu_{g0}\}\right]\{g(X_i)-\mu_{g0}\},\\
	\xi_{i3}&=H_\theta^{\T}G_\theta^{-1}C(X_i,A_i,Y_i;\theta^*),\\
	\xi_{i4}&=H_\eta^{\T}G_\eta^{-1}S(X_i,A_i;\eta^*).
\end{align*}
$\xi_{i1}, \xi_{i2}$, $\xi_{i3}$, and $\xi_{i4}$ are i.i.d. mean zero variables. Then we have,
\begin{align*}\sqrt{n}\left\{\widehat{V}^{\rm{c}}(\beta^*)-V^+(\beta^*)\right\} \longrightarrow  N(0,\sigma^2_1), \quad \text{in distribution},\tag{S9}\label{S9}
\end{align*}
where $\sigma^2_1=E(\xi_{i1}+\xi_{i2}+\xi_{i3}+\xi_{i4})^2$.

By (\ref{S8}) and (\ref{S9}), we have 
\begin{align*}
	&\sqrt{n}\left\{\widehat{V}^{\rm{c}}(\widehat{\beta}^{\rm{c}})-V^+(\beta^*)\right\}= \sqrt{n}\left\{\widehat{V}^{\rm{c}}(\widehat{\beta}^{\rm{c}})-\widehat{V}^{\rm{c}}(\beta^*)+\widehat{V}^{\rm{c}}(\beta^*)-V^+(\beta^*)\right\}\\
	=&\sqrt{n}\left\{\widehat{V}^{\rm{c}}(\widehat{\beta}^{\rm{c}})-\widehat{V}^{\rm{c}}(\beta^*)\right\}+\sqrt{n}\left\{\widehat{V}^{\rm{c}}(\beta^*)-V^+(\beta^*)\right\}\\
	=&o_p(1)+\sqrt{n}\left\{\widehat{V}^{\rm{c}}(\beta^*)-V^+(\beta^*)\right\} \longrightarrow N(0,\sigma^2_1), \quad \text{in distribution}.
\end{align*}
\QEDB

\section{Proof of Theorem \ref{thm2}}
\label{sec:thm2}

The proof of Theorem \ref{thm2} consists of three steps as follows.

\noindent\textit{Step 1:} We show
$\widehat{V}^{\rm{c}}(\beta)=\tilde{V}^*_n(\beta)+o_p(n^{-1/2})$,
where 
$$\tilde{V}^*_n(\beta)=\frac{1}{n}\sum_{i=1}^n W(X_i;\widehat{\lambda})\left[\frac{I\{A_i=d(X_i;\beta)\}}{\pi(X_i)A_i+\{1-\pi(X_i)\}(1-A_i)}\{Y_i-\mu_d(X_i;\beta)\}+\mu_d(X_i;\beta)\right].$$
Define
\begin{align*}
	&\psi(Y,A,X;\beta)=\frac{I\{A=d(X;\beta)\}}{\pi(X)A+\{1-\pi(X)\}(1-A)}\{Y-\mu_d(X;\beta)\}+\mu_d(X;\beta),\\
	&\widehat{\psi}(Y,A,X;\beta)=\frac{I\{A=d(X;\beta)\}}{\widehat{\pi}(X)A+\{1-\widehat{\pi}(X)\}(1-A)}\{Y-\widehat{\mu}_d(X;\beta)\}+\widehat{\mu}_d(X;\beta),\\
	&\tilde{\psi}(Y,A,X;\beta)=\frac{I\{A=d(X;\beta)\}}{\pi(X)A+\{1-\pi(X)\}(1-A)}\{Y-\widehat{\mu}_d(X;\beta)\}+\widehat{\mu}_d(X;\beta).
\end{align*}
We assume $W(X;\lambda)\psi(Y,A,X;\beta)$ belongs to a Donsker class. By Theorem 2.1 in \cite{van2007empirical}, we have 
\[
\widehat{V}^{\rm{c}}(\beta)-\tilde{V}^*_n(\beta)=P\{W(X;\widehat{\lambda})\widehat{\psi}(Y,A,X;\beta)-W(X;\widehat{\lambda})\psi(Y,A,X;\beta)\}+o_p(n^{-1/2}).
\]
To show $\widehat{V}^{\rm{c}}(\beta)=\tilde{V}^*_n(\beta)+o_p(n^{-1/2})$, we only need to show
$$
P\{W(X;\widehat{\lambda})\widehat{\psi}(Y,A,X;\beta)-W(X;\widehat{\lambda})\psi(Y,A,X;\beta)\}=o_p(n^{-1/2}).
$$
Note that
\begin{align*}
	&P\{W(X;\widehat{\lambda})\widehat{\psi}(Y,A,X;\beta)-W(X;\widehat{\lambda})\psi(Y,A,X;\beta)\}\\
	=&P\{W(X;\widehat{\lambda})\widehat{\psi}(Y,A,X;\beta)-W(X;\widehat{\lambda})\tilde{\psi}(Y,A,X;\beta)\}\\
	&+P\{W(X;\widehat{\lambda})\tilde{\psi}(Y,A,X;\beta)-W(X;\widehat{\lambda})\psi(Y,A,X;\beta)\}.\tag{S10}\label{S10}
\end{align*}
Define 
\begin{align*}
	&\varrho(A\mid X) = \pi(X)A+\{1-\pi(X)\}(1-A),\\
	&\widehat{\varrho}(A\mid X) = \widehat{\pi}(X)A+\{1-\widehat{\pi}(X)\}(1-A).
\end{align*}
\begin{align*}
	&P\{W(X;\widehat{\lambda})\widehat{\psi}(Y,A,X;\beta)-W(X;\widehat{\lambda})\tilde{\psi}(Y,A,X;\beta)\}\\
	=&P\left( W(X;\widehat{\lambda})(2A-1)\{\pi(X)-\widehat{\pi}(X)\}\left[\frac{I\{A=d(X;\beta)\}}{\widehat{\varrho}(A\mid X)\varrho(A\mid X)}\{Y-\mu_d(X,\beta)\}\right]\right)\\
	& + P\left( W(X;\widehat{\lambda})(2A-1)\{\pi(X)-\widehat{\pi}(X)\}\left[\frac{I\{A=d(X;\beta)\}}{\widehat{\varrho}(A\mid X)\varrho(A\mid X)}\{\mu_d(X;\beta)-\widehat{\mu}_d(X;\beta)\}\right]\right).\tag{S11}\label{S11}
\end{align*}
The first term in (\ref{S11}) is 0  since we have
$$P[I\{A=d(X;\beta)\}\{Y-\mu_d(X;\beta)\}]=0.$$
By (A3) and (A9), $W(X,\widehat{\lambda})$ and $\{\widehat{\varrho}(A\mid X)\varrho(A\mid X)\}^{-1}$ are bounded. By Cauchy-Schwarz inequality, the absolute value of the second term in (\ref{S11}) is bounded by 
$$
\left[{P}\{\widehat{\pi}(X)-\pi(X)\}^2\right]^{\frac{1}{2}}\sum_{a=0}^1\left[{P}\{\widehat{\mu}(X,a)-\mu(X,a)\}^2\right]^{\frac{1}{2}}=o_p(n^{-1/2}),
$$
under (A11'). Therefore, 
\begin{align*}
	P\{W(X;\widehat{\lambda})\widehat{\psi}(Y,A,X;\beta)-W(X;\widehat{\lambda})\tilde{\psi}(Y,A,X;\beta)\}=o_p(n^{-1/2}).\tag{S12}\label{S12}
\end{align*}
Note that  $$P\left[\frac{I\{A=d(X;\beta)\}}{\varrho(A\mid X)}\right]=1.$$ Thus, we have
\begin{align*}
	&P\{W(X;\widehat{\lambda})\tilde{\psi}(Y,A,X;\beta)-W(X;\widehat{\lambda})\psi(Y,A,X;\beta)\}\\
	=&P\left(W(X;\widehat{\lambda})\left[\frac{I\{A=d(X;\beta)\}}{\varrho(A\mid X)}-1\right]\{\mu_d(X;\beta)-\widehat{\mu}_d(X;\beta)\}\right)\\
	=&0. \tag{S13}\label{S13}
\end{align*}
By (\ref{S10}), (\ref{S12}) and (\ref{S13}), we have 
$$P\{W(X;\widehat{\lambda})\widehat{\psi}(Y,A,X;\beta)-W(X;\widehat{\lambda})\psi(Y,A,X;\beta)\}=o_p(n^{-1/2}),$$
and then
\begin{align*}
	\widehat{V}^{\rm{c}}(\beta)=\tilde{V}^*_n(\beta)+o_p(n^{-1/2}).\tag{S14}\label{S14}
\end{align*}

\noindent \textit{Step 2:} We show that $n^{1/3}\|\widehat{\beta}^{\rm{c}}-\beta^*\|_2=O_p(1)$, where $\|\cdot\|_2$ is the $L_2$ norm.

(A.) First, we show that $\widehat{\beta}^{\rm{c}}$ converges in probability to $\beta^*$ as $n \to \infty$, by checking three conditions for the Argmax Theorem:

(a1.) By  (A9), the true value function $V^+(\beta)$ is twice continuously differentiable in a neighborhood of  $\beta^*$.

(a2.) By  (A11'), it can be easily shown
$$\widehat{V}^{\rm{c}}(\beta)=V^+(\beta)+o_p(1).$$

(a3.) Since $\widehat{\beta}^{\rm{c}}=\argmax\limits_{\beta:\|\beta\|_2=1} \widehat{V}^{\rm{c}}(\beta)$, we have the estimated ITR as $d(X,\widehat{\beta}^{\rm{c}})=I(\tilde{X}^{\T}\widehat{\beta}^{\rm{c}}>0)$ and the corresponding value function $\widehat{V}^{\rm{c}}(\widehat{\beta}^{\rm{c}})$ such that
$$\widehat{V}^{\rm{c}}(\widehat{\beta}^{\rm{c}})\geq\sup\limits_{\beta:\|\beta\|_2=1 }\widehat{V}^{\rm{c}}(\beta).$$

Thus we have $\widehat{\beta}^{\rm{c}} \rightarrow \beta^*$, in probability, as $n\to\infty$.

(B). Next, we show that the convergence rate of $\widehat{\beta}^{\rm{c}}$ is $n^{1/3}$, i.e., $n^{1/3}\|\widehat{\beta}-\beta^*\|_2=O_p(1)$. We check three conditions of the Theorem 14.4: Rate of convergence in \cite{kosorok2008introduction}:

(b1.) For every $\beta$ in a neighborhood of $\beta^*$, i.e., $\| \beta-\beta^*\|_2 < \varepsilon$ for some $\varepsilon>0$, by  (A9), we take the second order Taylor expansion on $V^+(\beta)$ at $\beta=\beta^*$,
\begin{align*}
	V^+(\beta)-V^+(\beta^*)&=(V^+)'(\beta^*)\| \beta-\beta^*\|_2+\frac{1}{2}(V^+)''(\beta^*)\| \beta-\beta^*\|_2^2+o\left(\|\beta-\beta^*\|_2^2\right)\\
	&=\frac{1}{2}(V^+)''(\beta^*)\| \beta-\beta^*\|_2^2+o\left(\|\beta-\beta^*\|_2^2\right) ~\ (\text{by}~\ (V^+)'(\beta^*)=0).
\end{align*}
Since $(V^+)''(\beta^*)<0$, there exists  $c_0=-\frac{1}{2}(V^+)''(\beta^*)>0$ such that $V^+(\beta)-V^+(\beta^*)<c_0\|\beta-\beta^*\|^2_2$ holds. 

(b2.) For all $n$ large enough and sufficiently small $\varepsilon$, the centered process $\widehat{V}^{\rm{c}}-V^+$ satisfies 
\begin{align*}
	&E^*\left[n^{1/2} \sup\limits_{\|\beta-\beta^*\|_2<\varepsilon}\bigg|\widehat{V}^{\rm{c}}(\beta)-V^+(\beta)-\{\widehat{V}^{\rm{c}}(\beta^*)-V^+(\beta^*)\}\bigg|\right]\\
	=&E^*\Bigg[n^{1/2} \sup\limits_{\|\beta-\beta^*\|_2<\varepsilon}\bigg|\widehat{V}^{\rm{c}}(\beta)-\tilde{V}^*_n(\beta)+\tilde{V}^*_n(\beta)-V^+(\beta)-\{\widehat{V}^{\rm{c}}(\beta^*)-\tilde{V}^*_n(\beta^*)+\tilde{V}^*_n(\beta^*)-V^+(\beta^*)\}\bigg|\Bigg]\\
	\leq&  \underbrace{E^*\left[n^{1/2} \sup\limits_{\|\beta-\beta^*\|_2<\varepsilon}\bigg|\widehat{V}^{\rm{c}}(\beta)-\tilde{V}_n^*(\beta)-\left\{\widehat{V}^{\rm{c}}(\beta^*)-\tilde{V}_n^*(\beta^*)\right\}\bigg|\right]}_{\zeta_1}\\
	&+\underbrace{E^*\left[n^{1/2} \sup\limits_{\|\beta-\beta^*\|_2<\varepsilon}\bigg|\tilde{V}_n^*(\beta)-V^+(\beta)-\left\{\tilde{V}_n^*(\beta^*)-V^+(\beta^*)\right\}\bigg|\right]}_{\zeta_2}.\tag{S15}\label{S15}
\end{align*}

By the result (\ref{S14}), $\widehat{V}^{\rm{c}}(\beta)-\tilde{V}_n^*(\beta)=o_p(n^{-1/2})$ and $\widehat{V}^{\rm{c}}(\beta^*)-\tilde{V}_n^*(\beta^*)=o_p(n^{-1/2})$. Thus, 
\begin{align*}
	\zeta_1=E^*\left[n^{1/2} \sup\limits_{\|\beta-\beta^*\|_2<\varepsilon}\bigg|\widehat{V}^{\rm{c}}(\beta)-\tilde{V}_n^*(\beta)-\left\{\widehat{V}^{\rm{c}}(\beta^*)-\tilde{V}_n^*(\beta^*)\right\}\bigg|\right]=o_p(1)\tag{S16}\label{S16}
\end{align*}

Define 
$$\bar{V}^*_n(\beta)=\frac{1}{n}\sum_{i=1}^n W(X_i;\lambda^*)\left\{\frac{I\{A_i=d(X_i;\beta)\}}{\pi(X_i)A_i+\{1-\pi(X_i)\}(1-A_i)}\{Y_i-\mu_d(X_i;\beta)\}+\mu_d(X_i;\beta)\right\}.$$
\begin{align*}
	\zeta_2&=E^*\left[n^{1/2} \sup\limits_{\|\beta-\beta^*\|_2<\varepsilon}\bigg|\tilde{V}_n^*(\beta)-V^+(\beta)-\left\{\tilde{V}_n^*(\beta^*)-V^+(\beta^*)\right\}\bigg|\right]\\
	&=E^*\left[n^{1/2} \sup\limits_{\|\beta-\beta^*\|_2<\varepsilon}\bigg|\tilde{V}_n^*(\beta)-\bar{V}^*_n(\beta)+\bar{V}^*_n(\beta)-V^+(\beta)-\left\{\tilde{V}_n^*(\beta^*)-\bar{V}^*_n(\beta^*)+\bar{V}^*_n(\beta^*)-V^+(\beta^*)\right\}\bigg|\right]\\
	&\leq \underbrace{E^*\left[n^{1/2} \sup\limits_{\|\beta-\beta^*\|_2<\varepsilon}\bigg|\tilde{V}_n^*(\beta)-\bar{V}^*_n(\beta)-\left\{\tilde{V}_n^*(\beta^*)-\bar{V}^*_n(\beta^*)\right\}\bigg|\right]}_{\omega_1}\\
	&\quad +\underbrace{E^*\left[n^{1/2} \sup\limits_{\|\beta-\beta^*\|_2<\varepsilon}\bigg|\bar{V}^*_n(\beta)-V^+(\beta)-\left\{\bar{V}^*_n(\beta^*)-V^+(\beta^*)\right\}\bigg|\right]}_{\omega_2}.
\end{align*}

We have 
\begin{align*}
	&\bar{V}_n^*(\beta)-\bar{V}_n^*(\beta^*)\\
	=&\frac{1}{n}\sum_{i=1}^nW(\lambda^*;X)\Bigg[\frac{I\{A_i=d(X_i;\beta)\}\{Y_i-\mu_d(X_i;\beta)\}-I\{A_i=d(X_i;\beta^*)\}\{Y_i-\mu_d(X_i;\beta^*)\}}{\pi(X_i)A_i+\{1-\pi(X_i)\}(1-A_i)}\\
	&+\mu_d(X_i;\beta)-\mu_d(X_i;\beta^*)\Bigg]\\
	=&\frac{1}{n}\sum_{i=1}^nW(\lambda^*;X_i)\Bigg[\frac{(2A_i-1)Y_i-\mu(X_i,1)A_i+\mu(X_i,0)(1-A_i)}{\pi(X_i)A_i+\{1-\pi(X_i)\}(1-A_i)}+\mu(X_i,1)-\mu(X_i,0)\Bigg]\\
	& \{I(\tilde{X}_i^{\T}\beta>0)-I(\tilde{X}_i^{\T}\beta^*>0)\}.
\end{align*}

We define a class of functions 
\begin{align*}
	\mathcal{F}^1_{\beta}(y,a,x)=\Bigg\{&W(\lambda^*;x)\left[\frac{(2a-1)y-\mu(x,1)a+\mu(x,0)(1-a)}{\pi(x)a+\{1-\pi(x)\}(1-a)}+\mu(x,1)-\mu(x,0)\right]\\
	&\left\{I\left(\tilde{x}^{\T}\beta>0\right)-I\left(\tilde{x}^{\T}\beta^*>0\right)\right\}: \|\beta-\beta^*\|_2<\varepsilon\Bigg\}.
\end{align*}

Let $K_1 = \sup \left| W(\lambda^*;x)\left[\frac{(2a-1)y-\mu(x,1)a+\mu(x,0)(1-a)}{\pi(x)a+\{1-\pi(x)\}(1-a)}+\mu(x,1)-\mu(x,0)\right]\right|$. By (A3), (A6), (A7) and (A8), $K_1<\infty$.
Then we can define the envelope of $\mathcal{F}^1_{\beta}(y,a,x)$ as $F_1=K_1I\left(-k_0\varepsilon\leq\tilde{x}^{\T}\beta^*\leq k_0\varepsilon\right)$. Similar to (b2.1) in the proof of Theorem \ref{thm1}, we have
$$\|F_1\|_{P,2}=K_1\sqrt{{\rm pr}\left(-k_0\varepsilon\leq\tilde{x}^{\T}\beta^*\leq k_0\varepsilon\right)}=K_1\sqrt{k_1\cdot 2k_0\varepsilon}=K_1\sqrt{2k_0k_1}\varepsilon^{1/2}<\infty.$$
Since $\mathcal{F}^1_{\beta}$ is a class of indicator functions, by the conclusion of Lemma 9.6 and Lemma 9.9 in \cite{kosorok2008introduction}, $\mathcal{F}_{\beta}^1$ is a VC class of functions. Thus, the entropy of $\mathcal{F}^1_{\beta}$, denoted as $J^*_{[]}(1,\mathcal{F}^1)$, is finite, i.e., $J^*_{[]}(1,\mathcal{F}^1)<\infty$.
Next, we consider the following empirical process indexed by $\beta$,
$$\mathbb{G}_n\mathcal{F}^1_{\beta}=n^{-1/2}\sum_{i=1}^n\left[\mathcal{F}^1_{\beta}\left(Y_i,A_i,X_i\right)-E\left\{\mathcal{F}^1_{\beta}\left(Y_i,A_i,X_i\right)\right\}\right].$$
Note that $\mathbb{G}_n\mathcal{F}^1_{\beta}=n^{1/2}\left[\bar{V}_n^*(\beta)-\bar{V}_n^*(\beta^*)-\left\{V^+(\beta)-V^+(\beta^*)\right\}\right]$. By applying Theorem 11.2 in \cite{kosorok2008introduction}, we have 
\begin{align*}
	\omega_2&=E^*\left[n^{1/2} \sup\limits_{\|\beta-\beta^*\|_2<\varepsilon}\bigg|\bar{V}_n^*(\beta)-V^+(\beta)-\left\{\bar{V}_n^*(\beta^*)-V^+(\beta^*)\right\}\bigg|\right]\\
	&=E^*\left( \sup\limits_{\|\beta-\beta^*\|_2<\varepsilon}\left|\mathbb{G}_n\mathcal{F}^1_{\beta}\right|\right)\leq c_1J^*_{[]}(1,\mathcal{F}^1)\|F_1\|_{P,2}=c_1J^*_{[]}(1,\mathcal{F}^1)K_1\sqrt{2k_0k_1}\varepsilon^{1/2},
\end{align*}
where $c_1$ is a finite constant.

Let $C_1^*\equiv c_1J^*_{[]}(1,\mathcal{F}^1)K_1\sqrt{2k_0k_1}$, since $c_1, J^*_{[]}(1,\mathcal{F}^1)$, $K_1$, $k_0$ and $k_1$ are bounded, we have $C_1^*<\infty$, i.e.,
\begin{align*}
	\omega_2\leq C_1^*\varepsilon^{1/2}.\tag{S17}\label{S17}
\end{align*}

We rewrite the form of $\tilde{V}_n^*(\beta)-\bar{V}_n^*(\beta)       -\left\{\tilde{V}_n^*(\beta^*)-\bar{V}_n^*(\beta^*)\right\}$ as 
\begin{align*}
	&\tilde{V}_n^*(\beta)-\bar{V}_n^*(\beta)       -\left\{\tilde{V}_n^*(\beta^*)-\bar{V}_n^*(\beta^*)\right\}\\
	=&\frac{1}{n}\sum_{i=1}^n\left\{W(X_i;\widehat{\lambda})-W(X_i;\lambda^*)\right\}\\
	&\left[\frac{I\{A_i=d(X_i;\beta)\}\{Y_i-\mu_d(X_i;\beta)\}-I\{A_i=d(X_i;\beta^*)\}\{Y_i-\mu_d(X_i;\beta^*)\}}{\pi(X_i)A_i+\{1-\pi(X_i)\}(1-A_i)}+\mu_d(X_i;\beta)-\mu_d(X_i;\beta^*)\right]\\
	=&\frac{1}{n}\sum_{i=1}^n\left\{W(X_i;\widehat{\lambda})-W(X_i;\lambda^*)\right\}\Bigg[\frac{(2A_i-1)Y_i-\mu(X_i,1)A_i+\mu(X_i,0)(1-A_i)}{\pi(X_i)A_i+\{1-\pi(X_i)\}(1-A_i)}+\mu(X_i,1)-\mu(X_i,0)\Bigg]\\
	& \{I(\tilde{X}_i^{\T}\beta>0)-I(\tilde{X}_i^{\T}\beta^*>0)\}.
\end{align*}

We take the Taylor expansion on the above equation at $\lambda^*$,
\begin{align*}
	&\tilde{V}_n^*(\beta)-\bar{V}_n^*(\beta)       -\left\{\tilde{V}_n^*(\beta^*)-\bar{V}_n^*(\beta^*)\right\}\\
	=&\frac{1}{n}\sum_{i=1}^n\left\{\frac{\partial W(X;\lambda^*)}{\partial\lambda}\right\}^{\T}\Bigg[\frac{(2A_i-1)Y_i-\mu(X_i,1)A_i+\mu(X_i,0)(1-A_i)}{\pi(X_i)A_i+\{1-\pi(X_i)\}(1-A_i)}+\mu(X_i,1)-\mu(X_i,0)\Bigg]\\
	& \{I(\tilde{X}_i^{\T}\beta>0)-I(\tilde{X}_i^{\T}\beta^*>0)\}\left(\widehat{\lambda}-\lambda^*\right)+o_p(n^{-1/2}).
\end{align*}

Next, we define a class of functions,
\begin{align*}
	\mathcal{F}^2_{\beta}(y,a,x)=\Bigg\{&\frac{\partial W(X;\lambda^*)}{\partial\lambda}\left[\frac{(2a-1)y-\mu(x,1)a+\mu(x,0)(1-a)}{\pi(x)a+\{1-\pi(x)\}(1-a)}+\mu(x,1)-\mu(x,0)\right]\\
	&\left\{I\left(\tilde{x}^{\T}\beta>0\right)-I\left(\tilde{x}^{\T}\beta^*>0\right)\right\}: \|\beta-\beta^*\|_2<\varepsilon\Bigg\}.
\end{align*}

Let $K_2 = \sup \left| \frac{\partial W(X;\lambda^*)}{\partial\lambda}\left[\frac{(2a-1)y-\mu(x,1)a+\mu(x,0)(1-a)}{\pi(x)a+\{1-\pi(x)\}(1-a)}+\mu(x,1)-\mu(x,0)\right]\right|$. By (A3), (A6), (A7) and (A8), $K_2<\infty$.
Then we can define the envelope of $\mathcal{F}^2_{\beta}(y,a,x)$ as $F_2=K_2I\left(-k_0\varepsilon\leq\tilde{x}^{\T}\beta^*\leq k_0\varepsilon\right)$. Similarly to above, we have 
$$\|F_2\|_{P,2}\leq K_2\sqrt{2k_0k_1}\varepsilon^{1/2}<\infty,$$
and $\mathcal{F}_{\beta}^2$ is a VC class of functions. Thus, the entropy of  $\mathcal{F}^2_{\beta}$, denoted as $J^*_{[]}(1,\mathcal{F}^2)$, is finite, i.e., $J^*_{[]}(1,\mathcal{F}^2)<\infty$.

Next, we consider the following empirical process indexed by $\beta$,
$$\mathbb{G}_n\mathcal{F}^2_{\beta}=n^{-1/2}\sum_{i=1}^n\left[\mathcal{F}^2_{\beta}\left(Y_i,A_i,X_i\right)-E\left\{\mathcal{F}^2_{\beta}\left(Y_i,A_i,X_i\right)\right\}\right].$$

By applying Theorem 11.2 in \cite{kosorok2008introduction}, we have 
$$E^*\left(n^{1/2} \sup\limits_{\|\beta-\beta^*\|_2<\varepsilon}\left\|\mathbb{G}_n\mathcal{F}^2_{\beta}\right\|_1\right)\leq c_2J^*_{[]}(1,\mathcal{F}^2)\|F_2\|_{P,2}\leq c_2J^*_{[]}(1,\mathcal{F}^2)K_2\sqrt{2k_0k_1(\beta^*)}\varepsilon^{1/2},$$
where $c_2$ is a finite constant. By Theorem 2.14.5 in \cite{van1996weak}, 
\begin{align*}
	\left\{E^*\left( \sup\limits_{\|\beta-\beta^*\|_2<\varepsilon}\left\|\mathbb{G}_n\mathcal{F}^2_{\beta}\right\|_2^2\right)\right\}^{1/2}&\leq l_2\left\{E^*\left( \sup\limits_{\|\beta-\beta^*\|_2<\varepsilon}\left\|\mathbb{G}_n\mathcal{F}^2_{\beta}\right\|_1\right)+\|F_2\|_{P,2}\right\}\\
	&\leq l_2\left\{c_2J^*_{[]}(1,\mathcal{F}^2)+1\right\}\|F_2\|_{P,2}\\
	&= l_2\left\{c_2J^*_{[]}(1,\mathcal{F}^2)+1\right\}K_2\sqrt{2k_0k_1}\varepsilon^{1/2},
\end{align*}
where $l_2$ is a finite constant.
Let 
$$C_2^*\equiv l_2\left\{c_2J^*_{[]}(1,\mathcal{F}^2)+1\right\}K_2\sqrt{2k_0k_1}<\infty,$$
and we have 
$$\left\{E^*\left( \sup\limits_{\|\beta-\beta^*\|_2<\varepsilon}\left\|\mathbb{G}_n\mathcal{F}^2_{\beta}\right\|_2^2\right)\right\}^{1/2}\leq C_2^*\varepsilon^{1/2}.$$

Notice that
\begin{align*}
	\omega_1&=E^*\left[n^{1/2} \sup\limits_{\|\beta-\beta^*\|_2<\varepsilon}\bigg|\tilde{V}_n^*(\beta)-\bar{V}_n^*(\beta)       -\left\{\tilde{V}_n^*(\beta^*)-\bar{V}_n^*(\beta^*)\right\}\bigg|\right]\\
	&=E^*\left\{ \sup\limits_{\|\beta-\beta^*\|_2<\varepsilon}(\widehat{\lambda}-\lambda^*)^\T\mathbb{G}_n\mathcal{F}^2_{\beta}+o_p(1)\right\}\\
	&\leq n^{-1/2}\left[E^*\left\{ \sup\limits_{\|\beta-\beta^*\|_2<\varepsilon}\left|n^{1/2}(\widehat{\lambda}-\lambda^*)^\T\mathbb{G}_n\mathcal{F}^2_{\beta}\right|\right\}\right].
\end{align*}
By the Cauchy-Schwarz Inequality, we have
\begin{align*}
	\omega_1&\leq n^{-1/2}\left\{E\left(n\|\widehat{\lambda}-\lambda^*\|_2^2\right)\right\}^{1/2}\left\{E^*\left( \sup\limits_{\|\beta-\beta^*\|_2<\varepsilon}\left\|\mathbb{G}_n\mathcal{F}^2_{\beta}\right\|_2^2\right)\right\}^{1/2}.\\
\end{align*}
By (A11'),  $\sqrt{n}(\widehat{\lambda}-\lambda^*)=O_p(1)$, we have  $M_{\lambda}\equiv \left\{E\left(n\|\widehat{\lambda}-\lambda^*\|_2^2\right)\right\}^{1/2}<\infty$, then
\begin{align*}
	\omega_1\leq n^{-1/2} M_\lambda C_2^*\varepsilon^{1/2}.\tag{S18}\label{S18}
\end{align*}
By (\ref{S17}) and (\ref{S18}), we have 
\begin{align*}
	\zeta_2\leq \omega_1+\omega_2\leq C_1^*\varepsilon^{1/2}+n^{-1/2} M_\lambda C_2^*\varepsilon^{1/2}.\tag{S19}\label{S19}
\end{align*}
By (\ref{S15}), (\ref{S16}), and (\ref{S19}), as $n\to\infty$, we have the centered process $\widehat{V}^{\rm{c}}-V^+$ satisfies 
\begin{align*}
	&E^*\left[n^{1/2} \sup\limits_{\|\beta-\beta^*\|_2<\varepsilon}\bigg|\widehat{V}^{\rm{c}}(\beta)-V^+(\beta)-\left\{\widehat{V}^{\rm{c}}(\beta^*)-V^+(\beta^*)\right\}\bigg|\right]\leq \zeta_1+\zeta_2\leq C_1^*\varepsilon^{1/2}.\tag{S20}\label{S20}
\end{align*}
Let $\phi_n(\varepsilon)=\varepsilon^{1/2}$, and $\alpha=\frac{3}{2}<2$, check $\frac{\phi_n(\varepsilon)}{\varepsilon^\alpha}=\varepsilon^{-1}$ is decreasing not depending on $n$. Therefore, the second condition holds.

(b3.) By $\widehat{\beta}^{\rm{c}} \rightarrow \beta^*$, in probability, as $n\to\infty$, and $\widehat{V}^{\rm{c}}(\widehat{\beta}^{\rm{c}})\geq\sup\limits_{\beta:\|\beta\|_2=1}\widehat{V}^{\rm{c}}(\beta)$ shown previously, choose $r_n=n^{1/3}$, then $r_n$ satisfies
\begin{align*}
	&\quad r_n^2\phi_n(r_n^{-1})=n^{2/3}\phi_n(n^{-1/3})\\
	&=n^{2/3}(n^{-1/3})^{1/2}=n^{1/2}.
\end{align*}
Thus, the third condition holds. 

By the Theorem 14.4 in \cite{kosorok2008introduction}, we have $n^{1/3}\|\widehat{\beta}^{\rm{c}}-\beta^*\|_2=O_p(1)$.

\noindent \textit{Step 3:} We derive the asymptotic distribution of $\widehat{V}^{\rm{c}}(\widehat{\beta}^{\rm{c}})$ in this step. Note that
\begin{align*}
	&\sqrt{n}\left\{\widehat{V}^{\rm{c}}(\widehat{\beta}^{\rm{c}})-V^+(\beta^*)\right\}= \sqrt{n}\left\{\widehat{V}^{\rm{c}}(\widehat{\beta}^{\rm{c}})-\widehat{V}^{\rm{c}}(\beta^*)+\widehat{V}^{\rm{c}}(\beta^*)-V^+(\beta^*)\right\}\\
	=&\sqrt{n}\left\{\widehat{V}^{\rm{c}}(\widehat{\beta}^{\rm{c}})-\widehat{V}^{\rm{c}}(\beta^*)\right\}+\sqrt{n}\left\{\widehat{V}^{\rm{c}}(\beta^*)-V^+(\beta^*)\right\}.
\end{align*}

(C.) First, we show 
$$\sqrt{n}\left\{\widehat{V}^{\rm{c}}(\widehat{\beta}^{\rm{c}})-\widehat{V}^{\rm{c}}(\beta^*)\right\}=o_p(1),$$
which is sufficient to show $\sqrt{n}\left[\left\{\widehat{V}^{\rm{c}}(\widehat{\beta}^{\rm{c}})-\widehat{V}^{\rm{c}}(\beta^*)\right\}-\left\{V^+(\widehat{\beta}^{\rm{c}})-V^+(\beta^*)\right\}\right]=o_p(1)$ and\\ $\sqrt{n}\left\{V^+(\widehat{\beta}^{\rm{c}})-V^+(\beta^*)\right\}=o_p(1)$.

(c1.) First, by $n^{1/3}\|\widehat{\beta}^{\rm{c}}-\beta^*\|_2=O_p(1)$ and (A9), we take the Taylor expansion on 
$V^+(\widehat{\beta}^{\rm{c}})$ at $\beta^*$, 
\begin{align*}
	\sqrt{n}\left\{V^+(\widehat{\beta}^{\rm{c}})-V^+(\beta^*)\right\}&=\sqrt{n}\left\{(V^+)'(\beta^*)\| \widehat{\beta}^{\rm{c}}-\beta^*\|_2+\frac{1}{2}(V^+)''(\beta^*)\| \widehat{\beta}^{\rm{c}}-\beta^*\|_2^2+o\left(\|\widehat{\beta}^{\rm{c}}-\beta^*\|_2^2\right)\right\}\\
	&=\sqrt{n}\left\{\frac{1}{2}(V^+)''(\beta^*)\| \widehat{\beta}^{\rm{c}}-\beta^*\|_2^2+o\left(\|\widehat{\beta}^{\rm{c}}-\beta^*\|_2^2\right)\right\} ~\ (\text{by}~\ (V^+)'(\beta^*)=0)\\
	&=\sqrt{n}\left\{\frac{1}{2}(V^+)''(\beta^*)O_p(n^{-2/3})+o_p(n^{-2/3})\right\}\\
	&=\frac{1}{2}(V^+)''(\beta^*)O_p(n^{-1/6})=o_p(1).\tag{S21}\label{S21}
\end{align*}

(c2.) Next, recall the result (\ref{S20}) that
\begin{align*}
	E^*\left[n^{1/2} \sup\limits_{\|\beta-\beta^*\|_2<\varepsilon}\bigg|\widehat{V}^{\rm{c}}(\beta)-V^+(\beta)-\left\{\widehat{V}^{\rm{c}}(\beta^*)-V^+(\beta^*)\right\}\bigg|\right]\leq C_1^*\varepsilon^{1/2},
\end{align*}
where $C_1^*$ is a finite constant. Since $\|\widehat{\beta}^{\rm{c}}-\beta^*\|_2=O_p(n^{-1/3})$, i.e., $\|\widehat{\beta}^{\rm{c}}-\beta^*\|_2=c_3n^{-1/3}$, where $c_3$ is a finite constant. We have
\begin{align*}
	&\sqrt{n}\left[\left\{\widehat{V}^{\rm{c}}(\widehat{\beta}^{\rm{c}})-\widehat{V}^{\rm{c}}(\beta^*)\right\}-\left\{V^+(\widehat{\beta}^{\rm{c}})-V^+(\beta^*)\right\}\right]\\
	\leq & E^*\left[n^{1/2} \sup\limits_{\|\beta-\beta^*\|_2<c_4n^{-1/3}}\bigg|\widehat{V}^{\rm{c}}(\beta)-V^+(\beta)-\left\{\widehat{V}^{\rm{c}}(\beta^*)-V^+(\beta^*)\right\}\bigg|\right]\\
	\leq& C_1^*\sqrt{c_3n^{-1/3}}=C_1^*\sqrt{c_3}n^{-1/6}=o_p(1).\tag{S22}\label{S22}
\end{align*}

(c3.) By (\ref{S21}) and (\ref{S22}), we have
\begin{align*}
	&\sqrt{n}\left\{\widehat{V}^{\rm{c}}(\widehat{\beta}^{\rm{c}})-\widehat{V}^{\rm{c}}(\beta^*)\right\}\\
	=&\sqrt{n}\left[\left\{\widehat{V}^{\rm{c}}(\widehat{\beta}^{\rm{c}})-\widehat{V}^{\rm{c}}(\beta^*)\right\}-\left\{V^+(\widehat{\beta}^{\rm{c}})-V^+(\beta^*)\right\}\right]+\sqrt{n}\left\{V^+(\widehat{\beta}^{\rm{c}})-V^+(\beta^*)\right\}\\
	=&o_p(1)+o_p(1)=o_p(1).\tag{S23}\label{S23}
\end{align*}

(D.) 
By the result (\ref{S14}) shown in Step 1, $\sqrt{n}\left\{\widehat{V}^{\rm{c}}(\beta^*)-\tilde{V}^*_n(\beta^*)\right\}=o_p(1).$ With (\ref{S23}), we have
\begin{align*}
	&\sqrt{n}\left\{\widehat{V}^{\rm{c}}(\widehat{\beta}^{\rm{c}})-V^+(\beta^*)\right\}\\
	=&\sqrt{n}\left\{\widehat{V}^{\rm{c}}(\widehat{\beta}^{\rm{c}})-\widehat{V}^{\rm{c}}(\beta^*)\right\}+\sqrt{n}\left\{\widehat{V}^{\rm{c}}(\beta^*)-\tilde{V}^*_n(\beta^*)\right\}+\sqrt{n}\left\{\tilde{V}^*_n(\beta^*)-V^+(\beta^*)\right\}\\
	=&o_p(1)+o_p(1)+\sqrt{n}\left\{\tilde{V}^*_n(\beta^*)-V^+(\beta^*)\right\}\\
	=&\sqrt{n}\left\{\tilde{V}^*_n(\beta^*)-V^+(\beta^*)\right\}+o_p(1).\tag{S24}\label{S24}
\end{align*}

Thus, we only need to show the asymptotic distribution of $\sqrt{n}\left\{\tilde{V}^*_n(\beta^*)-V^+(\beta^*)\right\}$.
By taking the Taylor expansion on $\tilde{V}^*_n(\beta^*)$ at $\lambda^*$, we have 
\begin{align*}
	\tilde{V}^*_n(\beta^*)=\bar{V}^*_n(\beta^*)+H_{\lambda}^{\T}(\widehat{\lambda}-\lambda^*)+o_p(n^{-1/2}),
\end{align*}
where 
$$H_\lambda=\lim\limits_{n\to\infty}\frac{1}{n}\sum_{i=1}^n\left\{\frac{\partial W(X_i;\lambda^*)}{\partial \lambda}\right\}\psi(Y_i,A_i,X_i;\beta^*).$$
Note that
$$\frac{1}{n}\sum_{i=1}^n\rho\left[\widehat{\lambda}^{\T}\{g(X_i)-\mu_{g0}\}\right]\left\{g(X_i)-\mu_{g0}\right\}=0.$$
By taking the Taylor expansion on the above equation at $\lambda^*$, we have
$$\sqrt{n}(\widehat{\lambda}-\lambda^*)=G_\lambda^{-1}\frac{1}{\sqrt{n}}\sum_{i=1}^n \rho\left[(\lambda^*)^{\T}\{g(X_i)-\mu_{g0}\}\right]\{g(X_i)-\mu_{g0}\}+o_p(1),$$
where 
$$G_\lambda=-E\left(\rho'\left[(\lambda^*)^{\T}\{g(X)-\mu_{g0}\}\right]\{g(X)-\mu_{g0}\}\{g(X)-\mu_{g0}\}^{\T}\right).$$
Thus,
$$\sqrt{n}\left\{\tilde{V}^*_n(\beta^*)-V^+(\beta^*)\right\}=\frac{1}{\sqrt{n}}\sum_{i=1}^n\left(\xi_{i1}+\xi_{i2}\right)+o_p(1),$$
where 
\begin{align*}
	\xi_{i1}&=W(X_i;\lambda^*)\psi(Y_i,A_i,X_i;\beta^*)-V^+(\beta^*),\\
	\xi_{i2}&=H_\lambda^{\T}G_\lambda^{-1}\rho'\left[(\lambda^*)^{\T}\{g(X_i)-\mu_{g0}\}\right]\{g(X_i)-\mu_{g0}\}.
\end{align*}
$\xi_{i1}$ and $\xi_{i2}$ are i.i.d. mean-zero random variables. Then,
\begin{align*}\sqrt{n}\left\{\tilde{V}^*_n(\beta^*)-V^+(\beta^*)\right\} \longrightarrow N(0,\sigma^2_2), \quad \text{in distribution},\tag{S25}\label{S25}
\end{align*}
where $\sigma^2_2=E\left\{(\xi_{i1}+\xi_{i2})^2\right\}$. By (\ref{S24}) and (\ref{S25}), we have 
\begin{align*}
	&\sqrt{n}\left\{\widehat{V}^{\rm{c}}(\widehat{\beta}^{\rm{c}})-V^+(\beta^*)\right\}
	=\sqrt{n}\left\{\tilde{V}^*_n(\beta^*)-V^+(\beta^*)\right\}+o_p(1) \longrightarrow N(0,\sigma^2_2), \quad \text{in distribution}.
\end{align*}
\QEDB

\section{Proof of Theorem \ref{thm3}}
\label{sec:thm3}

When $\mathbb{P}^{\rm{s}}=\mathbb{P}^{\rm{t}}$, $f^{\rm{s}}(X)=f^{\rm{t}}(X)$, $\widehat{\lambda} \overset{p}{\to} \lambda^*=0$ and $W(X;\lambda^*)=W(X;0)=1=f^{\rm{t}}(X)/f^{\rm{s}}(X)$. Thus, $\mathbb{P}^+=\mathbb{P}^{\rm{s}}=\mathbb{P}^{\rm{t}}$. Therefore, for a fixed ITR $d(X;\beta)$, $V^+(\beta)=V^{\rm{t}}(\beta).$  Meanwhile, we have $E\left\{g(X)\right\}=E^{\rm{t}}\left\{g(X)\right\}=\mu_{g0}$. By Theorem \ref{thm2},  we have
$$\sqrt{n}\left\{\widehat{V}^{\rm{c}}(\beta)-V^{\rm{t}}(\beta)\right\} \longrightarrow N(0,(\sigma^{\rm{c}})^2), \quad \text{in distribution},$$
where 
$$(\sigma^{\rm{c}})^2=E\left\{(\xi_{1}+\xi_{2})^2\right\},$$
$$\xi_{1}=W(X;\lambda^*)\psi(Y,A,X;\beta)-V^+(\beta)=\psi(Y,A,X;\beta)-V^+(\beta),$$
$$\xi_{2}=H_\lambda^{\T}G_\lambda^{-1}\rho\left[(\lambda^*)^{\T}\{g(X)-\mu_{g0}\}\right]\{g(X)-\mu_{g0}\}=H_\lambda^{\T}G_\lambda^{-1}\{g(X)-\mu_{g0}\} \quad (\rho(0)=1),$$
$$H_\lambda=\lim\limits_{n\to\infty}\frac{1}{n}\sum_{i=1}^n\left\{\frac{\partial W(X_i;\lambda^*)}{\partial \lambda}\right\}\psi(Y_i,A_i,X_i;\beta^*),$$
\begin{align*}
	G_\lambda&=-E\left(\rho'\left[(\lambda^*)^{\T}\{g(X)-\mu_{g0}\}\right]\{g(X)-\mu_{g0}\}\{g(X)-\mu_{g0}\}^{\T}\right)\\
	&=E\left[\{g(X)-\mu_{g0}\}\{g(X)-\mu_{g0}\}^{\T}\right] \quad (\rho'(0)=1)\\
	&=var\{(g(X)\}.
\end{align*}

Notice 
\begin{align*}\frac{\partial W(X_i;\lambda^*)}{\partial \lambda}=  n\Bigg\{&\frac{\rho'\left[(\lambda^*)^{\T}\{g(X_i)-\mu_{g0}\}\right]\{g(X_i)-\mu_{g0}\}\left(\sum_{i=1}^n\rho\left[(\lambda^*)^{\T}\{g(X_i)-\mu_{g0}\}\right]\right)}{\left(\sum_{i=1}^n\rho\left[(\lambda^*)^{\T}\{g(X_i)-\mu_{g0}\}\right]\right)^2}\\
	&- \frac{\rho\left[(\lambda^*)^{\T}\{g(X_i)-\mu_{g0}\}\right]\left(\sum_{i=1}^n\rho'\left[(\lambda^*)^{\T}\{g(X_i)-\mu_{g0}\}\right]\{g(X_i)-\mu_{g0}\}\right)}{\left(\sum_{i=1}^n\rho\left[(\lambda^*)^{\T}\{g(X_i)-\mu_{g0}\}\right]\right)^2}\Bigg\}\\
	=\{g(&X_i)-\mu_{g0}\}-\frac{\sum_{i=1}^n\{g(X_i)-\mu_{g0}\}}{n}.
\end{align*}

Therefore, we have 
\begin{align*}
	H_\lambda&=\lim\limits_{n\to\infty}\frac{1}{n}\sum_{i=1}^n\left\{\frac{\partial W(X_i;\lambda^*)}{\partial \lambda}\right\}\psi(Y_i,A_i,X_i;\beta^*)\\
	&=\lim\limits_{n\to\infty}\frac{1}{n}\sum_{i=1}^n\left[\{g(X_i)-\mu_{g0}\}-\frac{\sum_{i=1}^n\{g(X_i)-\mu_{g0}\}}{n}\right]\psi(Y_i,A_i,X_i;\beta^*)\\
	&=\lim\limits_{n\to\infty}\frac{1}{n}\sum_{i=1}^n\left\{g(X_i)-\mu_{g0}\right\}\psi(Y_i,A_i,X_i;\beta^*)\\
	&\quad -\lim\limits_{n\to\infty}\frac{1}{n}\sum_{i=1}^n\left[\lim\limits_{n\to\infty}\frac{1}{n}\sum_{i=1}^n \frac{\sum_{i=1}^n\{g(X_i)-\mu_{g0}\}}{n}\right]\psi(Y_i,A_i,X_i;\beta^*)\\
	& = E\left[\left\{g(X_i)-\mu_{g0}\right\}\psi(Y_i,A_i,X_i;\beta^*)\right].
\end{align*}

The asymptotic variance of the original AIPW estimator is $\sigma^2=E(\xi_1^2)$.
\begin{align*}
	(\sigma^{\rm{c}})^2-\sigma^2 &= E\left\{(\xi_{1}+\xi_{2})^2\right\}-E(\xi_1^2)\\
	&= E\left(2\xi_1\xi_2+\xi_2^2\right)\\
	&=  -2H_\lambda^{\T}G_\lambda^{-1}E\left[\left\{\psi(Y,A,X;\beta)\right\}\{g(X)-\mu_{g0}\}\right]+2H_\lambda^{\T}G_\lambda^{-1}V^+(\beta)E\{g(X)-\mu_{g0}\}\\
	& \quad + H_\lambda^{\T}G_\lambda^{-1}var\{g(X)\}G_\lambda^{-1}H_\lambda\\
	&=-H_\lambda^{\T}G_\lambda^{-1}H_\lambda.
\end{align*}

Since $G_\lambda=var\{(g(X)\}$ is positive semidefinite, $(\sigma^{\rm{c}})^2-\sigma^2=-H_\lambda^{\T}G_\lambda^{-1}H_\lambda\leq 0$. Therefore, the calibrated AIPW estimator has the same or smaller asymptotic variance than the original AIPW estimator.
\QEDB

\section{Additional Simulation Results}
We implemented the $CR(1)$ method under the simulation setting in Section \ref{sec:simulation}. We call the $CR(1)$ method as the least squares method since minimizing $CR(1)$ is equivalent to minimizing the sum of squares $\sum_{i=1}^n(w_i-n^{-1})^2.$ In Scenarios 3 and 4, the calibration weights obtained by the least squares method can be negative, and thus the pseudo populations are no longer well-defined. Therefore, for a given linear ITR $d(X;\beta)$, we define the value function for the least squares method as 
$V_{LS}^+(\beta)=E\{W^*(X;\lambda^*)Y(d(X;\beta))\}$. Then, the true optimal linear ITR  is $d(X;\beta_{LS}^*)$, where $\beta_{LS}^*=\argmax_{\beta}V_{LS}^+(\beta)$. For a given linear ITR $d(X;\beta)$, we denote the calibrated value function estimator using the least squares method as  $\widehat{V}_{LS}^{\rm{c}}(\beta)$. The estimated optimal linear ITR is  $d(X;\widehat{\beta}_{LS}^{\rm{c}})$, where $\widehat{\beta}_{LS}^{\rm{c}}=\argmax_{\beta}\widehat{V}_{LS}^{\rm{c}}(\beta)$. We report the simulation results of the least squares method together with the other methods for comparison.

\subsection{Value and Percentage of Correct Decisions Results of Estimated Optimal ITRs}

\begin{figure}[!h]
	\centering
	\includegraphics[width=0.75\textwidth]{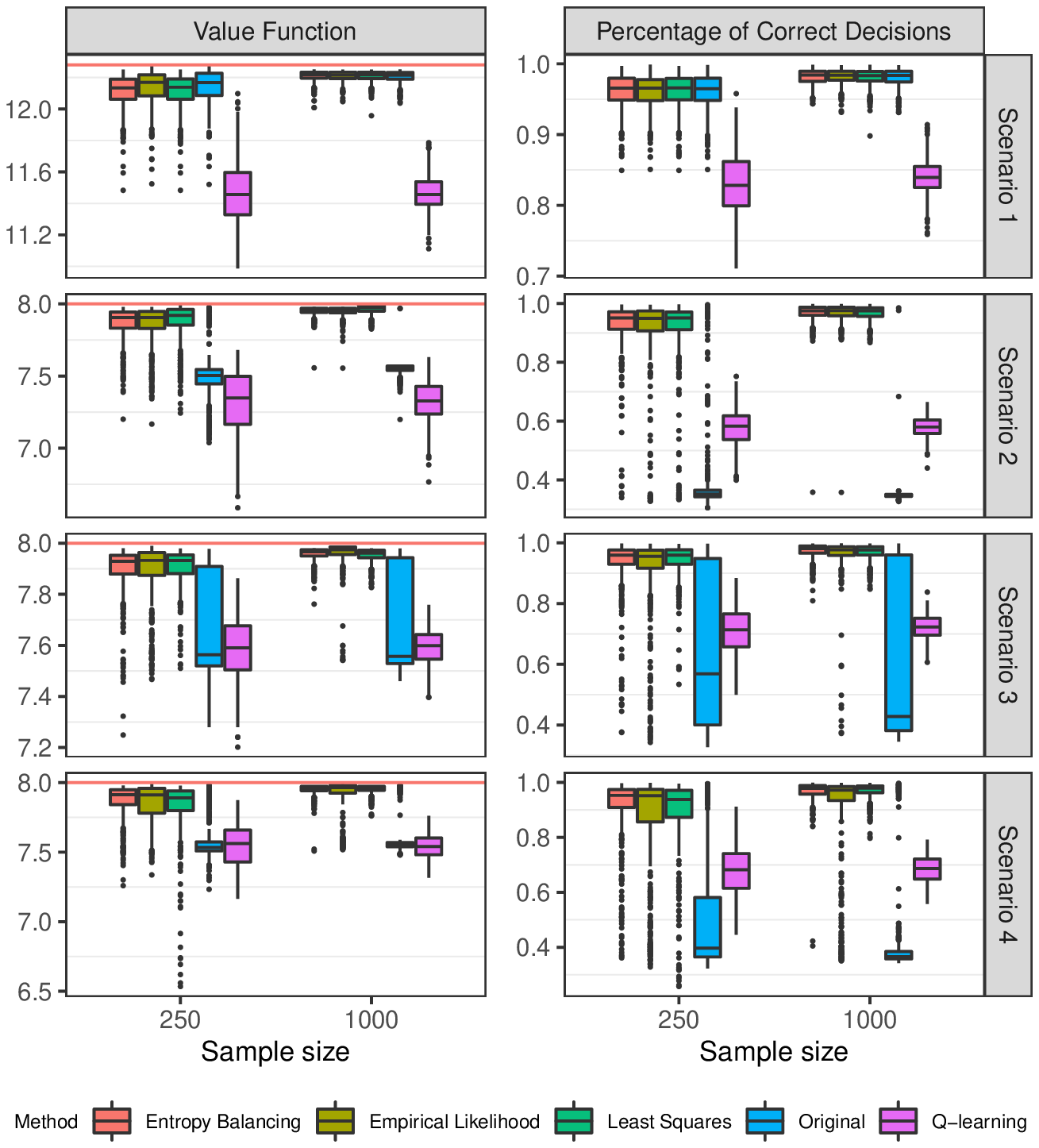}
	\caption{The value and percentage of correct decisions results of estimated optimal ITRs for the randomization study with implementation method I. The red lines are the values of the true optimal ITRs for the target population.}
	\label{figS1}
\end{figure}

\begin{figure}[!h]
	\centering
	\includegraphics[width=0.75\textwidth]{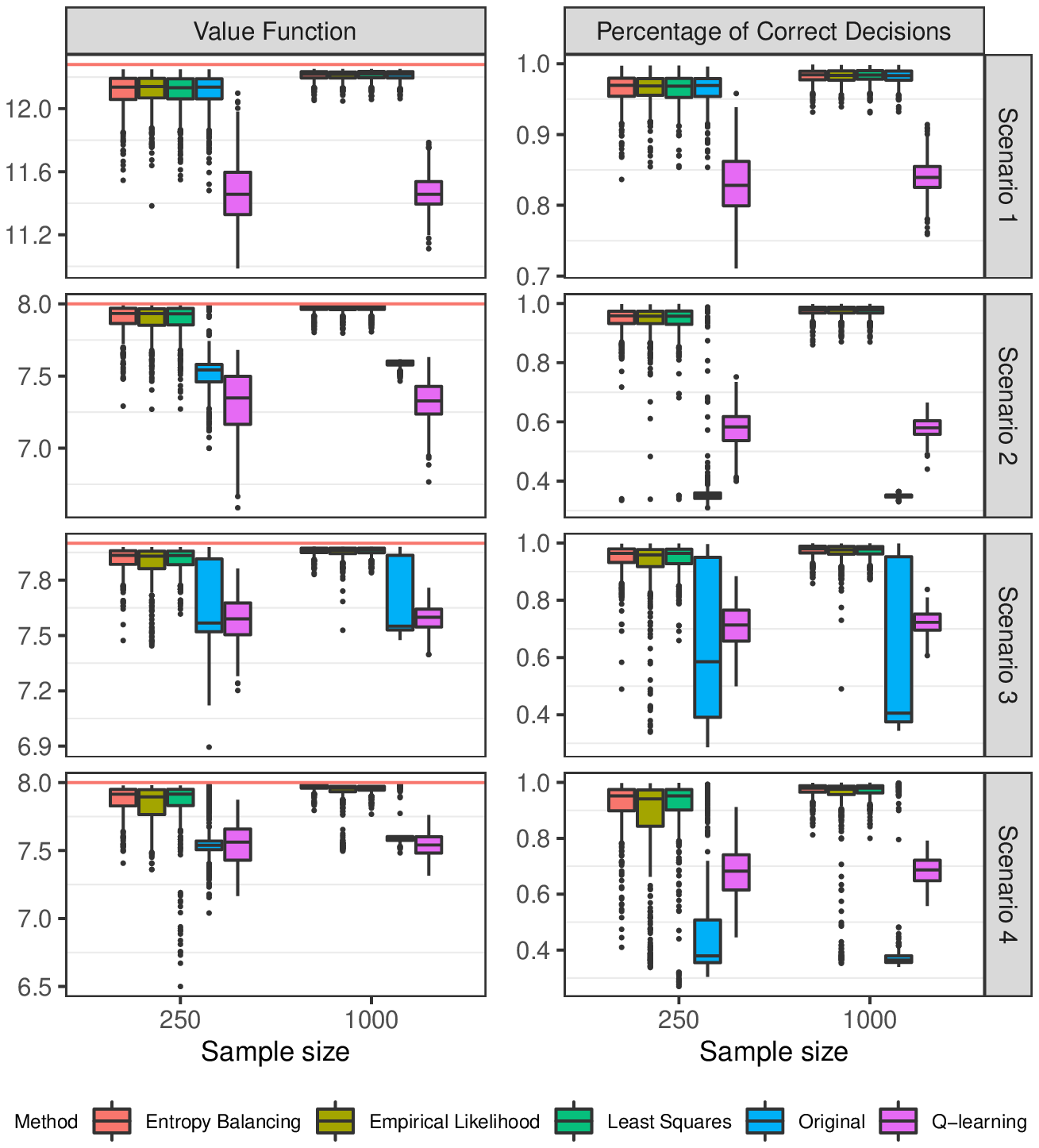}
	\caption{The value and percentage of correct decisions results of estimated optimal ITRs for the randomization study with implementation method II. The red lines are the values of the true optimal ITRs for the target population.}
	\label{figS2}
\end{figure}

\begin{figure}[!h]
	\centering
	\includegraphics[width=0.75\textwidth]{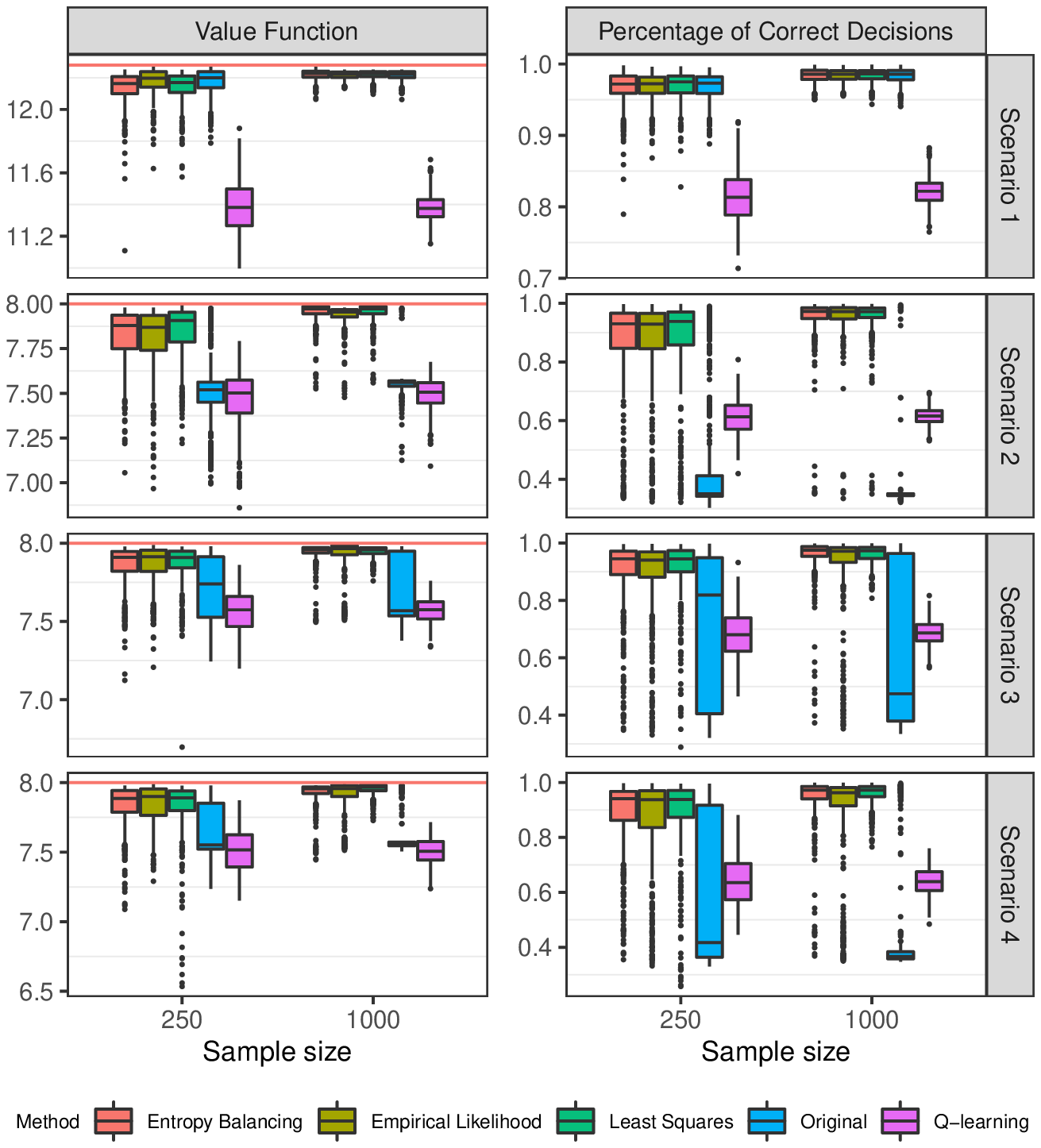}
	\caption{The value and percentage of correct decisions results of estimated optimal ITRs for the  observational study with implementation method I. The red lines are the values of the true optimal ITRs for the target population.}
	\label{figS3}
\end{figure}

\begin{figure}[!h]
	\centering
	\includegraphics[width=0.75\textwidth]{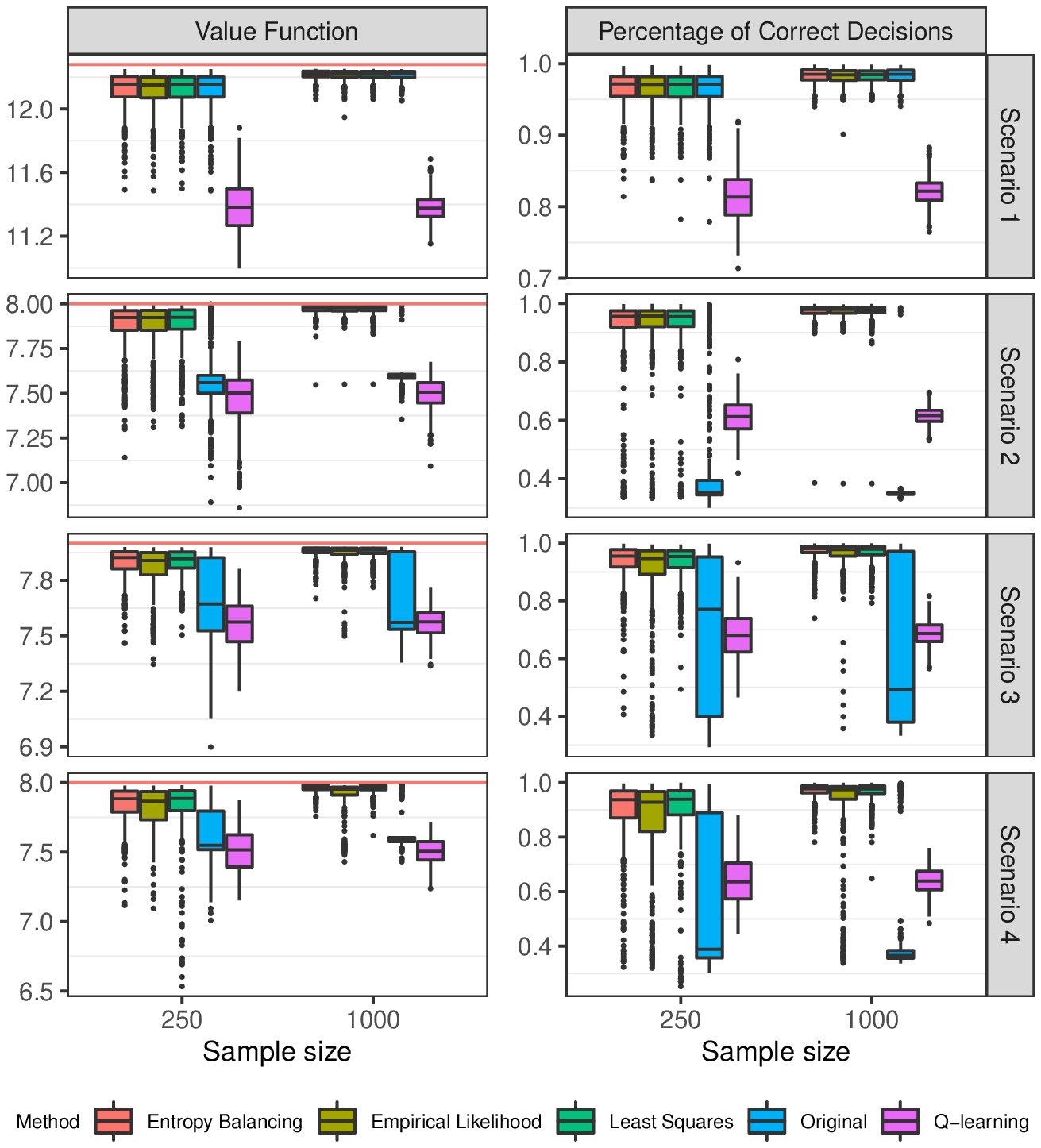}
	\caption{The value and percentage of correct decisions results of estimated optimal ITRs for the  observational study with implementation method II. The red lines are the values of the true optimal ITRs for the target population.}
	\label{figS4}
\end{figure}

The value and percentage of correct decisions results of estimated optimal ITRs obtained by different methods for the randomization study are summarized in Figure \ref{figS1} (method I) and Figure \ref{figS2} (method II).  The similar results for the observational study are summarized in Figure \ref{figS3} (method I) and Figure \ref{figS4} (method II). 

\clearpage

\clearpage

\subsection{Estimation and Inference Results of Optimal Value Estimators}
The estimation and inference results of $\widehat{V}^{\rm{c}}_{EB}(\widehat{\beta}^{\rm{c}}_{EB})$, $\widehat{V}^{\rm{c}}_{EL}(\widehat{\beta}^{\rm{c}}_{EL})$, and $\widehat{V}^{\rm{c}}_{LS}(\widehat{\beta}^{\rm{c}}_{LS})$ for the randomization study  are summarized in Table \ref{tableS1} with implementation method I and Table \ref{tableS2} with implementation method II. 
The similar results for the observational study are summarized in Table \ref{tableS3} with implementation method I and Table \ref{tableS4} with implementation method II. 

\begin{table}[!h]
	\def~{\hphantom{0}}
	\spacingset{1.45}
	\centering
	\caption{Simulation results for the randomization study with implementation method I.}
		\vspace{6pt}
	\begin{tabular}{cccccccccc}
		\hline 
		\multirow{3}{*}{Method} & Scenario & \multicolumn{2}{c}{1} & \multicolumn{2}{c}{2} & \multicolumn{2}{c}{3} & \multicolumn{2}{c}{4}\tabularnewline
		& $n$ & 250 & 1000 & 250 & 1000 & 250 & 1000 & 250 & 1000\tabularnewline
		& $V^{{\rm {t}}}(\beta^{{\rm {t}}})$ & \multicolumn{2}{c}{12.28} & \multicolumn{2}{c}{8.00} & \multicolumn{2}{c}{8.00} & \multicolumn{2}{c}{8.00}\tabularnewline
		\hline 
		\multirow{6}{*}{\shortstack{Entropy\\Balancing}} & $V_{EB}^{+}(\beta_{EB}^{*})$ & \multicolumn{2}{c}{12.28} & \multicolumn{2}{c}{8.00} & \multicolumn{2}{c}{7.99} & \multicolumn{2}{c}{7.99}\tabularnewline
		& Mean & 12.38 & 12.32 & 8.13 & 8.05 & 8.13 & 8.06 & 8.14 & 8.06\tabularnewline
		& SD & 0.41 & 0.20 & 0.37 & 0.18 & 0.32 & 0.15 & 0.37 & 0.18\tabularnewline
		& SE & 0.47 & 0.23 & 0.41 & 0.20 & 0.34 & 0.17 & 0.40 & 0.19\tabularnewline
		& $\text{CP}^{+}$ & 96.6 & 96.4 & 94.8 & 95.0 & 94.8 & 95.0 & 95.4 & 95.2\tabularnewline
		& $\text{CP}^{\text{t}}$ & 96.6 & 96.4 & 94.8 & 95.0 & 94.8 & 95.2 & 95.8 & 95.2\tabularnewline
		\hline 
		\multirow{6}{*}{\shortstack{Empirical\\Likelihood}} & $V_{EL}^{+}(\beta_{EL}^{*})$ & \multicolumn{2}{c}{12.28} & \multicolumn{2}{c}{8.00} & \multicolumn{2}{c}{8.14} & \multicolumn{2}{c}{8.16}\tabularnewline
		& Mean & 12.40 & 12.30 & 8.11 & 8.04 & 8.02 & 8.16 & 7.86 & 8.18\tabularnewline
		& SD & 0.40 & 0.20 & 0.34 & 0.18 & 0.37 & 0.18 & 0.43 & 0.25\tabularnewline
		& SE & 0.38 & 0.19 & 0.40 & 0.20 & 0.33 & 0.19 & 0.38 & 0.25\tabularnewline
		& $\text{CP}^{+}$ & 93.0 & 93.8 & 96.4 & 96.0 & 88.6 & 95.4 & 78.0 & 94.0\tabularnewline
		& $\text{CP}^{\text{t}}$ & 93.0 & 93.8 & 96.4 & 96.0 & 92.2 & 89.2 & 85.8 & 92.0\tabularnewline
		\hline 
		\multirow{6}{*}{\shortstack{Least\\Squares}} & $V_{LS}^{+}(\beta_{LS}^{*})$ & \multicolumn{2}{c}{12.28} & \multicolumn{2}{c}{8.00} & \multicolumn{2}{c}{8.13} & \multicolumn{2}{c}{8.15}\tabularnewline
		& Mean & 12.38 & 12.31 & 8.15 & 8.03 & 8.27 & 8.17 & 8.40 & 8.20\tabularnewline
		& SD & 0.41 & 0.20 & 0.37 & 0.19 & 0.34 & 0.18 & 0.44 & 0.21\tabularnewline
		& SE & 0.38 & 0.19 & 0.49 & 0.24 & 0.45 & 0.22 & 0.60 & 0.30\tabularnewline
		& $\text{CP}^{+}$ & 92.2 & 93.4 & 98.6 & 98.6 & 98.2 & 98.2 & 98.4 & 99.6\tabularnewline
		& $\text{CP}^{\text{t}}$ & 92.2 & 93.4 & 98.6 & 98.6 & 96.6 & 93.2 & 97.2 & 96.8\tabularnewline
		\hline 
	\end{tabular}
	\label{tableS1}
\end{table}

\begin{table}[!h]
	\def~{\hphantom{0}}
	\caption{Simulation results for the randomization study with implementation method II.}
	\spacingset{1.45}
	\vspace{6pt}
	\centering
	\begin{tabular}{cccccccccc}
		\hline 
		\multirow{3}{*}{Method} & Scenario & \multicolumn{2}{c}{1} & \multicolumn{2}{c}{2} & \multicolumn{2}{c}{3} & \multicolumn{2}{c}{4}\tabularnewline
		& $n$ & 250 & 1000 & 250 & 1000 & 250 & 1000 & 250 & 1000\tabularnewline
		& $V^{{\rm {t}}}(\beta^{{\rm {t}}})$ & \multicolumn{2}{c}{12.28} & \multicolumn{2}{c}{8.00} & \multicolumn{2}{c}{8.00} & \multicolumn{2}{c}{8.00}\tabularnewline
		\hline 
		\multirow{6}{*}{\shortstack{Entropy\\Balancing}} & $V_{EB}^{+}(\beta_{EB}^{*})$ & \multicolumn{2}{c}{12.28} & \multicolumn{2}{c}{8.00} & \multicolumn{2}{c}{7.99} & \multicolumn{2}{c}{7.99}\tabularnewline
		& Mean & 12.39 & 12.31 & 8.12 & 8.02 & 8.14 & 8.05 & 8.14 & 8.06\tabularnewline
		& SD & 0.40 & 0.19 & 0.33 & 0.16 & 0.28 & 0.14 & 0.35 & 0.17\tabularnewline
		& SE & 0.47 & 0.23 & 0.39 & 0.19 & 0.32 & 0.15 & 0.38 & 0.18\tabularnewline
		& $\text{CP}^{+}$ & 96.8 & 97.4 & 96.2 & 96.6 & 95.4 & 94.6 & 95.2 & 95.4\tabularnewline
		& $\text{CP}^{\text{t}}$ & 96.8 & 97.4 & 96.2 & 96.6 & 95.4 & 95.2 & 95.2 & 95.2\tabularnewline
		\hline 
		\multirow{6}{*}{\shortstack{Empirical\\Likelihood}} & $V_{EL}^{+}(\beta_{EL}^{*})$ & \multicolumn{2}{c}{12.28} & \multicolumn{2}{c}{8.00} & \multicolumn{2}{c}{8.14} & \multicolumn{2}{c}{8.16}\tabularnewline
		& Mean & 12.38 & 12.31 & 8.12 & 8.02 & 8.04 & 8.16 & 7.82 & 8.17\tabularnewline
		& SD & 0.40 & 0.19 & 0.33 & 0.16 & 0.32 & 0.18 & 0.45 & 0.24\tabularnewline
		& SE & 0.38 & 0.18 & 0.39 & 0.19 & 0.31 & 0.18 & 0.36 & 0.25\tabularnewline
		& $\text{CP}^{+}$ & 93.2 & 94.6 & 95.6 & 96.6 & 92.2 & 95.2 & 73.0 & 93.4\tabularnewline
		& $\text{CP}^{\text{t}}$ & 93.2 & 94.6 & 95.6 & 96.6 & 94.4 & 88.8 & 81.0 & 94.0\tabularnewline
		\hline 
		\multirow{6}{*}{\shortstack{Least\\Squares}} & $V_{LS}^{+}(\beta_{LS}^{*})$ & \multicolumn{2}{c}{12.28} & \multicolumn{2}{c}{8.00} & \multicolumn{2}{c}{8.13} & \multicolumn{2}{c}{8.15}\tabularnewline
		& Mean & 12.39 & 12.31 & 8.13 & 8.02 & 8.26 & 8.16 & 8.31 & 8.19\tabularnewline
		& SD & 0.40 & 0.19 & 0.34 & 0.17 & 0.32 & 0.16 & 0.41 & 0.20\tabularnewline
		& SE & 0.39 & 0.19 & 0.48 & 0.23 & 0.43 & 0.21 & 0.58 & 0.28\tabularnewline
		& $\text{CP}^{+}$ & 93.8 & 94.4 & 98.8 & 99.0 & 98.8 & 98.8 & 99.6 & 99.6\tabularnewline
		& $\text{CP}^{\text{t}}$ & 93.8 & 94.4 & 98.8 & 99.0 & 96.8 & 94.8 & 98.0 & 96.6\tabularnewline
		\hline 
	\end{tabular}
	\label{tableS2}
\end{table}

\newpage

\begin{table}[!h]
	\def~{\hphantom{0}}
	\caption{Simulation results for the observational study with implementation method I.}
	\spacingset{1.45}
	\vspace{6pt}
	\centering
	\begin{tabular}{cccccccccc}
		\hline 
		\multirow{3}{*}{Method} & Scenario & \multicolumn{2}{c}{1} & \multicolumn{2}{c}{2} & \multicolumn{2}{c}{3} & \multicolumn{2}{c}{4}\tabularnewline
		& $n$ & 250 & 1000 & 250 & 1000 & 250 & 1000 & 250 & 1000\tabularnewline
		& $V^{{\rm {t}}}(\beta^{{\rm {t}}})$ & \multicolumn{2}{c}{12.28} & \multicolumn{2}{c}{8.00} & \multicolumn{2}{c}{8.00} & \multicolumn{2}{c}{8.00}\tabularnewline
		\hline 
		\multirow{6}{*}{\shortstack{Entropy\\Balancing}} & $V_{EB}^{+}(\beta_{EB}^{*})$ & \multicolumn{2}{c}{12.28} & \multicolumn{2}{c}{8.00} & \multicolumn{2}{c}{7.99} & \multicolumn{2}{c}{7.99}\tabularnewline
		& Mean & 12.34 & 12.32 & 8.20 & 8.06 & 8.22 & 8.08 & 8.24 & 8.08\tabularnewline
		& SD & 0.42 & 0.21 & 0.36 & 0.17 & 0.31 & 0.16 & 0.45 & 0.17\tabularnewline
		& SE & 0.47 & 0.24 & 0.41 & 0.19 & 0.35 & 0.17 & 0.43 & 0.19\tabularnewline
		& $\text{CP}^{+}$ & 96.8 & 95.6 & 95.4 & 95.6 & 95.0 & 95.2 & 94.6 & 94.8\tabularnewline
		& $\text{CP}^{\text{t}}$ & 96.8 & 95.6 & 95.4 & 95.6 & 95.2 & 96.0 & 94.8 & 96.0\tabularnewline
		\hline 
		\multirow{6}{*}{\shortstack{Empirical\\Likelihood}} & $V_{EL}^{+}(\beta_{EL}^{*})$ & \multicolumn{2}{c}{12.28} & \multicolumn{2}{c}{8.00} & \multicolumn{2}{c}{8.14} & \multicolumn{2}{c}{8.16}\tabularnewline
		& Mean & 12.36 & 12.31 & 8.17 & 8.06 & 8.09 & 8.19 & 7.90 & 8.22\tabularnewline
		& SD & 0.41 & 0.22 & 0.37 & 0.16 & 0.35 & 0.20 & 0.43 & 0.26\tabularnewline
		& SE & 0.38 & 0.20 & 0.41 & 0.19 & 0.32 & 0.20 & 0.37 & 0.27\tabularnewline
		& $\text{CP}^{+}$ & 93.6 & 96.0 & 96.4 & 97.0 & 92.0 & 96.8 & 81.2 & 94.0\tabularnewline
		& $\text{CP}^{\text{t}}$ & 93.6 & 96.0 & 96.4 & 97.0 & 93.2 & 88.4 & 87.0 & 93.4\tabularnewline
		\hline 
		\multirow{6}{*}{\shortstack{Least\\Squares}} & $V_{LS}^{+}(\beta_{LS}^{*})$ & \multicolumn{2}{c}{12.28} & \multicolumn{2}{c}{8.00} & \multicolumn{2}{c}{8.13} & \multicolumn{2}{c}{8.15}\tabularnewline
		& Mean & 12.37 & 12.30 & 8.19 & 8.05 & 8.34 & 8.19 & 8.40 & 8.23\tabularnewline
		& SD & 0.42 & 0.21 & 0.37 & 0.17 & 0.35 & 0.17 & 0.44 & 0.20\tabularnewline
		& SE & 0.39 & 0.20 & 0.48 & 0.23 & 0.46 & 0.22 & 0.60 & 0.29\tabularnewline
		& $\text{CP}^{+}$ & 93.4 & 95.8 & 98.4 & 98.2 & 97.8 & 98.4 & 98.4 & 98.8\tabularnewline
		& $\text{CP}^{\text{t}}$ & 93.4 & 95.8 & 98.4 & 98.2 & 94.0 & 93.8 & 97.2 & 95.4\tabularnewline
		\hline 
	\end{tabular}
	\label{tableS3}
\end{table}

\begin{table}[!h]
	\def~{\hphantom{0}}
	\caption{Simulation results  for the observational study with implementation method II.}
	\spacingset{1.45}
	\vspace{6pt}
	\centering
	\begin{tabular}{cccccccccc}
		\hline 
		\multirow{3}{*}{Method} & Scenario & \multicolumn{2}{c}{1} & \multicolumn{2}{c}{2} & \multicolumn{2}{c}{3} & \multicolumn{2}{c}{4}\tabularnewline
		& $n$ & 250 & 1000 & 250 & 1000 & 250 & 1000 & 250 & 1000\tabularnewline
		& $V^{{\rm {t}}}(\beta^{{\rm {t}}})$ & \multicolumn{2}{c}{12.28} & \multicolumn{2}{c}{8.00} & \multicolumn{2}{c}{8.00} & \multicolumn{2}{c}{8.00}\tabularnewline
		\hline 
		\multirow{6}{*}{\shortstack{Entropy\\Balancing}} & $V_{EB}^{+}(\beta_{EB}^{*})$ & \multicolumn{2}{c}{12.28} & \multicolumn{2}{c}{8.00} & \multicolumn{2}{c}{7.99} & \multicolumn{2}{c}{7.99}\tabularnewline
		& Mean & 12.42 & 12.32 & 8.12 & 8.03 & 8.15 & 8.06 & 8.19 & 8.06\tabularnewline
		& SD & 0.50 & 0.22 & 0.33 & 0.15 & 0.29 & 0.13 & 0.34 & 0.15\tabularnewline
		& SE & 0.54 & 0.25 & 0.38 & 0.18 & 0.30 & 0.14 & 0.37 & 0.17\tabularnewline
		& $\text{CP}^{+}$ & 97.4 & 97.0 & 96.8 & 96.2 & 93.8 & 95.4 & 94.6 & 95.4\tabularnewline
		& $\text{CP}^{\text{t}}$ & 97.4 & 97.0 & 96.8 & 96.2 & 94.6 & 95.6 & 95.0 & 96.0\tabularnewline
		\hline 
		\multirow{6}{*}{\shortstack{Empirical\\Likelihood}} & $V_{EL}^{+}(\beta_{EL}^{*})$ & \multicolumn{2}{c}{12.28} & \multicolumn{2}{c}{8.00} & \multicolumn{2}{c}{8.14} & \multicolumn{2}{c}{8.16}\tabularnewline
		& Mean & 12.42 & 12.32 & 8.11 & 8.03 & 8.06 & 8.18 & 7.86 & 8.16\tabularnewline
		& SD & 0.50 & 0.23 & 0.33 & 0.15 & 0.31 & 0.17 & 0.43 & 0.23\tabularnewline
		& SE & 0.46 & 0.21 & 0.37 & 0.17 & 0.29 & 0.17 & 0.35 & 0.24\tabularnewline
		& $\text{CP}^{+}$ & 95.4 & 95.4 & 96.6 & 96.0 & 91.6 & 95.4 & 74.4 & 93.2\tabularnewline
		& $\text{CP}^{\text{t}}$ & 95.4 & 95.4 & 96.6 & 96.0 & 94.0 & 84.0 & 83.4 & 96.2\tabularnewline
		\hline 
		\multirow{6}{*}{\shortstack{Least\\Squares}} & $V_{LS}^{+}(\beta_{LS}^{*})$ & \multicolumn{2}{c}{12.28} & \multicolumn{2}{c}{8.00} & \multicolumn{2}{c}{8.13} & \multicolumn{2}{c}{8.15}\tabularnewline
		& Mean & 12.42 & 12.32 & 8.13 & 8.03 & 8.31 & 8.17 & 8.38 & 8.21\tabularnewline
		& SD & 0.50 & 0.23 & 0.34 & 0.16 & 0.32 & 0.15 & 0.40 & 0.18\tabularnewline
		& SE & 0.46 & 0.22 & 0.46 & 0.22 & 0.42 & 0.21 & 0.56 & 0.27\tabularnewline
		& $\text{CP}^{+}$ & 95.8 & 95.6 & 98.8 & 98.4 & 98.0 & 99.2 & 98.6 & 99.6\tabularnewline
		& $\text{CP}^{\text{t}}$ & 95.8 & 95.6 & 98.8 & 99.0 & 96.0 & 95.0 & 97.6 & 96.2\tabularnewline
		\hline 
	\end{tabular}
	\label{tableS4}
\end{table}

\clearpage
\subsection{Comparison between Calibration with All Covariates and with Only \boldmath{$X_1$}}

We conducted additional simulation studies under the setting in Section \ref{sec:simulation}, but only use  mean of $X_1$ as the summary statistics from the target population for calibration. The value and percentage of correct decisions results of  the entropy balancing, empirical likelihood, and least squares methods using different covariates for calibration for the randomization study are summarized in Table \ref{tableS5} (method I) and Table \ref{tableS6} (method II).  The similar results for the observational study are summarized in Table \ref{tableS7} (method I) and Table \ref{tableS8} (method II).

\begin{table}[!h]
	\def~{\hphantom{0}}
	\caption{Mean and standard deviations (in parenthesis) of the values and percentages of correct decisions (PCD) using different covariates for calibration for the randomization study with implementation method I.}
	\vspace{6pt}
		\resizebox{\textwidth}{!}{%
			\begin{tabular}{ccccccccc}
				\hline 
				\multirow{2}{*}{Scenario} & \multirow{2}{*}{$n$} & \multirow{2}{*}{Covariate} & \multicolumn{2}{c}{Entropy Balancing} & \multicolumn{2}{c}{Empirical Likelihood} & \multicolumn{2}{c}{Least Squares}\tabularnewline
				&  &  & Value & PCD & Value & PCD & Value & PCD\tabularnewline
				\hline 
				\multirow{4}{*}{1} & \multirow{2}{*}{250} & $X_{1}$ & 12.12(0.10) & 0.96(0.02) & 12.12(0.10) & 0.96(0.02) & 12.12(0.10) & 0.96(0.02)\tabularnewline
				&  & $X_{1},X_{2},X_{3}$ & 12.11(0.10) & 0.96(0.02) & 12.14(0.11) & 0.96(0.02) & 12.11(0.11) & 0.96(0.02)\tabularnewline
				& \multirow{2}{*}{1000} & $X_{1}$ & 12.21(0.03) & 0.98(0.01) & 12.21(0.03) & 0.98(0.01) & 12.21(0.03) & 0.98(0.01)\tabularnewline
				&  & $X_{1},X_{2},X_{3}$ & 12.21(0.03) & 0.98(0.01) & 12.21(0.03) & 0.98(0.01) & 12.21(0.03) & 0.98(0.01)\tabularnewline
				\hline 
				\multirow{4}{*}{2} & \multirow{2}{*}{250} & $X_{1}$ & 7.86(0.13) & 0.91(0.11) & 7.86(0.13) & 0.91(0.11) & 7.86(0.13) & 0.91(0.11)\tabularnewline
				&  & $X_{1},X_{2},X_{3}$ & 7.87(0.11) & 0.92(0.09) & 7.86(0.13) & 0.91(0.13) & 7.88(0.13) & 0.91(0.12)\tabularnewline
				& \multirow{2}{*}{1000} & $X_{1}$ & 7.95(0.04) & 0.97(0.02) & 7.95(0.03) & 0.97(0.02) & 7.95(0.04) & 0.97(0.02)\tabularnewline
				&  & $X_{1},X_{2},X_{3}$ & 7.95(0.03) & 0.97(0.04) & 7.95(0.04) & 0.97(0.04) & 7.96(0.03) & 0.97(0.02)\tabularnewline
				\hline 
				\multirow{4}{*}{3} & \multirow{2}{*}{250} & $X_{1}$ & 7.68(0.19) & 0.65(0.25) & 7.68(0.19) & 0.65(0.25) & 7.68(0.19) & 0.65(0.25)\tabularnewline
				&  & $X_{1},X_{2},X_{3}$ & 7.90(0.10) & 0.93(0.09) & 7.89(0.12) & 0.91(0.14) & 7.91(0.07) & 0.94(0.05)\tabularnewline
				& \multirow{2}{*}{1000} & $X_{1}$ & 7.70(0.20) & 0.62(0.28) & 7.69(0.20) & 0.62(0.28) & 7.70(0.20) & 0.62(0.28)\tabularnewline
				&  & $X_{1},X_{2},X_{3}$ & 7.96(0.03) & 0.97(0.03) & 7.96(0.06) & 0.96(0.07) & 7.95(0.03) & 0.97(0.03)\tabularnewline
				\hline 
				\multirow{4}{*}{4} & \multirow{2}{*}{250} & $X_{1}$ & 7.60(0.16) & 0.52(0.23) & 7.60(0.17) & 0.53(0.23) & 7.60(0.16) & 0.52(0.23)\tabularnewline
				&  & $X_{1},X_{2},X_{3}$ & 7.87(0.12) & 0.91(0.12) & 7.84(0.16) & 0.85(0.20) & 7.83(0.20) & 0.89(0.13)\tabularnewline
				& \multirow{2}{*}{1000} & $X_{1}$ & 7.58(0.12) & 0.43(0.18) & 7.57(0.11) & 0.42(0.16) & 7.58(0.12) & 0.43(0.18)\tabularnewline
				&  & $X_{1},X_{2},X_{3}$ & 7.95(0.04) & 0.97(0.04) & 7.90(0.15) & 0.88(0.20) & 7.95(0.03) & 0.97(0.03)\tabularnewline
				\hline 
			\end{tabular}
	}
	\label{tableS5}
\end{table}

\begin{table}[!h]
	\def~{\hphantom{0}}
	\caption{Mean and standard deviations (in parenthesis) of the values and percentages of correct decisions (PCD) using different covariates for calibration for the randomization study with implementation method II.}
	\vspace{6pt}
	\resizebox{\textwidth}{!}{%
		\begin{tabular}{ccccccccc}
			\hline 
			\multirow{2}{*}{Scenario} & \multirow{2}{*}{$n$} & \multirow{2}{*}{Covariate} & \multicolumn{2}{c}{Entropy Balancing} & \multicolumn{2}{c}{Empirical Likelihood} & \multicolumn{2}{c}{Least Squares}\tabularnewline
			&  &  & Value & PCD & Value & PCD & Value & PCD\tabularnewline
			\hline 
			\multirow{4}{*}{1} & \multirow{2}{*}{250} & $X_{1}$ & 12.11(0.11) & 0.96(0.02) & 12.11(0.11) & 0.96(0.02) & 12.11(0.11) & 0.96(0.02)\tabularnewline
			&  & $X_{1},X_{2},X_{3}$ & 12.11(0.11) & 0.96(0.02) & 12.11(0.11) & 0.96(0.02) & 12.10(0.12) & 0.96(0.02)\tabularnewline
			& \multirow{2}{*}{1000} & $X_{1}$ & 12.21(0.04) & 0.98(0.01) & 12.21(0.03) & 0.98(0.01) & 12.21(0.04) & 0.98(0.01)\tabularnewline
			&  & $X_{1},X_{2},X_{3}$ & 12.21(0.03) & 0.98(0.01) & 12.21(0.03) & 0.98(0.01) & 12.21(0.03) & 0.98(0.01)\tabularnewline
			\hline 
			\multirow{4}{*}{2} & \multirow{2}{*}{250} & $X_{1}$ & 7.86(0.12) & 0.93(0.07) & 7.86(0.12) & 0.93(0.08) & 7.86(0.12) & 0.93(0.07)\tabularnewline
			&  & $X_{1},X_{2},X_{3}$ & 7.90(0.10) & 0.94(0.06) & 7.90(0.10) & 0.94(0.06) & 7.89(0.11) & 0.94(0.07)\tabularnewline
			& \multirow{2}{*}{1000} & $X_{1}$ & 7.96(0.03) & 0.98(0.02) & 7.95(0.03) & 0.98(0.02) & 7.96(0.03) & 0.98(0.02)\tabularnewline
			&  & $X_{1},X_{2},X_{3}$ & 7.97(0.03) & 0.97(0.02) & 7.97(0.03) & 0.97(0.02) & 7.97(0.03) & 0.97(0.02)\tabularnewline
			\hline 
			\multirow{4}{*}{3} & \multirow{2}{*}{250} & $X_{1}$ & 7.69(0.21) & 0.66(0.26) & 7.68(0.20) & 0.66(0.26) & 7.69(0.21) & 0.66(0.26)\tabularnewline
			&  & $X_{1},X_{2},X_{3}$ & 7.91(0.07) & 0.95(0.05) & 7.89(0.10) & 0.92(0.11) & 7.91(0.06) & 0.95(0.05)\tabularnewline
			& \multirow{2}{*}{1000} & $X_{1}$ & 7.70(0.20) & 0.62(0.28) & 7.70(0.20) & 0.62(0.28) & 7.70(0.20) & 0.62(0.28)\tabularnewline
			&  & $X_{1},X_{2},X_{3}$ & 7.96(0.02) & 0.97(0.02) & 7.95(0.04) & 0.97(0.04) & 7.96(0.02) & 0.97(0.02)\tabularnewline
			\hline 
			\multirow{4}{*}{4} & \multirow{2}{*}{250} & $X_{1}$ & 7.59(0.18) & 0.51(0.24) & 7.58(0.18) & 0.51(0.24) & 7.59(0.18) & 0.51(0.24)\tabularnewline
			&  & $X_{1},X_{2},X_{3}$ & 7.88(0.10) & 0.92(0.09) & 7.83(0.15) & 0.86(0.18) & 7.85(0.21) & 0.91(0.13)\tabularnewline
			& \multirow{2}{*}{1000} & $X_{1}$ & 7.58(0.11) & 0.42(0.17) & 7.58(0.11) & 0.41(0.16) & 7.58(0.11) & 0.42(0.17)\tabularnewline
			&  & $X_{1},X_{2},X_{3}$ & 7.97(0.03) & 0.97(0.02) & 7.92(0.11) & 0.93(0.14) & 7.95(0.03) & 0.97(0.02)\tabularnewline
			\hline 
		\end{tabular}
	}
	\label{tableS6}
\end{table}

\begin{table}[!h]
	\def~{\hphantom{0}}
	\caption{Mean and standard deviations (in parenthesis) of the values and percentages of correct decisions (PCD) using different covariates for calibration for the observational study with implementation method I.}
	\vspace{6pt}
	\resizebox{\textwidth}{!}{%
		\begin{tabular}{ccccccccc}
			\hline 
			\multirow{2}{*}{Scenario} & \multirow{2}{*}{$n$} & \multirow{2}{*}{Covariate} & \multicolumn{2}{c}{Entropy Balancing} & \multicolumn{2}{c}{Empirical Likelihood} & \multicolumn{2}{c}{Least Squares}\tabularnewline
			&  &  & Value & PCD & Value & PCD & Value & PCD\tabularnewline
			\hline 
			\multirow{4}{*}{1} & \multirow{2}{*}{250} & $X_{1}$ & 12.14(0.10) & 0.97(0.02) & 12.14(0.09) & 0.97(0.02) & 12.14(0.10) & 0.97(0.02)\tabularnewline
			&  & $X_{1},X_{2},X_{3}$ & 12.14(0.10) & 0.97(0.02) & 12.18(0.08) & 0.97(0.02) & 12.14(0.09) & 0.97(0.02)\tabularnewline
			& \multirow{2}{*}{1000} & $X_{1}$ & 12.22(0.03) & 0.98(0.01) & 12.22(0.03) & 0.98(0.01) & 12.22(0.03) & 0.98(0.01)\tabularnewline
			&  & $X_{1},X_{2},X_{3}$ & 12.22(0.03) & 0.98(0.01) & 12.22(0.03) & 0.98(0.01) & 12.22(0.03) & 0.98(0.01)\tabularnewline
			\hline 
			\multirow{4}{*}{2} & \multirow{2}{*}{250} & $X_{1}$ & 7.82(0.15) & 0.87(0.16) & 7.82(0.15) & 0.87(0.16) & 7.82(0.15) & 0.87(0.16)\tabularnewline
			&  & $X_{1},X_{2},X_{3}$ & 7.82(0.16) & 0.87(0.17) & 7.81(0.17) & 0.86(0.16) & 7.84(0.16) & 0.87(0.17)\tabularnewline
			& \multirow{2}{*}{1000} & $X_{1}$ & 7.94(0.06) & 0.95(0.07) & 7.94(0.05) & 0.96(0.07) & 7.94(0.06) & 0.95(0.07)\tabularnewline
			&  & $X_{1},X_{2},X_{3}$ & 7.95(0.06) & 0.95(0.08) & 7.93(0.06) & 0.95(0.08) & 7.95(0.06) & 0.95(0.07)\tabularnewline
			\hline 
			\multirow{4}{*}{3} & \multirow{2}{*}{250} & $X_{1}$ & 7.72(0.19) & 0.71(0.26) & 7.73(0.19) & 0.71(0.26) & 7.72(0.19) & 0.71(0.26)\tabularnewline
			&  & $X_{1},X_{2},X_{3}$ & 7.86(0.13) & 0.90(0.12) & 7.86(0.13) & 0.89(0.14) & 7.87(0.12) & 0.91(0.10)\tabularnewline
			& \multirow{2}{*}{1000} & $X_{1}$ & 7.73(0.20) & 0.66(0.28) & 7.73(0.20) & 0.66(0.28) & 7.73(0.20) & 0.66(0.28)\tabularnewline
			&  & $X_{1},X_{2},X_{3}$ & 7.94(0.07) & 0.96(0.07) & 7.92(0.12) & 0.91(0.15) & 7.95(0.03) & 0.96(0.03)\tabularnewline
			\hline 
			\multirow{4}{*}{4} & \multirow{2}{*}{250} & $X_{1}$ & 7.64(0.18) & 0.57(0.26) & 7.64(0.18) & 0.57(0.26) & 7.64(0.18) & 0.57(0.26)\tabularnewline
			&  & $X_{1},X_{2},X_{3}$ & 7.83(0.16) & 0.89(0.13) & 7.84(0.15) & 0.86(0.17) & 7.83(0.20) & 0.89(0.13)\tabularnewline
			& \multirow{2}{*}{1000} & $X_{1}$ & 7.61(0.14) & 0.45(0.21) & 7.61(0.14) & 0.46(0.21) & 7.61(0.14) & 0.45(0.21)\tabularnewline
			&  & $X_{1},X_{2},X_{3}$ & 7.93(0.09) & 0.95(0.09) & 7.89(0.15) & 0.88(0.20) & 7.95(0.04) & 0.96(0.04)\tabularnewline
			\hline 
		\end{tabular}
	}
	\label{tableS7}
\end{table}

\begin{table}[!h]
	\def~{\hphantom{0}}
	\caption{Mean and standard deviations (in parenthesis) of the values and percentages of correct decisions (PCD) using different covariates for calibration for the observational study with implementation method II.}
	\vspace{6pt}
	\resizebox{\textwidth}{!}{%
		\begin{tabular}{ccccccccc}
			\hline 
			\multirow{2}{*}{Scenario} & \multirow{2}{*}{$n$} & \multirow{2}{*}{Covariate} & \multicolumn{2}{c}{Entropy Balancing} & \multicolumn{2}{c}{Empirical Likelihood} & \multicolumn{2}{c}{Least Squares}\tabularnewline
			&  &  & Value & PCD & Value & PCD & Value & PCD\tabularnewline
			\hline 
			\multirow{4}{*}{1} & \multirow{2}{*}{250} & $X_{1}$ & 12.12(0.12) & 0.97(0.03) & 12.12(0.11) & 0.97(0.02) & 12.12(0.12) & 0.97(0.03)\tabularnewline
			&  & $X_{1},X_{2},X_{3}$ & 12.12(0.11) & 0.96(0.02) & 12.12(0.12) & 0.97(0.02) & 12.12(0.12) & 0.96(0.03)\tabularnewline
			& \multirow{2}{*}{1000} & $X_{1}$ & 12.21(0.04) & 0.98(0.01) & 12.21(0.03) & 0.98(0.01) & 12.21(0.04) & 0.98(0.01)\tabularnewline
			&  & $X_{1},X_{2},X_{3}$ & 12.22(0.03) & 0.98(0.01) & 12.21(0.03) & 0.98(0.01) & 12.21(0.03) & 0.98(0.01)\tabularnewline
			\hline 
			\multirow{4}{*}{2} & \multirow{2}{*}{250} & $X_{1}$ & 7.86(0.12) & 0.91(0.13) & 7.86(0.12) & 0.92(0.13) & 7.86(0.12) & 0.91(0.13)\tabularnewline
			&  & $X_{1},X_{2},X_{3}$ & 7.87(0.14) & 0.91(0.14) & 7.88(0.13) & 0.91(0.14) & 7.88(0.13) & 0.91(0.15)\tabularnewline
			& \multirow{2}{*}{1000} & $X_{1}$ & 7.95(0.04) & 0.97(0.04) & 7.95(0.04) & 0.97(0.04) & 7.95(0.04) & 0.97(0.04)\tabularnewline
			&  & $X_{1},X_{2},X_{3}$ & 7.97(0.03) & 0.97(0.03) & 7.97(0.03) & 0.97(0.03) & 7.97(0.03) & 0.97(0.03)\tabularnewline
			\hline 
			\multirow{4}{*}{3} & \multirow{2}{*}{250} & $X_{1}$ & 7.71(0.21) & 0.68(0.27) & 7.71(0.21) & 0.69(0.27) & 7.71(0.21) & 0.68(0.27)\tabularnewline
			&  & $X_{1},X_{2},X_{3}$ & 7.89(0.09) & 0.93(0.07) & 7.87(0.12) & 0.90(0.13) & 7.90(0.07) & 0.93(0.06)\tabularnewline
			& \multirow{2}{*}{1000} & $X_{1}$ & 7.71(0.20) & 0.64(0.28) & 7.72(0.20) & 0.64(0.28) & 7.71(0.20) & 0.64(0.28)\tabularnewline
			&  & $X_{1},X_{2},X_{3}$ & 7.96(0.03) & 0.97(0.03) & 7.95(0.06) & 0.96(0.06) & 7.95(0.03) & 0.97(0.03)\tabularnewline
			\hline 
			\multirow{4}{*}{4} & \multirow{2}{*}{250} & $X_{1}$ & 7.62(0.19) & 0.56(0.26) & 7.62(0.19) & 0.57(0.26) & 7.62(0.19) & 0.56(0.26)\tabularnewline
			&  & $X_{1},X_{2},X_{3}$ & 7.84(0.14) & 0.89(0.12) & 7.81(0.16) & 0.85(0.18) & 7.82(0.22) & 0.89(0.14)\tabularnewline
			& \multirow{2}{*}{1000} & $X_{1}$ & 7.60(0.14) & 0.45(0.20) & 7.59(0.12) & 0.43(0.18) & 7.60(0.14) & 0.45(0.20)\tabularnewline
			&  & $X_{1},X_{2},X_{3}$ & 7.96(0.03) & 0.97(0.03) & 7.90(0.13) & 0.90(0.17) & 7.96(0.04) & 0.97(0.03)\tabularnewline
			\hline 
		\end{tabular}
	}
	\label{tableS8}
\end{table}

\clearpage
\subsection{Mean Squared Error Results for Density Ratio Estimation by Different Calibration Methods}
We evaluate the calibration methods by the difference between the estimated calibration weights and density ratios of the target and source populations.  One way to measure such difference is the mean squared error, which is defined as 
\[
\text{mean squared error} =\frac{1}{N} \sum_{i=1}^N \left\{W(X;\widehat{\lambda}) - f^{\text{t}}(X)/f^{\text{s}}(X)\right\}^2.
\]
Specifically, for Scenarios 3 and 4 in Section 5, we generate a sample of covariates
$X=(X_{1},X_{2},X_{3})^{\T}$ with sample size $N=10^{5}$ from the
source population.  We consider using means of all covariates and only mean of $X_1$ as the summary statistics from the target population. We compute the calibration weights using different methods and the corresponding mean squared errors. The results are summarized in Table \ref{tableS9}.
\begin{table}[!h]
	\def~{\hphantom{0}}
	\caption{Mean squared error results for different calibration methods using different covariates.}
	\vspace{6pt}
		\spacingset{1.45}
		\centering
	\begin{tabular}{ccccccc}
		\hline 
		\multirow{2}{*}{Scenario} & \multirow{2}{*}{Covariate} & Entropy Balancing & Empirical Likelihood & Least Squares\tabularnewline
		&  & &  Mean squared error\tabularnewline
		\hline 
		\multirow{2}{*}{3} & $X_{1}$ & 4.07 & 3.02 & 3.02\tabularnewline
		& $X_{1},X_{2},X_{3}$ & 1.96 & 2.75 & 2.67\tabularnewline
		\hline 
		\multirow{2}{*}{4} & $X_{1}$ & 5.27 & 4.11 & 4.11\tabularnewline
		& $X_{1},X_{2},X_{3}$ & 1.90 & 3.62 & 3.57\tabularnewline
		\hline 
	\end{tabular}
	\label{tableS9}
\end{table}

\end{document}